\documentclass[aps,pre,superscriptaddress,twocolumn]{revtex4-2}
\usepackage{times}
\usepackage{amsmath,amssymb}
\usepackage{graphicx}
\graphicspath{{figures/}{/}}
\usepackage{verbatim}
\usepackage{color}
\usepackage{physics}
\usepackage{float}
\usepackage{placeins}  
\usepackage{flafter}
\usepackage{dsfont} 
\usepackage[utf8]{inputenc}
\usepackage{mathtools}
\usepackage{etoolbox}
\usepackage{braket}
\usepackage{physics}
\usepackage{bbold}
\usepackage{dutchcal}

\newcommand{\pvec}[1]{\vec{#1}\mkern2mu\vphantom{#1}}

\begin{document}

\title{Reduced Dimensional Monte Carlo Method: Preliminary Integrations}
\author{Jarod Tall}
\affiliation{Department of Physics and Astronomy, Washington State University, Pullman, WA 99164-2814 USA}
\author{Steven Tomsovic}
\affiliation{Department of Physics and Astronomy, Washington State University, Pullman, WA 99164-2814 USA}
\date{11/10/2023}
\begin{abstract}

A technique for reducing the number of integrals in a Monte Carlo calculation is introduced.  For integrations relying on classical or mean-field trajectories with local weighting functions, it is possible to integrate analytically at least half of the integration variables prior to setting up the particular Monte Carlo calculation of interest, in some cases more. Proper accounting of invariant phase space structures shows the system's dynamics is reducible into composite stable and unstable degrees of freedom.  Stable degrees of freedom behave locally in the reduced dimensional phase space exactly as an analogous integrable system would. Classification of the unstable degrees of freedom is dependent upon the degree of chaos present in the dynamics. The techniques for deriving the requisite canonical coordinate transformations are developed and shown to block diagonalize the stability matrix into irreducible parts. In doing so, it is demonstrated how to reduce the amount of sampling directions necessary in a Monte Carlo simulation. The technique is illustrated by calculating return probabilities and expectation values for different dynamical regimes of a two-degree-of-freedom coupled quartic oscillator within a classical Wigner method framework. 
	
\end{abstract}

\maketitle 
	
\section{Introduction}
\label{intro}

A considerable body of physics research relies on classical Monte Carlo and molecular dynamics methods to evaluate a wide variety of physical observables~\cite{Hammersley64, Binder86, Newman99, Binder10, Kroese11, Berne77, Berne77b, Alder59, Pathria11}. The applicability of these methods for many-body systems has motivated the extension of their use to approximating the analogous quantum mechanical problem, which is notoriously difficult due to the exponential increase in the size of the Hilbert space with increasing degrees of freedom. For example, molecular dynamics techniques have been applied in approximations of quantum statistical mechanics using an imaginary time path integral representation of the density matrix~\cite{Ceperley95, Cao94, Cao94b, Jang99, Jang99b, Craig04, Craig05, Habershon13, Berne86}. In ultracold atomic physics, mean field approximations such as the truncated Wigner approximation (TWA)~\cite{Polkovnikov10, Steel98, Sinatra01, Sinatra02, Blakie08} involve Hamiltonian-based Monte Carlo methods with incoherently summed classical trajectory contributions.

In addition, there are semiclassical methods that also depend upon classical trajectories, which have the ability to incorporate quantum interference effects~\cite{Keller58, Heller81b, Maslov81}. However, semiclassical methods typically involve root searches (two-point boundary value problems), which in many-degree-of-freedom systems can render the search untenable even with sophisticated techniques~\cite{Tomsovic18b}. Attempts to circumvent this problem, such as Herman-Kluk (HK) propagation~\cite{Herman84}, use a semiclassical initial value representation (SC-IVR)~\cite{Miller70, Miller01}, which converts the root search into a ``shooting problem'' that is ideally suited to a Monte Carlo approach.  This can be carried out coherently or incoherently~\cite{Kay94,Miller01}, but the integrals are more complicated in the coherent case due to the necessity of incorporating the phases. The incoherent case that can be derived from the SC-IVR is a classical Wigner method ~\cite{Miller74, Heller76b, Wang98, Sun98, Hernandez98, Poulsen03}, which is mathematically identical to the TWA. In practice, the implementation of the classical Wigner method amounts to a replacement of the Boltzmann distribution with the initial Wigner function~\cite{Wigner32} and functions of phase space variables with the Weyl symbol of the associated quantum operators~\cite{Ozorio98}, as well as classical propagation of trajectories. 

The intent of this paper is to demonstrate how to exploit invariant structural properties of classical phase space dynamics, e.g.~surfaces determined by constants of the motion, unstable/stable manifolds, and individual trajectories, to determine which degrees of freedom can be integrated analytically before setting up the Monte Carlo calculation. This work is similar in spirit to Kocia and Heller's directed HK propagator~\cite{Kocia15, Kocia14b}, which reduced their calculation to a single integral aligned along the most unstable direction of the local dynamics. Here, significant use is made of a stability analysis along with canonical transformations to establish a more general way of identifying classically invariant structures influence on transport.  The essential method emerges from a previously employed technique to reduce the dimensions in a search of saddles in a complex phase space for a multi-degree-of-freedom semiclassical theory of coherent state propagation~\cite{Tomsovic18b}.  

A key element enabling the analytic integrations is the presence of local weight functions in the expressions to be evaluated.  An important example, to be studied here, are weight functions determined by Wigner transforms of minimum uncertainty quantum states, though the foundational ideas are more general and have the potential to be extended to more general contexts.  Both Gaussian wave packets~\cite{Schrodinger26b} and Glauber coherent states~\cite{Glauber63} for many-body bosonic systems in a quadrature representation lead to localized Gaussian phase space densities and provide ideal examples.  For wave packets, in the short wavelength limit (colloquially, $\hbar\rightarrow 0)$ the density becomes increasingly localized, and similarly for coherent states as particle number tends to infinity ($N\rightarrow \infty$).  Gaussian phase space densities resulting from a purely classical origin would also admit the same techniques, but the volumes of the density would not be determined effectively by the values of $\hbar$ or $N$.

The advantage provided by problems involving local weight functions is directly connected to the presence of dynamically invariant structures in the phase space of the system under the Hamiltonian flow.  Perhaps the most evident example arises from constants of the motion.  The set of all phase points of a given value of some constant of the motion clearly forms an invariant set under the flow.  For systems with multiple constants, subsets with fixed values for the multiple constants also form invariant sets under the flow.  These sets can have quite complicated surfaces in phase space, but over a sufficiently localized region, they can be approximated by their differentials, which gives a linearized description of the local flow characteristics.  

It turns out to be unnecessary to sample canonically conjugate coordinates to constants of the motion, and, in certain cases, the coordinates themselves related to the constants of motion; these are referred to as the stable degrees of freedom. By invoking local canonically conjugate coordinate pairs that behave proportionally or similarly to action-angle coordinates, the stable degrees of freedom decouple from the rest of the phase space. The stable degrees of freedom are completely defined by their shearing rates and once these are known the local dynamics can be understood analytically.  

Another critical example is provided by unstable and stable manifolds of a system's short periodic orbits~\cite{Poincare99}.  Having to self-avoid, they foliate the available chaotic phase space on an arbitrarily fine scale due to ergodicity, and are incredibly complicated from a global structural perspective.  Nevertheless, within a localized phase space region, it is frequently possible to approximate the manifolds locally by their tangents; i.e.~they pass through that particular very localized region in a nearly hyper-planar manner.  The tangent directions are identifiable with the information contained in the stability matrix as discussed in the latter half of Sec.~\ref{ideal}. As initial conditions along stable manifolds must converge towards each other, it is clearly unnecessary to Monte Carlo sample along such directions and they can be analytically integrated prior to the Monte Carlo calculation in all cases considered in this paper. The essential question then is how to identify the specific directions, associated with the stable degrees of freedom, and the stable and unstable manifolds. 
In some cases, the local directions may be knowable locally in an analytic form, e.g.~the Hamiltonian and analytically known constants of the motion may directly provide certain directions.  Otherwise, as these directions are respected by the local linearized flow, the necessary information must be somehow encoded in the stability matrices related to the centroid phase points of the initial and final Wigner densities.  
Indeed, there exist canonical coordinate transformations that block diagonalize the stability matrices.  The space associated with constants of the motion can be separated from the remaining space associated with unstable motion~\cite{Oseledec68}. 

The decoupling of subspaces and block diagonalization procedure just mentioned have the additional benefit of reducing the search space to find the stable/unstable manifolds, since only the reduced stability matrix of the unstable degrees of freedom needs to be considered. The reduced stability matrix can then be approximately diagonalized, with eigenvalues corresponding to contraction and stretching rates of the stable/unstable manifolds. It should be noted that in general transforming the stability matrix requires information of both the final and initial points, and is not done through a similarity transformation as the initial and final coordinate transformations are distinct. The end result of these dynamical considerations is that the directions aligned with each stable manifold's local tangents (in the neighborhood of the localized initial Wigner density) do not need to be sampled in a Monte Carlo simulation. Whether or not stable degrees of freedom need to be sampled will depend on if the dynamics are shearing or rotational and if contributions from the integral are localized in a region of phase space.

Depending on the physical system of study, say perhaps ultracold bosons or atomic, molecular, and optical systems, different classes of observables are most relevant.  The two classes treated here are expectation values, presumably of simple, low rank operators (like a number operator), and transport coefficients~\cite{Zwanzig65}, which go by a variety of names depending on the field of study.  An important example of such a coefficient is the diagonal case that effectively encodes a return probability.  The integrations have to be handled differently for these two classes and ahead they are dealt with separately.

This paper is organized as follows: the next section contains a brief background on Glauber coherent states and Gaussian wave packets, expectation values and transport coefficients within the classical Wigner method (or TWA), and the two-degree-of-freedom coupled quartic oscillator Hamiltonian, which will be used as a model to compare Monte Carlo calculations with and without prior integrations of several variables. This Hamiltonian has a tunable parameter that allows integrable, mixed, and chaotic dynamical regimes to be explored. The chaotic case and integrable case are treated in full, whereas the mixed case creates additional complications that are discussed. The homogeneity of the potential is useful for simplifying the identification of stable coordinates. Section~\ref{ideal} derives the identification of ideal coordinates and a transformation that block diagonalizes the stability matrix relying on canonically conjugate pairs of generalized coordinates. The entire formulation is applied to expectation values in Sec.~\ref{expval} and to transport coefficients in Sec.~\ref{transport}. The formulas are directly tested using the model Hamiltonian and it is shown how the original four dimensional integrals can be reduced depending on the dynamical regime and type of integral being considered. The last section summarizes the results and discusses the outlook of the method to problems with more degrees of freedom. 

\section{Background}
\label{bg}

Two extremely important classes of quantum states leading to localized weight functions are provided by Gaussian wave packets in a Schr\"odinger mechanics context~\cite{Schrodinger26b} and Glauber coherent states in a discrete bosonic field theory context~\cite{Glauber63}.  The former leads to general Gaussian phase space forms under Wigner transformation whose volume inside the two standard deviation surface is given by $h^D$, where $D$ is the number of degrees of freedom; minimum uncertainty states are included as a special class.  The latter constructed in a quadrature variable representation also leads to Gaussian phase space forms after Wigner transformation, although squeezing is necessary to make use of the natural shape parameters; e.g. see the appendix of~\cite{Tomsovic18b}.  Through appropriate scaling an effective $\hbar$ can be introduced, usually most conveniently proportional to the inverse total number of bosons, $N$ (or number per degree of freedom, $N/D$, such as a filling factor). 

Due to the strong dependence of the dimensional reduction procedure on the nature of the dynamics, it is useful to illustrate the technique with a simple model system that possesses at least one constant of the motion, and which has a coupling parameter that can be smoothly tuned to generate any dynamics from integrable to fully chaotic.  It is also helpful to have intermediate coupling values generate various mixed phase space regimes in which significant amounts of both regular and chaotic dynamics exist.  The coupled purely quartic oscillators of~\cite{Bohigas93} provide an ideal example for a number of reasons referred to ahead.  Thus, an introduction to notation, a few basic state properties, and the model system follows.

\subsection{Wave packets, coherent states, and the TWA}
\label{cswp}

In the context of Schr\"odinger quantum mechanics in which the degrees of freedom are those of the particles involved, a general form for a multidimensional Gaussian wave packet can be written as 
\begin{equation}
\begin{aligned}
\phi_\alpha(\vec{q}) = & \braket{\vec q | \alpha}  \qquad \qquad \Bigl{(}\ket{\alpha} = \ket{\vec p_\alpha, \vec q_\alpha, {\bf b}_\alpha}\Bigr{)} \\
= & {\cal N}_\alpha^0 \text{exp}\left[-(\vec{q} - \vec{q}_\alpha)^T\cdot \frac{\mathbf{b}_\alpha}{2\hbar} \cdot(\vec{q} - \vec{q}_\alpha) + \frac{i\vec p_\alpha}{\hbar} \cdot(\vec{q} - \vec{q}_\alpha)\right] \\
& \left( {\cal N}_\alpha^0 = \left[\frac{{\rm Det}\left({\bf b}_\alpha+{\bf b}^*_\alpha\right)}{(2\pi\hbar)^D}\right]^{1/4} \exp\left[\frac{i}{2\hbar}\vec p_\alpha \cdot\vec{q}_\alpha\right] \right)
\end{aligned}
\end{equation}
where $\alpha$ is a convenient shorthand label for the full set of centroid and shape parameters, $\vec p_\alpha$ is the momentum centroid, $\vec q_\alpha$ is the position centroid, the global phase is chosen to match that of the coherent state ahead, and ${\bf b}_\alpha$ is a $D$-dimensional  positive definite, potentially complex, symmetric matrix from which the shape parameters of a hyper-ellipse in the Wigner transform emerge.  In fact, the Wigner transform is given by:
\begin{equation}
\label{wtwp}
\begin{aligned}
&{} \rho_\alpha(\vec{p}, \vec{q}) \\
& = \frac{1}{(2\pi\hbar)^D}\int^{\infty}_{-\infty}
d\vec{x}\hspace{.1cm}\text{e}^{i\vec{p}\cdot\vec{x}/\hbar}\phi_\alpha\Bigr{(}\vec{q} - \frac{\vec{x}}{2}\Big{)}\phi^*_\alpha\Bigr{(}\vec{q} + \frac{\vec{x}}{2}\Bigr{)} \\[6pt]
& = \frac{1}{(\pi\hbar)^{D}}\text{exp}\left[-\begin{pmatrix} \vec{p} - \vec{p}_\alpha \\ \vec{q} - \vec{q}_\alpha \end{pmatrix}^T \cdot \frac{\mathbf{A_\alpha}}{\hbar}\cdot \begin{pmatrix} \vec{p} - \vec{p}_\alpha \\ \vec{q} - \vec{q}_\alpha \end{pmatrix} \right]
\end{aligned}
\end{equation}
where $\mathbf{A}_\alpha$ is
\begin{equation}
\mathbf{A}_\alpha = \begin{pmatrix} \mathbf{c}^{-1} & \mathbf{c}^{-1} \cdot \mathbf{d} \\
\mathbf{d}\cdot\mathbf{c}^{-1} &   \mathbf{c} + \mathbf{d} \cdot \mathbf{c}^{-1} \cdot \mathbf{d} \end{pmatrix}
\end{equation}
with the association
\begin{equation}
\label{symmatrix}
\mathbf{b}_\alpha = \mathbf{c} + i\mathbf{d}
\end{equation}
It follows immediately using $2$ x $2$ block 
relations~\cite{Lu02} that ${\rm Det}\left(\mathbf{A}_\alpha\right) = 1$.  

In the context of second quantization of bosonic fields in which the degrees of freedom relate to the fields and not the particles themselves, it is possible to introduce Glauber coherent states~\cite{Glauber63}. Their usual form for a single degree of freedom is given by:
\begin{equation}
\ket{z} = \exp\left(-\frac{|z|^2}{2} + z\hat{a}^\dagger\right)\ket{0}
\end{equation}
where the ground state represents the absence of any bosons in that field degree of freedom. In an $\hbar_{\rm eff}$ scaled quadrature representation with an index $j$ for each of the $D$-degrees-of-freedom, one has
\begin{equation}
i\sqrt{\frac{2}{\hbar_{\rm eff}}} \hat p_j = \hat a_j - \hat a_j^\dagger \qquad \sqrt{\frac{2}{\hbar_{\rm eff}}}\hat q_j = \hat a_j + \hat a_j^\dagger \ .
\end{equation}
The quadrature operators, $\{\hat p_j, \hat q_j\}$, obey a usual looking commutation relation $[\hat q_j, \hat p_j] = i \hbar_{\rm eff}$. In the mean field limit, the $\{p_j, q_j\}$ correspond to the imaginary and real parts of the mean field, respectively, and play the role of canonically conjugate variables in a Hamiltonian dynamics.  Unlike the Schr\"odinger context $\{\hat p_j, \hat q_j\}$, which correspond to the $j^{th}$ particle's momentum and position, in the second quantized context, despite notational similarity, they have no such interpretation as a particle's momentum or position. Making the association
\begin{equation}
\vec z = \frac{\vec q_\alpha + i \vec p_\alpha}{\sqrt{2\hbar_{\rm eff}}}    
\end{equation}
 and recognizing $\mathbf{b}_\alpha = \mathbb{1}$ leads to the same mathematical relation, Eq.~\eqref{wtwp}, with $\hbar$ replaced by $\hbar_{\rm eff}$ for the coherent state. Irrespective of the different context and interpretations, the localized weight functions for problems involving either wave packets or coherent states have the same forms, and thus all of the results derived ahead apply equally well in either context.

The classical Wigner method (or TWA) are quasi-classical or mean field approximations for quantum observables. For expectation values, it can be expressed as
\begin{align}
\label{wmexp}
\langle f(\mathbf{\hat{p}}, \mathbf{\hat{q}})\rangle (t) & \approx  {\cal F}_W(t)\nonumber \\
{\cal F}_W(t) & = \int  d\vec{p}_0 d\vec{q}_0 \rho_\alpha(\vec{p}_0, \vec{q}_0)f_W(\vec{p}_t, \vec{q}_t)
\end{align}
where ${\cal F}_W(t)$ represents the quasi-classical or mean field approximation, and $f_W(\vec{p}, \vec{q})$ is the Weyl symbol of a general function of the quadrature operators, $f(\mathbf{\hat{p}}, \mathbf{\hat{q}})$. Each individual trajectory starts at some $(\vec{p}_0, \vec{q}_0)$ and arrives at $(\vec{p}_t, \vec{q}_t)$ after a time $t$ of propagation using Hamilton's equations.  For separable functions of $\mathbf{\hat{p}}$ and $\mathbf{\hat{{q}}}$ the Weyl symbol is found by direction substitution of the phase space coordinates: $\mathbf{\hat{{p}}} \rightarrow \vec{p}$ and $\mathbf{\hat{{q}}} \rightarrow \vec{q}$. The substitutions can also be made within small $\hbar$ corrections for general functions \cite{Schlagheck19}.  Similarly, applying the Wigner method to transport coefficients gives,
\begin{align}
\label{wmtc}
|\braket{\beta| \alpha(t)}|^2 & = |\braket{\beta| \hat U(t)| \alpha}|^2 \approx {\cal C}_{\alpha\beta}(t)\nonumber \\
{\cal C}_{\alpha\beta}(t)& = (2\pi\hbar)^D\int d\vec{p}_0 d\vec{q}_0  \rho_\alpha(\vec{p}_0, \vec{q}_0)\rho_\beta(\vec{p}_t, \vec{q}_t)
\end{align}
where ${\cal C}_{\alpha\beta}(t)$ represents the quasi-classical or mean field approximation. The case of $\alpha = \beta$ gives the return probabilities. 

Where invoked, Monte Carlo methods are generally applied to all of the integrals in the Wigner method relations, with the Wigner function providing a natural object for importance sampling. Using the ideas presented in Secs.~\ref{ideal} and \ref{derive}, results are derived for transport coefficients in Sec.~\ref{transport}. It is demonstrated that it is possible, and a good approximation, to integrate analytically all the directions excluding the tangent directions of the unstable manifold, i.e.~the directions of maximal exponential stretching.  Thus, only those must necessarily be sampled in the Monte Carlo simulation.  As these directions are associated with the unstable manifold, local initial conditions sampled on this manifold maximize the exponential rate of exploration of the available (ergodic part of the) phase space.  In a sense, this maximizes the efficiency of the Monte Carlo simulation as no effort is wasted on sampling initial condition variations that can be pre-integrated.

On the other hand, expectation values of simple observables, such as occupation numbers, require a modified approach as compared with the aforementioned transport coefficients. Mathematically, this is due to the presence of $\hbar$ in the exponential of $\rho_\beta(\vec{p}_t, \vec{q}_t)$ not admitting a power series expansion in the phase space variables in the same way as the function $f_W(\vec{p}, \vec{q})$, which is straightforwardly expanded.  This distinction is behind the fact that transport coefficients only have contributions to the integral in local regions of phase space, whereas expectation values contribute globally. As a result, it is demonstrated in Sec.~\ref{expval} that for expectation values fewer degrees of freedom can be approximated by analytical integration beforehand.  In this case, if the stable degrees of freedom correspond to shearing dynamics, rather than rotational dynamics, the coordinates corresponding to the constants of the motion must also be sampled in the Monte Carlo simulation. 

\subsection{Two coupled pure quartic oscillators}
\label{tcqo}

The two-degree-of-freedom coupled quartic oscillator discussed in Ref.~\cite{Bohigas93} provides a suitable toy model for illustrating the method. It includes an example of both a stable degree of freedom defined by the energy, and in the chaotic dynamical case, an unstable degree of freedom.  The Hamiltonian is given as follows:
\begin{equation}
\label{ham}
H = \frac{p_1^2}{2m} + \frac{p_2^2}{2m}  + q^4_1 / b + q^4_2b + 2\lambda q_1^2q_2^2
\end{equation} 
The parameter $\lambda$ can be continuously adjusted from the integrable case of $\lambda = 0$ to the completely chaotic case  (in the sense that the vast majority of trajectories explore the entire energy surface) of $\lambda = -.60$. The choice of $b \neq 1$ reduces the symmetries of the system. This Hamiltonian has the additional benefit of having a homogeneous potential which allows for a simple analytic technique to be used to derive the necessary coordinate transformations. 

The phase space of this system has four dimensions, which is the dimensionality that the usual Monte Carlo calculation would sample.  For the integrable case, $\lambda=0$, when dealing with transport coefficients, it is possible to integrate out all the variables leading to a `zero-dimensional' Monte Carlo.  For the fully chaotic case, the two variables associated with the energy constant of the motion can be integrated out as well as the variable tangent locally to the stable manifold, leaving only a single direction to sample, a dimensional reduction from four to one. For expectation values, only the directions along stable manifold and the coordinate conjugate to the energy constant of the motion can be integrated beforehand if the dynamics are shearing. In the case of rotational motion the shearing rate is zero and the constant of the motion coordinate can also be pre-integrated. Likewise, for mixed phase space systems, localized densities in the chaotic region would be similar to the chaotic case, whereas inside a regular island would be like the integrable case. However, the presence of transport barriers in the `chaotic sea' and unknown constants of motion make this case significantly more difficult to handle.

To approximate the integrals, a small $\hbar$ approximation is made. However, as described in~\cite{Bohigas93}, there is an equivalence between a small $\hbar$ expansion and large energy expansion. For this Hamiltonian the equivalence has the form, $\hbar E^{3/4}= $  constant. Therefore, for many of the calculations $\hbar$ is set to unity with the understanding that the energy is large enough for the approximations (linearizations) to be valid.  Nevertheless, $\hbar\ne 1$ is taken in several of the calculations to illustrate directly the effect of changing $\hbar$ on the accuracy of the approximations.

\section{Local dynamics and homological decomposition of the stability matrix}
\label{ideal}

Consider the phase points in the neighborhood of the initial localized density centroid: $\delta\vec{q}_0 = \vec{q}_0 - \vec{q}_\alpha$, and similarly for $\vec p_\alpha$. How this local point propagates relative to the orbit starting at $(\vec{q}_\alpha, \vec{p}_\alpha)$ is dictated by the stability matrix, $\mathbf{M}_{\alpha,t}$. Thus,
\begin{equation}
\begin{pmatrix}
\delta \vec{p}_t \\ \delta \vec{q}_t
\end{pmatrix} = \mathbf{M}_{\alpha,t} \begin{pmatrix}
\delta \vec{p}_0 \\ \delta \vec{q}_0
\end{pmatrix} = \begin{pmatrix} \mathbf{M_{11}} & \mathbf{M_{12}} \\ \mathbf{M_{21}} & \mathbf{M_{22}} \end{pmatrix}_{\alpha,t} \begin{pmatrix}
\delta \vec{p}_0 \\ \delta \vec{q}_0
\end{pmatrix}
\end{equation}
where $\delta\vec{q}_t = \vec{q}_t(\vec q_0, \vec p_0) - \vec{q}_t(\vec q_\alpha, \vec p_\alpha)$, and the stability matrix, defined by a local linearization, depends on the properties of the central orbit and propagation time only. The stability matrix is a multiplicative cocycle, and $\mathbf{M}'_{\alpha,t}$ and $\mathbf{M}_{\alpha,t}$ are said to be cohomologous if canonical transformations $\mathbf{S}_{\alpha}$ exist such that~\cite{Arnold99}:
\begin{equation}
\label{lh}
\mathbf{M}'_{\alpha,t} = \mathbf{S}_{\alpha,t} \mathbf{M}_{\alpha,t} \mathbf{S}^{-1}_{\alpha,0}
\end{equation}
This transformation is also referred to as a Lyapunov homology \cite{Gaspard98}. It should be emphasized that Eq.~\eqref{lh} is not a similarity transformation, since it is necessary to evaluate $\mathbf{S}_{\alpha}$ at both the initial and final points. The local position and momenta in the new coordinate system are related to the old ones as follows:
\begin{equation}
\label{eq:trans}
\begin{pmatrix}
\delta \vec{P}_t \\ \delta \vec{Q}_t
\end{pmatrix} = \mathbf{S}_{\alpha,t} \begin{pmatrix}
 \delta \vec{p}_t \\ \delta \vec{q}_t
\end{pmatrix} \hspace{.5cm} \begin{pmatrix}
\delta \vec{P}_0 \\ \delta \vec{Q}_0
\end{pmatrix} = \mathbf{S}_{\alpha,0} \begin{pmatrix}
 \delta \vec{p}_0 \\ \delta \vec{q}_0
\end{pmatrix}  
\end{equation}
In order for Eq~\eqref{eq:trans} to be a canonical transformation, $\mathbf{S}_{\alpha}$ is required to be a symplectic matrix. $\mathbf{S}_{\alpha}$ is then the Jacobian between the coordinate systems and Poisson brackets between the local position and momenta are given in terms of its matrix elements.
	
The Multiplicative Ergodic Theorem of Oseledec~\cite{Oseledec68} implies the tangent space of the stability matrix can be reduced into subspaces of stable and unstable dynamics, with vanishing and non vanishing Lyapunov exponents respectively (see also \cite{Gaspard98} and references therein). More recent refinements prove that it is always possible to find a Lyapunov homology of the stability matrix that is at least an upper triangular matrix \cite{Arnold99}. One of the aims of this paper is to construct a local action-angle-like coordinate system that is decoupled from the locally unstable dynamics, resulting in a block diagonal stability matrix. The standard method of decomposing the stability matrix is to first recognize that every constant of the motion implies two directions of vanishing Lyapunov exponents. For a definite example, consider the case where the energy is a constant of the motion, then the directions along the trajectory and perpendicular to the energy surface have vanishing Lyapunov exponents. The components of these directions are given by $\mathbf{\Omega} \cdot \gradient H$ and $\gradient H$, where $\mathbf{\Omega}$ is the symplectic matrix,
\[\mathbf{\Omega} = \begin{pmatrix} \mathbf{0} & -\mathbf{1} \\ \mathbf{1} & \mathbf{0} \end{pmatrix}. \]
However, these two directions are insufficient for constructing a local action-angle-like coordinate system. To understand this, consider making a global canonical coordinate transformation where a conjugate pair of position and momentum are replaced by the time and energy. This can be done by taking the action as a generating function; see Chap.~9 of~\cite{Haake10} for a proof. Locally, these coordinates can be written in terms of the original coordinates as:
\begin{align}
\label{TE1}
& \delta E = \frac{\partial{E}}{\partial{\vec{p}}}\delta \vec{p} + \frac{\partial{E}}{\partial{\vec{q}}}\delta \vec{q} \\
\label{TE2}
& \delta t = \frac{\partial{t}}{\partial{\vec{p}}}\delta \vec{p} + \frac{\partial{t}}{\partial{\vec{q}}}\delta \vec{q}
\end{align}
If the local deviations are taken to be the basis vectors, then the components of $\delta E$ are given by $\gradient H$ and the energy coordinate represents the direction perpendicular to the energy surface. On the contrary, the components of the conjugate coordinate, $\delta t$, do not correspond to $\mathbf{\Omega} \cdot \gradient H$. In fact, it can be considered the dual to this direction, as
\begin{align}
\frac{\partial}{\partial{t}} & = \frac{\partial{\vec{p}}}{\partial{{t}}}\Bigr|_E \frac{\partial{}}{\partial{\vec{p}}} + \frac{\partial{\vec{q}}}{\partial{{t}}}\Bigr|_E \frac{\partial{}}{\partial{\vec{q}}} \\
& = -\frac{\partial{H}}{\partial{\vec{q}}}\frac{\partial{}}{\partial{\vec{p}}} + \frac{\partial{H}}{\partial{\vec{p}}}\frac{\partial{}}{\partial{\vec{q}}} \nonumber
\end{align}
does have components that correspond to $\mathbf{\Omega} \cdot \gradient H$. From Eqs.~\eqref{TE1} and \eqref{TE2}, it is evident that knowledge of both $E(\vec{p}, \vec{q})$ and $t(\vec{p}, \vec{q})$ are necessary to derive the desired coordinate system for the $(\delta E, \delta t)$ pair. For a general chaotic system, $t(\vec{p}, \vec{q})$ would not be known analytically making it necessary to construct the local canonical coordinate system numerically. In Sec.~\ref{derive}, a method is developed to construct the coordinate system analytically for systems with homogeneous potentials, thus simplifying the application to the chaotic two-dimensional quartic oscillator considered in this paper.

The preceding discussion can be generalized to a system with $n$ constants of the motion. For every constant, $I_j(\vec{p}, \vec{q})$, a conjugate pair of local coordinates, $(\delta I_j, \delta \theta_j)$, can be constructed. The stability matrix in this coordinate system would contain $n$ $2\times 2$ blocks, corresponding to each local coordinate pair, and a block with no specific structure that contains all the exponentially stretching and contracting directions. Each $2\times 2$ block would have the following form: 
\begin{equation}
\label{shear}
\begin{pmatrix}
\delta {I}_t \\ \delta \theta_t
\end{pmatrix} = \begin{pmatrix} 1 & 0 \\ \omega' t & 1 \end{pmatrix} \begin{pmatrix}
\delta I_0 \\ \delta \theta_0
\end{pmatrix}
\end{equation}
A stability matrix of this form, with $\omega' \neq 0$, is said to represent a shearing degree of freedom and $\omega'$ gives the shearing rate. The case of $\omega' = 0$ gives a rotational degree of freedom. 

Once a transformation that decouples stable and unstable degrees of freedom is found, the reduced dimensional stability matrix of the unstable block can be handled seperately. This subspace contains all directions of exponential stretching and contraction and is locally hyperbolic, defined by its non-zero finite-time stability exponents. These exponents can be found by diagonalizing $\mathbf{M}^h_{t}(\mathbf{M}^h_t)^T$, where $\mathbf{M}^h_{t}$ is the stability matrix of the hyperbolic block:

\begin{equation}
\mathbf{O}_t\mathbf{M}^h_{t}(\mathbf{M}^h_t)^T\mathbf{O}_t^T = \mathbf{\Lambda}_t \ .
\end{equation}
Due to the symplectic nature of the stability matrix, $\mathbf{O}_t$ is an orthogonal-symplectic matrix and the eigenvalues come in pairs, such that the $j^{th}$ block of $\mathbf{\Lambda}_t$ has the form:
\begin{equation}
\mathbf{\Lambda}_{j,t} = \begin{pmatrix} \text{e}^{-2\lambda_jt} & 0 \\ 0 & \text{e}^{2\lambda_jt} \end{pmatrix}
\end{equation}
where the $\lambda_j$ are positive definite.
For a large class of dynamical systems, although the $\lambda_j$ exhibit potentially large finite-time fluctuations~\cite{Wolfson01,Tomsovic07}, in the long time limit they converge to the Lyapunov exponent spectrum associated with the unstable block;  the time, for the purposes of this part of the discussion, is assumed long enough that the $\lambda_j$ have sufficiently well converged. In order to restrict Monte Carlo integrations only to the unstable manifolds, it is necessary to have a transformation of the form of Eq.~\eqref{lh} that diagonalizes the hyperbolic stability matrix. One approach is to follow~\cite{Tomsovic18b}, where directions of initial conditions that evolve into the eigenvectors of $\mathbf{M}^h_{t}(\mathbf{M}^h_t)^T$ can be identified by multiplying the eigenvectors by the inverse of the stability matrix and normalizing the directions:
\begin{equation}
\label{direct}
\mathbf{O}^T_0 = 
(\mathbf{M}^h_t)^{-1} \mathbf{O}_t^T \sqrt{\mathbf{\Lambda}_t} 
\end{equation}
$\sqrt{\mathbf{\Lambda}_t}$ gives the normalization of the directions. The sign of $\sqrt{\mathbf{\Lambda}_t}$ is determined by the condition that the stability matrix is the identity at $t=0$. It is clear that $\mathbf{O}_0$ diagonalizes $(\mathbf{M}^h_{t})^T\mathbf{M}^h_t$ and is also an orthogonal-symplectic matrix. Applying Eq.~\eqref{lh} then gives:
\begin{equation}
\label{direct2}
\mathbf{O}_{t} \mathbf{M}^h_{t} \mathbf{O}^{T}_{0} = \sqrt{\mathbf{\Lambda}_t}
\end{equation}
This decomposition is not an exact Lyapunov homology because $\mathbf{O}_{t}$ and $\mathbf{O}_{0}$ are only related through a linearization of the dynamics and  because $\mathbf{O}_{0}$, as given by Eq.~\eqref{direct}, changes with time. The former restriction is not an issue as this paper is only concerned with situations where a linearization of the local dynamics is appropriate. The latter restriction is an issue, since it implies the canonical coordinate transformation made at $t=0$ is changing with time. However, the directions given by $\mathbf{O}_{0}$ stabilize quickly in numerical calculations in a number of different dynamical systems.  Presumably, they are stabilized on the actual invariant manifolds defined by the ``infinite-time'' dynamics. Any time greater than the time at which this stabilization takes place, $t_c$ generates the same results. It is at this time the initial directions are defined:
\begin{equation}
\label{tcf}
\mathbf{O}^T_0 = 
(\mathbf{M}^h_{t_c})^{-1} \mathbf{O}_{t_c}^T \sqrt{\mathbf{\Lambda}_{t_c}} 
\end{equation}
Furthermore, it is expected that any deviation from the unstable manifold, unless directly along the stable manifold, soon collapses back on to the unstable manifold. This logic leads to the conclusion that even for times $t < t_c$, Eq.~\eqref{tcf} may still be a suitable approximation. Numerical results from the preliminary integrations presented in later sections support this idea.

\section{Local canonical coordinates for homogeneous constants of the motion}
\label{derive}

There is a straightforward method for constructing the desired local canonical transformation of the coordinates if the constant of the motion is a sum of separable homogeneous functions. This is the case for the coupled quartic oscillators considered ahead. As briefly mentioned in the background section, given a constant of the motion $I_j(\vec{p},\vec{q})$, a local deviation is defined by its gradient:

\begin{equation}
\label{action}
\delta I_j = \frac{\partial{I_j}}{\partial{\vec{p}}} \delta \vec{p} + \frac{\partial{I_j}}{\partial{\vec{q}}} \delta \vec{q} 
\end{equation}
For each constant there is a conjugate coordinate, $\delta \theta_j$, that satisfies $[\delta \theta_j, \delta I_j] = 1$.  In general, the direction that gives the conjugate coordinate is not trivial as it would require knowledge of $\theta_j$ as a function of the trajectories; note that generating functions are implicit, not explicit, as they involve a combination of old and new coordinates. However, for the case when $I_j$ can be written as a sum of separable homogeneous functions, there exists a method for identifying this direction. The idea makes use of Euler's homogeneous function theorem, which states for a homogeneous function $g(\vec{x})$ of degree $k$ the following is true:
\begin{equation}
\label{euler}
\frac{\vec{x}}{k} \cdot \gradient g(\vec{x}) = g(\vec{x})
\end{equation}
This equation can be put into the language of Poisson brackets using the symplectic matrix.  Let the notation, $\vec{x}$ denote all the phase space variables: $\vec{x} = (p_1,...,p_N,q_1,...,q_N)$. Define a new function, ${X}$, such that its variation is:
\begin{equation}
\delta {X} =  \frac{\vec{x}}{k} \cdot \mathbf{\Omega}  \cdot \delta \vec{x}
\end{equation}
Equation \eqref{euler} can then be written as:
\begin{equation}
\label{euler2}
[\delta {X}, \delta g] = g 
\end{equation} 
Assume the constant of the motion is a sum of separable homogeneous functions of degree $k_i$, such that, $\delta I_j = \sum_i \delta g_i$. $\delta X_j$ can be defined in a similar fashion: $\delta X_j = \sum_i\mathbf{\Omega} \frac{\vec{x}_i}{k_i} \cdot \delta \vec{x}$, where $\vec{x}_i$ has zeroes for coordinates irrelevant to $g_i$. Then using Eq. (~\ref{euler2}) and summing over $i$ gives,
\begin{equation}
\begin{aligned}
&{} [\delta X_j, \delta I_j] = \sum_i[\delta X_i, \delta g_i]  = \sum_i g_i = I_j \rightarrow \\[.25cm] 
& \left[\frac{1}{I_j}\delta X_j, \delta I_j\right] = 1\ ,
\end{aligned}
\end{equation} 
where the separability conditions of $\delta I_j$ and $\delta X_j$ are used. Thus, $\delta \theta_j$ is given by:
\begin{equation}
\label{angle}
\delta \theta_j = \frac{1}{I_j}\delta X_j = \frac{1}{I_j}\sum_i\mathbf{\Omega} \frac{\vec{x}_i}{k_i} \cdot \delta \vec{x} \ .
\end{equation} 

It is instructive to look at a uni-dimensional quartic oscillator example for which
\begin{equation}
\label{1dquartic}
H = \frac{p^2}{2} + q^4 
\end{equation}
This system can be solved exactly in action angle coordinates, $(\delta I, \delta \theta)$, and the transformed stability matrix is known analytically. Constructing the local action-angle coordinate system requires the linear transformation from $(\delta p, \delta q) \rightarrow (\delta I, \delta \theta)$ that gives this stability matrix.  However, it is convenient to work in the time-energy coordinate system given by a dilation with the frequency,
\begin{equation}
(\delta E, \delta t) = \left(\omega\delta I, \frac{\delta \theta}{\omega} \right) \ .
\end{equation}
The energy-time coordinate system can be implemented whether or not an analytic form of the action coordinate is known, and is later used for the chaotic two-dimensional quartic oscillator. The direction that gives the energy coordinate is just the gradient of the Hamiltonian,
\begin{equation}
\delta E = \frac{\partial E}{\partial p}\delta p + \frac{\partial E}{\partial q}\delta q = {p}\delta p + 4q^3\delta q 
\end{equation}
It is understood that the variations are evaluated along the central trajectory. The time coordinate can be found by applying Eq. (\ref{angle}), 
\begin{equation}
\begin{aligned}
&{} \delta t = \frac{1}{E} \begin{pmatrix} \frac{p}{2}, & 0  \end{pmatrix} \begin{pmatrix} 0 & 1 \\ -1 & 0 \end{pmatrix} \begin{pmatrix} \delta p \\ \delta q  \end{pmatrix} + \frac{1}{E}\begin{pmatrix} 0,& \frac{q}{4}  \end{pmatrix} \begin{pmatrix} 0 & 1 \\ -1 & 0 \end{pmatrix} \begin{pmatrix} \delta p \\ \delta q  \end{pmatrix}   \\[.5cm]
& = - \frac{q}{4E}\delta p + \frac{p}{2E}\delta q 
\end{aligned}
\end{equation}
The transformation that takes $(\delta p, \delta q)$ to $(\delta E, \delta t)$ is then given by:

\begin{equation}
\label{1d}
\mathbf{S}_\alpha= \begin{pmatrix}
p & 4q^3 \\[3pt] -\frac{q}{4E} & \frac{p}{2E} \end{pmatrix} 
\end{equation}
Using Eq. (\ref{lh}) the new stability matrix is: 
\begin{equation}
\mathbf{M}'_{\alpha,t} = \begin{pmatrix} p_t & 4q_t^3 \\[3pt] -\frac{q_t}{4E} & \frac{p_t}{2E} \end{pmatrix} \mathbf{M}_{\alpha,t} \begin{pmatrix} \frac{p_0}{2E} & -4q_0^3 \\[3pt] \frac{q_0}{4E} & p_0 \end{pmatrix}
\end{equation}
By finding $\mathbf{M}_{\alpha,t}$ numerically, it can be shown that $\mathbf{M}'_{\alpha,t}$ is exactly in the form of Eq. (\ref{shear}) with $\omega'$ being the dilated shearing rate,
\begin{equation}
\omega' = \frac{1}{\omega^2}\frac{\partial^2{H}}{\partial{I}^2} \ .
\end{equation}
For a system with additional degrees of freedom, it is necessary to isolate the subspace that defines the unstable block of the stability matrix. In Appendix~\ref{transformation} one such transformation is given for the two-dimensional quartic oscillator, which renders its stability matrices block diagonal with the time and energy coordinates explicitly given.  It remains only to diagonalize the portion of the stability matrix in the hyperbolic subspace as discussed in Sec.~\ref{ideal}.

\section{Expectation Values}
\label{expval}

Generally speaking, expectation values lead to integrals over phase space functions which vary smoothly on a classical scale, often low order polynomials in momentum and position variables.  This distinguishes them from transport coefficients treated ahead, in which the final state density leads to structure on the finer scale, $\hbar$. It turns out to be possible to pre-integrate analytically at least half of the variables prior to setting up the Monte Carlo calculation.  In some cases, mentioned just ahead, it is possible to integrate even more variables.  

Consider first an integrable system possessing $D$ constants of the motion, for which it is convenient to canonically transform the system to action-angle coordinates.  Any small change in any one of the initial angle coordinates leads to a similar small change in the corresponding final angle coordinate and it turns out possible to integrate all of the angle coordinates analytically.  In essence, there is no spreading in the final endpoints of the angle variables.  

If there are fewer constants of the motion, it is still possible to integrate their canonically conjugate variables, but in addition, it is also possible to integrate the stable manifold associated with the hyperbolic degrees of freedom.  Because this manifold contracts exponentially, its integral leads to a constant plus an exponentially decaying contribution.  Unless the finite-time stability exponents are quite small, it is quickly safe to ignore the exponentially decaying contribution, and only the overall normalization is affected by the stable manifold integrations.  The total number of conjugate variables plus the dimension of the stable manifold adds up to half the total phase space dimension.  Finally, for constants of the motion leading to local rotational motion, where the shearing of the dynamics vanishes, it is also possible to integrate the coordinate associated with each such constant's gradient.

\subsection{Derivation of reduced dimensionality formula for expectation values}

Expectation values within a classical Wigner method framework are given by Eq.~\eqref{wmexp}:
\begin{equation}
\begin{aligned}
\label{int}
&{} \langle f(\mathbf{\hat{p}}, \mathbf{\hat{q}})\rangle (t) = \int d\vec{q}_0d\vec{p}_0 f_W(\vec{p}_t,\vec{q}_t) \rho_{\alpha}(\vec{p}_0, \vec{q}_0) \\
& \rho_{\alpha}(\vec{p}_0, \vec{q}_0) = \\
& \frac{1}{(\pi\hbar)^{D}}\text{exp}\Bigr{[}-(\vec{p}_0 - \vec{q}_\alpha, \vec{q}_0 - \vec{q}_\alpha)\cdot \frac{\mathbf{A_\alpha}}{\hbar} \cdot \begin{pmatrix} \vec{p}_0 - \vec{p}_\alpha \\ \vec{q}_0 - \vec{q}_\alpha \end{pmatrix} \Bigr{]}
\end{aligned}
\end{equation}
Due to the presence of $\hbar$ in $\rho_{\alpha}(\vec{p}_0, \vec{q}_0)$, the significant contributions come from trajectory initial conditions highly localized around the central trajectory, $(\vec{p}_\alpha, \vec{q}_\alpha)$.  This suggests first shifting the coordinate system to local coordinates as follows:
\begin{equation}
(\vec{p}_0 - \vec{p}_\alpha, \vec{q}_0 - \vec{q}_\alpha) = (\delta \vec{p}_0, \delta \vec{q}_0), \hspace{.5cm} d\vec{q}_0d\vec{p}_0 \rightarrow d\delta \vec{q}_0 d\delta \vec{p}_0
\end{equation}
For expectation values there may be circumstances in which it is worth understanding the coordinate system in which $\mathbf{A}_\alpha$ transforms to unity; see Appendix~\ref{cc}.

As discussed in Secs.~\ref{ideal} and \ref{derive}, there are canonical coordinate transformations defined by directions associated with coordinate pairs for each constant of the motion and stable/unstable manifolds; see Eq.~\eqref{eq:trans}.  The goal is to evaluate the integrals along the  directions corresponding to the stable manifolds and the local coordinates conjugate to the constants of the motion. Given that the number of coordinates, $j$, that cannot be integrated analytically, i.e. coordinates along unstable manifolds or associated with constants of motion with shearing, that leaves $2D - j$ coordinates to integrate analytically.  Recall that for rotational dynamics, $\omega' = 0$, both coordinates remain localized and can be analytically integrated.  

For clarity and reordering the coordinates, the following change of notation is made: $(\delta \vec{P}, \delta \vec{Q}) \rightarrow  \delta \vec{X} = (\delta \vec{X}^{2D-j}, \delta \vec{X}^j)$, where $\delta \vec{X}^{2D-j}$ are the $2D-j$ coordinates to be integrated analytically, and $\delta \vec{X}^j$ contains the remaining $j$ coordinates to be used in the final Monte Carlo method. The reordering of the coordinates is done simply by appropriately exchanging rows in $\mathbf{S}_{\alpha}$ to generate an $\mathbf{S'}_{\alpha}$. In this coordinate system the equation for the expectation value becomes
\begin{equation}
\label{eq:int2}
{\cal F}_W(t) = \int {\rm d}\delta \vec{X}_0 f_W\left(\vec{p}_t(\delta \vec{X}_0),\vec{q}_t(\delta \vec{X}_0)\right) \rho_{\alpha}(\delta \vec{X}_0) 
\end{equation}
with the Wigner function:
\begin{equation}
\begin{aligned}
\label{rho}
\rho_\alpha(\delta \vec{X}_0) = & \frac{1}{(\pi\hbar)^{D}}\text{exp}\Bigr{[}-(\delta \vec{X}_0)\cdot \frac{\mathbf{A'_\alpha}}{\hbar} \cdot (\delta \vec{X}_0) \Bigr{]} \\
& \text{with} \hspace{.3cm} \mathbf{A}'_\alpha = (\mathbf{S'}^{-1}_{\alpha,0})^T\mathbf{A}_{\alpha} \mathbf{S'}^{-1}_{\alpha,0}
\end{aligned}
\end{equation}
Although $\vec{p}_t$ and $\vec{q}_t$ explicitly depend on the central orbit initial conditions and the variations to be integrated analytically, $\delta \vec{X}^{2D-j}_0$, this coordinate system allows for these coordinate integrations.  First, note that variations in $\delta \vec{X}^{2D-j}_0$ do not induce any later time variations in $\delta \vec{X}^{j}_t$. Second the variations later in time, $\delta \vec{X}^{2D-j}_t$, remain small and thus the function $f_W(\vec{q}_t, \vec{p}_t)$ can be expanded about $\delta \vec{X}^{2D-j}_0$ at a time $t$, while holding $\delta \vec{X}^{j}_0$ fixed, in two steps. First expand $f_W$ to quadratic order in variations at a fixed trajectory's endpoint at time $t$,
\begin{align} 
\label{expand}
f_W \approx f_j + & \left(\frac{\partial f_W}{\partial \vec{X}^{2D-j}_t}\right)^T \cdot \delta \vec{X}^{2D-j}_t \nonumber \\
& + (\delta \vec{X}^{2D-j}_t)^T \cdot \frac{\partial^2 f_W}{2\partial\vec{X}^{2D-j}_t\partial\vec{X}^{2D-j}_t} \cdot \delta \vec{X}^{2D-j}_t
\end{align}
where the shorthand $f_j$ is introduced,
\begin{equation}
f_j \equiv f_W\left(\vec{p}_t(\delta \vec{X}^{2D-j}_0 =0, \delta \vec{X}^{j}_0), \vec{q}_t(\delta \vec{X}^{2D-j}_0 =0, \delta \vec{X}^{j}_0)\right)
\end{equation}
This should be thought of as there being a trajectory with the initial conditions $(\vec{p}_j, \vec{q}_j)$ that differs from $(\vec{p}_{\alpha}, \vec{q}_{\alpha})$ only by variations associated with $\delta \vec{X}^j_0$. 

Under many circumstances it may be possible to evaluate the derivatives as expressed. Note that it is also possible to evaluate the derivatives using the local canonical coordinate transformation at the trajectory endpoint (the $j$ subscript indicates that it is one of the trajectories chosen with $\delta \vec{X}^{j}_0$ fixed), i.e.
\begin{equation}
\delta \vec{X}_t = \mathbf{S'}_{j,t}    \begin{pmatrix}
\delta \vec{p}_t \\
\delta \vec{q}_t
\end{pmatrix}
\end{equation}
with the appropriate ordering of the coordinates.  That leads to the alternate expression
\begin{equation}
\frac{\partial f_W}{\partial \vec{X}_t} = (\mathbf{S'}_{j,t}^{-1})^T \mathbb{\nabla}_{p_t,q_t} f_W
\end{equation}
In a similar way, but accounting for the chain rule, the full $2D \times 2D$ matrix of second derivatives is given by
\begin{equation}
\frac{\partial^2 f_W}{\partial \vec{X}_t \partial \vec{X}_t} = (\mathbf{S'}_{j,t}^{-1})^T  \nabla_{p_t,q_t} \left[ (\mathbf{S'}^{-1}_{j,t})^{T} \cdot \nabla_{p_t,q_t}\right]^T f_W 
\end{equation}
where the second derivative acts on the transformation as well.
Ahead the $(2D-j) \times (2D-j)$ submatrix of these derivatives appearing in Eq.~\eqref{expand} are denoted by $\mathbf{F}$ for brevity.

The second step is to connect the integration variables, $\delta \vec{X}^{2D-j}_0$, to the variations, $\delta \vec{X}^{2D-j}_t$, found in Eq.~\eqref{expand}.  Following Eq.~\eqref{lh}, the subsequent discussion, and noting that the coefficients in Eq.~\eqref{expand} depend only on $\delta \vec{X}^{j}_0$, it turns out that there is a very simple relationship.  The stability matrix decouples the block associated with $\delta \vec{X}^{2D-j}_0$ from the block associated with $\delta \vec{X}^{j}_0$.  Furthermore, the $(2D-j) \times (2D-j)$ stability matrix block is diagonal, taking on the value of unity for each of the canonically conjugate variables to some constant of the motion (and unity for any additional variables for rotational motion) and $e^{-\lambda_n t}$ for the $n^{th}$ component of the stable manifold, where $\lambda_n$ is the finite time stability exponent.  This diagonal matrix is denoted $\mathbf{M}_t$ such that:
\begin{align}
\label{diagM}
    \delta\vec{X}^{2D-j}_t = \mathbf{M}_t \delta \vec{X}^{2D-j}_0
\end{align}
and this substitution is made in Eq.~\eqref{expand}. Note the prime indicating it is the transformed stability matrix has been dropped for notational simplicity. The subscript indicating on which trajectory it is to be evaluated has also been dropped because going forward it is either independent of the trajectory or always dependent on the central trajectory.  Given the $\hbar$-dependence for the argument of the exponential in Eq.~\eqref{rho}, after integration of the $\delta \vec{X}^{2D-j}_0$ variations ahead, the expansion of Eq.~\eqref{expand} is generally seen to be an expansion in powers of $\hbar^{1/2}$, but for the stable manifold variations, it turns out to be an expansion in $\hbar^{1/2}e^{-\lambda_n t}$.  For these directions, unless $\lambda_n$ is exceedingly small, only the leading term, $f_j$, remains after a very short propagation time.  Presumably, those correction terms can usually be safely ignored.  However, it is long known that finite-time stability exponents (or finite-time Lyapunov exponents) exhibit rather large fluctuations~\cite{Ott02,Sepulveda89,Grassberger88}, and therefore for some of the trajectories in the Monte Carlo, those corrections terms could be potentially relevant.  However, if the measure of such terms over the $\delta \vec{X}^{j}_0$ variations is small, the correction terms can continue to be dropped and the only critical correction contributions come from the conjugate variables to the constants of motion. For short times, the approximation $e^{-\lambda_n t} \approx 0$ is not valid. Though these corrections are of order $\hbar^{1/2}$ and in the results ahead the short time error was not seen to be noticeably larger than the error at longer times. Furthermore, if it was desired to have extremely accurate short term behavior, it can be approximated by a linearization of the dynamics in the $(\vec{p}, \vec{q})$ coordinate system, in which case all the integrals can be performed. This technique is employed to capture the initial decay of return probabilities in Sec.~\ref{transport}.

Introduction of the stability matrix into Eq.~\ref{expand} gives:
\begin{equation}
f_W = f_j + \pvec{f}_1 \cdot \delta \vec{X}^{2D-j}_0 + (\delta \vec{X}^{2D-j}_0)^T \cdot \frac{\mathbf{F'}}{2} \cdot \delta \vec{X}^{2D-j}_0
\end{equation}
with
\begin{equation}
\pvec{f}_1 = \left(\frac{\partial f_W}{\partial \vec{X}^{2D-j}_t}\right)^T \cdot \mathbf{M}_t \ \text{and}\ \ \mathbf{F'} = \mathbf{M}_t^T\mathbf{F}\mathbf{M}_t \ .
\end{equation}
The integral for the expectation value then becomes:
\begin{equation}
\begin{aligned}
\label{int2}
& {\cal F}_W(t) =  \frac{1}{(\pi \hbar)^D} \int d \delta \vec{X}^{j}_0  \int d \delta \vec{X}^{2D-j}_0  \\[5pt]& \times \hspace{.1cm} \text{exp}\Bigr{[}-(\delta \vec{X}^{2D-j}_0, \delta \vec{X}^j_0)\cdot \frac{\mathbf{A'_\alpha}}{\hbar} \cdot (\delta \vec{X}^{2D-j}_0, \delta \vec{X}^j_0) \Bigr{]} \\[5pt]
&\times \hspace{.1cm} \left(f_j + \pvec{f}_1 \cdot \delta \vec{X}^{2D-j}_0 + (\delta \vec{X}^{2D-j}_0) \cdot \frac{\mathbf{F'}}{2} \cdot \delta \vec{X}^{2D-j}_0\right)
\end{aligned}
\end{equation}
The integrals over $\delta \vec{X}^{2D-j}_0$ can be performed by decomposing the matrix $\mathbf{A'_\alpha}$ into blocks relating to the $2D-j$ and $j$ dimensional subspaces, i.e., 
\begin{equation}
\label{cov}
\mathbf{A'_\alpha} = \begin{pmatrix} \mathbf{A'_{11}} & \mathbf{A'_{12}} \\ \mathbf{A'_{21}} & \mathbf{A'_{22}}  
\end{pmatrix}
\end{equation}\\
If the off-diagonal blocks, $\mathbf{A'_{12}}$ and its transpose $\mathbf{A'_{21}}$, do not vanish, then they introduce terms in the argument of the exponential which are linear in $\delta \vec{X}^{2D-j}_0$.  The shift 
\begin{equation}
\label{eq:shift}
\delta \vec{X}^{2D-j}_0 \rightarrow \delta \vec{X}^{2D-j}_0 + \mathbf{A^{'-1}_{11}}\mathbf{A'_{12}} \cdot \delta \vec{X}^{j}_0 
\end{equation}
can be used to eliminate these terms, but completing the square modifies the quadratic $\delta \vec{X}^{j}_0$ term to,
\begin{equation}
\label{eq:square}
 \mathbf{A'_{22}} - \mathbf{A'_{21}}\mathbf{A^{'-1}_{11}}\mathbf{A'_{12}}
\end{equation}
\begin{widetext}
The expression for the expectation value takes the form,
\begin{align}
\label{int1}
 {\cal F}_W(t)  = & \frac{1}{(\pi\hbar)^{D}}\int d \delta \vec{X}_0^{j} \text{exp}\Bigr{[}-(\delta \vec{X}^{j}_0)^T \cdot \frac{\mathbf{A'_{22}}-\mathbf{A'_{21}}\mathbf{A^{'-1}_{11}}\mathbf{A'_{12}}}{\hbar} \cdot \delta \vec{X}^{j}_0 \Bigr{]} \int d \delta \vec{X}_0^{2D-j} \text{exp}\Bigr{[}-(\delta \vec{X}^{2D-j}_0)^T \cdot \frac{\mathbf{A'_{11}}}{\hbar} \cdot \delta \vec{X}^{2D-j}_0 \Bigr{]}\nonumber \\[6pt]
&  \times \Bigr{(} f_j - \pvec{f}_1 \cdot \mathbf{A^{'-1}_{11}}\mathbf{A'_{12}} \cdot \delta \vec{X}^{j}_0 + (\delta \vec{X}^j_0)^T \cdot \frac{\mathbf{A'_{21}}\mathbf{A^{'-1}_{11}}\mathbf{F'}\mathbf{A^{'-1}_{11}}\mathbf{A'_{12}}}{2} \cdot \delta \vec{X}^{j}_0 + (\delta \vec{X}^{2D-j}_0)^T \cdot \frac{\mathbf{F'}}{2} \cdot \delta \vec{X}^{2D-j}_0 \Bigr{)}
\end{align}
The terms linear in $\delta \vec{X}^{2D-j}_0 $ are dropped because the integral vanishes. The first three terms in the $f_W$-expansion have no dependence on $\delta \vec{X}^{2D-j}_0 $ and the integrals only account for a change in the normalization. The last term can be integrated using the identity,
\begin{equation}
\frac{1}{(\hbar\pi)^{D-j/2}} \int d\vec{x}\hspace{.1cm} (\vec{x} \cdot \frac{\mathbf{F'}}{2} \cdot \vec{x}) \text{exp}[-\vec{x}\cdot\frac{\mathbf{{A'}_{11}}}{\hbar}\cdot \vec{x}] = \frac{\hbar\text{ Tr}(\mathbf{F'{A'}_{11}^{-1}})}{4\sqrt{\text{Det}(\mathbf{{A'}_{11}})}}
\end{equation}
After doing the $2D-j$ integrals in Eq.~\ref{int1}, the remaining $j$ dimensional integral for the expectation value Monte Carlo calculation becomes,
\begin{equation}
\begin{aligned}
\label{int3}
 {\cal F}_W(t) & = \frac{1}{(\hbar\pi)^{j/2}\sqrt{\text{Det}(\mathbf{A'_{11}}})} \int d \delta \vec{X}^{j}_0 \text{exp}\Bigr{[}- \delta \vec{X}^{j}_0 \cdot \frac{\mathbf{A'_{22}} - \mathbf{A'_{21}}\mathbf{A^{'-1}_{11}}\mathbf{A'_{12}}}{\hbar} 
\cdot \delta\vec{X}^j_0 \Bigr{]} \\[6pt]
& \times \Bigr{(} f_j - \pvec{f}_1 \cdot \mathbf{A^{'-1}_{11}}\mathbf{A'_{12}} \cdot \delta \vec{X}^{j}_0 +
(\delta \vec{X}^j_0)^T \cdot \frac{\mathbf{A'_{21}}(\mathbf{A^{'-1}_{11}})^T\mathbf{F'}\mathbf{A^{'-1}_{11}}\mathbf{A'_{12}}}{2} \cdot \delta \vec{X}^{j}_0+ \frac{\hbar\text{ Tr}(\mathbf{F'}\mathbf{A^{'-1}_{11}})}{4} \Bigr{)}
\end{aligned}
\end{equation}
\end{widetext}
Recall that
\begin{equation}
\label{eq:det22}
\text{Det}(\mathbf{A'_{22}} - \mathbf{A'_{21}}\mathbf{A^{'-1}_{11}}\mathbf{A'_{12}})\text{Det}(\mathbf{A'_{11}}) =1 
\end{equation}
from properties of $2 \times 2$-block determinants~\cite{Lu02} and the density is still properly normalized. The reduced dimensional Wigner function can now be defined as: 
\begin{align}
\label{reduced_wigner}
 & \rho'_\alpha(\delta \vec{X}^j_0) = \frac{\sqrt{ \text{Det}(\mathbf{\Sigma_{\alpha}})}}{(\hbar \pi)^{j/2}}\text{exp}[-\delta \vec{X}^j_0 \cdot \frac{\mathbf{\Sigma_{\alpha}}}{\hbar} \cdot \delta \vec{X}^j_0] \\[5pt]
 & \text{where} \hspace{3mm} \mathbf{\Sigma_{\alpha}} = \mathbf{A'_{22}} - \mathbf{A'_{21}}\mathbf{A^{'-1}_{11}}\mathbf{A'_{12}}
\end{align}
The form of the integral remains invariant, with the coherent state Wigner function being replaced by a unit-normalized Gaussian distribution dependent only on the subspace of the coordinates corresponding to the unstable manifolds and constants of the motion.  All that is needed to evaluate Eq.~\eqref{int3} via Monte Carlo methods is the initial coordinate transformation, $\mathbf{S}_{\alpha,0}$, and first and second order derivatives evaluated at the endpoint of the trajectories. The direct evaluation consists of sampling initial conditions for the $j$ coordinates from $\rho'_\alpha(\delta \vec{X}^j_0)$ and using zero for the initial conditions of the other $2D-j$ coordinates that have already been integrated. The initial conditions are then reverted back to the original coordinate system to run the trajectories and evaluate $f_W(\delta \vec{X}^j_0)$. 

We emphasize that this exhausts the number of integrals that can be performed analytically before setting up the Monte Carlo calculation.  The functions, $f_j$, $\pvec{f}_1$, and $\mathbf{F'}$ all depend in some complicated unknown way on the variations, $\delta \vec{X}^j_0$. The last three terms in Eq.~(\ref{int3}), are referred to as correction terms. Note that for strongly chaotic systems the quantity, $e^{-\lambda_n t}$, for each chaotic degree of freedom will rapidly decay to zero. Approximating this quantity to be zero for each chaotic degree of freedom leads the matrix  $\mathbf{M}_t$ to only be a diagonal matrix of unity or zeros. This results in the correction terms only depending on derivatives of $f_W(\vec{q}_t, \vec{p}_t)$ with respect to the conjugate coordinate of each constant of the motion, and thus can be found by using the Poisson bracket, rendering the terms straightforward to evaluate numerically. For example, if the the constant of the motion is the energy all that is needed is Hamilton's equations. 

\subsection{Application to the purely quartic oscillator}
\label{1dqo}

It is worthwhile illustrating this procedure with as simple an example as possible.  The above equations may appear more complicated than they really are.  The purely quartic oscillator with a single degree of freedom provides an excellent case study.  It is an integrable system and $\delta \vec{X}$ could be taken to be variations in the action-angle variables, $(\delta I, \delta \theta)$. However, it is possible to work directly with the energy and its conjugate coordinate, which in this case is $(\delta E, \delta t)$, and it is valuable to illustrate the approach with these coordinates.  

Consider the classical Hamiltonian of Eq.~\eqref{1dquartic} and the expectation value of the Wigner transform of a number operator ($n(t)$) in quadratures, 
\begin{equation}
\label{nop}
f_W(p,q)=\frac{p^2+q^2}{2}
\end{equation}
For the illustration, take $\mathbf{A}_\alpha =\mathbb{1}$, and $(p_\alpha, q_\alpha) = (10,0)$.  The period of motion for the central orbit $(p_\alpha, q_\alpha)$ is given by $\tau_\alpha = 2K(1/2)/E_\alpha^{1/4}$, where $K(x)$ is a complete elliptic integral of the first kind. Due to the fact that this is a low-dimensional case, it is possible to create a uniform grid of initial conditions in $(\delta E, \delta t)$, each one weighted by the appropriate value of the Wigner transform as opposed to choosing random initial conditions with a probability given by the Wigner transform.  This is done here to isolate the effects of $f_j$ from its correction term, and study $\hbar$-dependence without having to consider sampling error Monte Carlo fluctuations.  A grid of $100\times 100$ initial conditions covering $\pm 5 \sigma$ in the $(\delta E, \delta t)$ Wigner transform density leads to essentially fluctuation free results in the unintegrated full two dimensional Monte Carlo as well as $100$ uniformly spaced initial conditions chosen $\pm 5\sigma$ along $\delta E$ for the integrated one dimensional Monte Carlo version.

The $S_\alpha$ matrix of Eq.~\eqref{1d} for this case is diagonal with values $(p_\alpha, 1/p_\alpha)$, respectively.  Thus, $\mathbf{A'}$ is also diagonal with values $(p_\alpha^2,1/p_\alpha^2)$ (reordered properly).  The two dimensional Monte Carlo grid is constructed directly using Eq.~\eqref{rho} and the $(\delta E, \delta t)$ coordinate system, but the trajectories are run in the $(p,q)$ coordinate system; the endpoints $(p_t,q_t)$ are directly substituted into $f_W$.  Thus to get the initial conditions for the trajectories, $\delta p = \delta E/p_\alpha$ and $\delta q = p_\alpha \delta t$ are used, consistent with Eq.~\eqref{1d}.  

For the one dimensional Monte Carlo, $\mathbf{A'}_{12} =\mathbf{A'}_{12} =0$, which makes two correction terms and the change due to the shift in the exponential vanish, leaving just the last correction term to evaluate (which depends on second derivatives), i.e., it is necessary to construct the $\mathbf{F'}$ matrix of second derivatives.  However, in this simple example, $2D-j=1$, and there is only one term to calculate.  The element $\mathbf{F'}=\mathbf{F}$ since $M_t=1$ for a conjugate variable to a constant of the motion.  Thus, 
\begin{eqnarray}
\mathbf{F} &=& \dot{p}^2 + p\ddot{p} + \dot{q}^2 + q\ddot{q} \nonumber \\
&=& 16 q_t^6 - 12 p_t^2 q_t^2 + p_t^2 - 4 q_t^4 \ ,
\end{eqnarray}
which follows from repeated use of Hamilton's equations.  For $f_j$ and $\mathbf{F}$
\begin{equation}
p_t = p_t(\delta E, 0)\ \text{and}\ q_t = q_t(\delta E, 0)
\end{equation}
unlike the $2d$ Monte Carlo for which
\begin{equation}
p_t = p_t(\delta E, \delta t)\ \text{and}\ q_t = q_t(\delta E, \delta t)\ .
\end{equation}
The constant multiplying $\mathbf{F}$ in the one dimensional Monte Carlo is $\hbar/(4\mathbf{A'}_{11}) =\hbar/ (4p_\alpha^2)$. To get the constant correctly, it must be remembered that  $\mathbf{A'}_{11}^{-1}$ in Eq.~\eqref{int3} relies on the row-reordered $\mathbf{S'}_{\alpha,0}$ coordinate transformation of Eq.\eqref{rho}, not the original $\mathbf{S}_{\alpha,0}$.

\begin{figure}[ht]
\includegraphics[width=\linewidth]{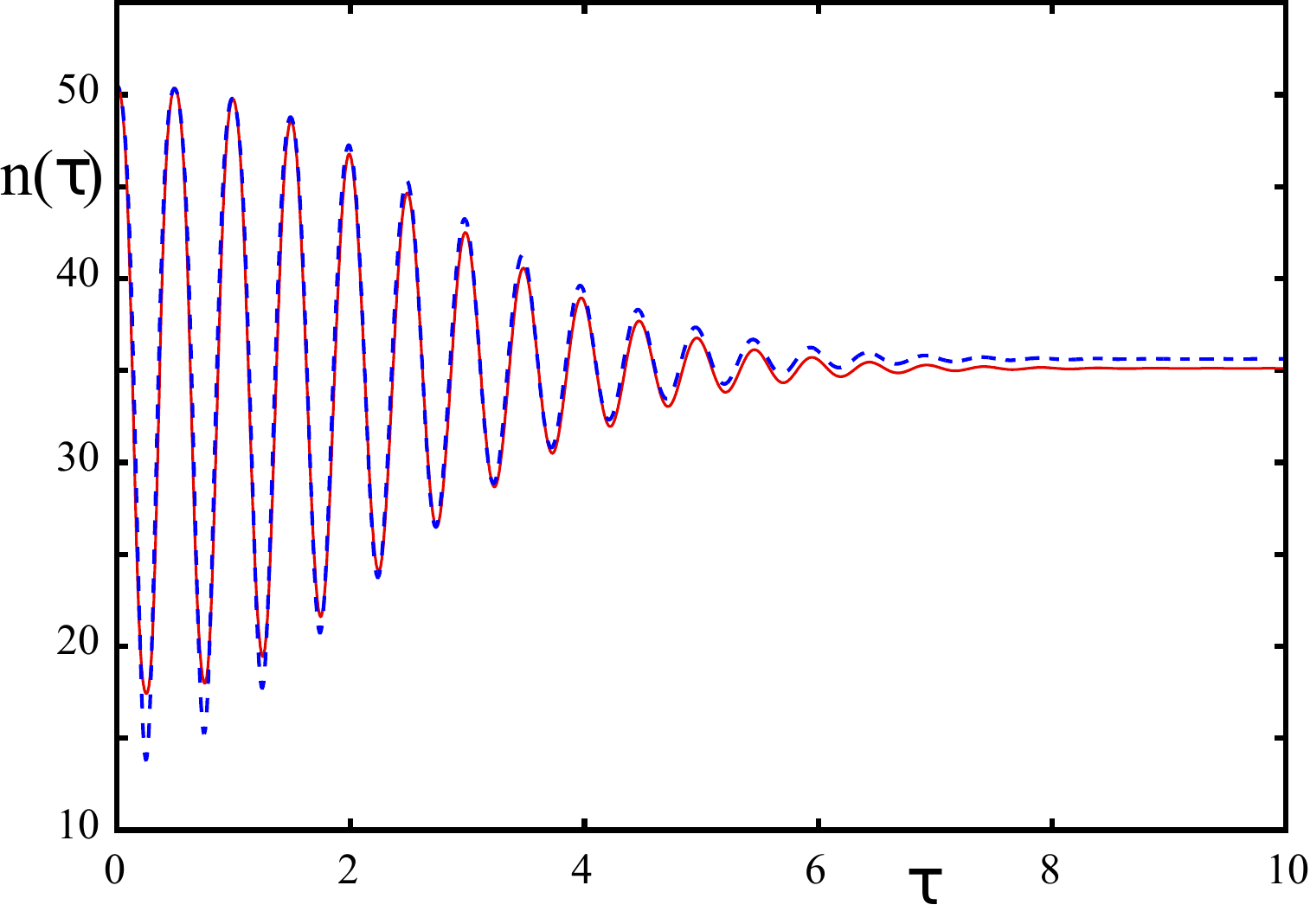}
\includegraphics[width=\linewidth]{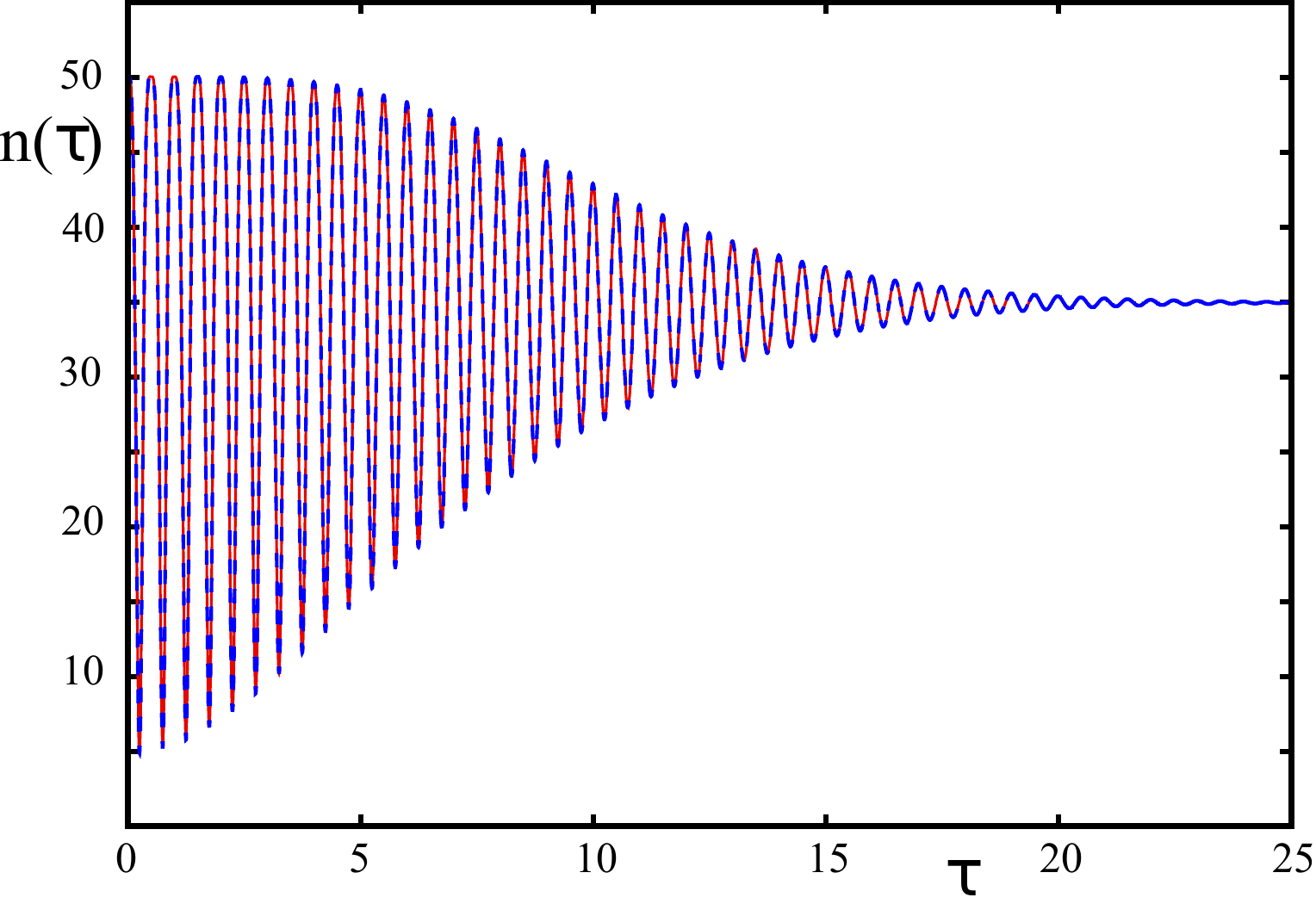}
\caption{The number expectation value as a function of scaled time $\tau$.  It has been scaled to unity for a single period of the motion for the trajectory with initial condition $(p_\alpha,q_\alpha)$, i.e. $\tau=t/\tau_\alpha$ (the value of $\tau_\alpha$ is given in the main text just following Eq.~\eqref{nop}).  Due to the shearing in the dynamics, the initial large oscillations die off leading to an equilibrated value after a number of periods of the motion.  For smaller $\hbar$ this transient period is much longer ($\approx 3 \times$) due it being an integrable system, having a smaller energy uncertainty, and thus having a smaller range of varying periods of the motion within the Wigner transform density. The blue dashed curve corresponds to the two dimensional Monte Carlo and the red solid line to the one dimensional version.}
\label{fig1}
\end{figure}
In Fig.~\ref{fig1} the two dimensional Monte Carlo and one dimensional Monte Carlo results (including the $O(\hbar)$ correction term) are compared.  In the upper panel, $\hbar=1$, and in the lower panel, $\hbar=0.1$.  Initially, there are large oscillations in $n(\tau)$, which damp out after a number of periods of the motion.
\begin{figure}[ht]
\includegraphics[width=\linewidth]{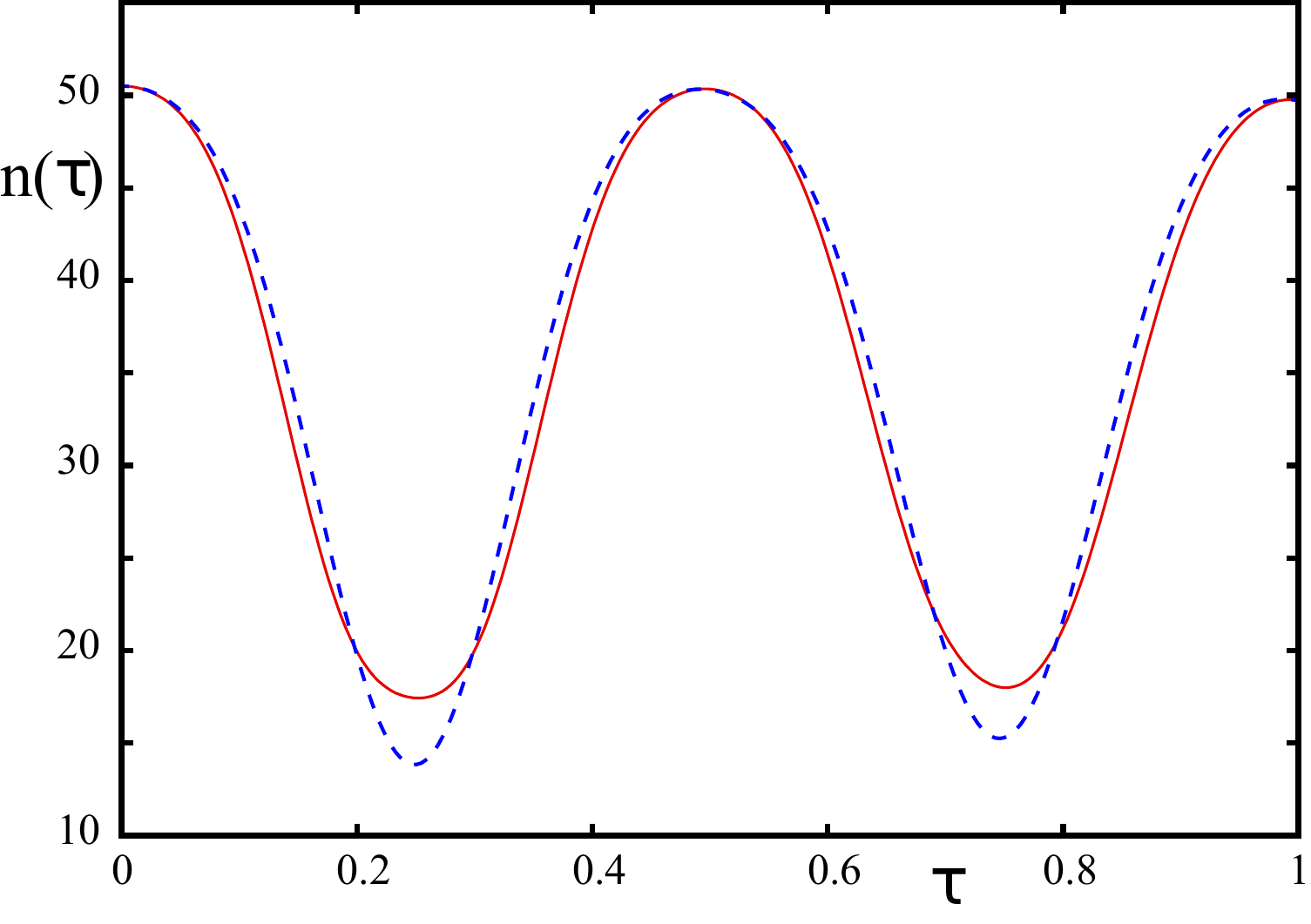}
\caption{The maxima and minima of $n(\tau)$ over the first period of the motion.  Here, $\hbar=1$.  Even in this expanded view, the differences between the two and one dimensional results are not that significant. The blue dashed curve corresponds to the two dimensional Monte Carlo and the red solid line to the one dimensional version.}
\label{fig2}
\end{figure}
Over the course of the first period of the motion, in Fig.~\ref{fig2} it is seen that there are two maxima and minima.  The maxima occur at $\tau=0,0.5$ where the momentum is at its maximum absolute values.  Likewise, the minima occur where the momentum vanishes and the position takes on its maximum absolute value, $\tau=0.25,0.75$. 

\begin{figure}[ht]
\includegraphics[width=\linewidth]{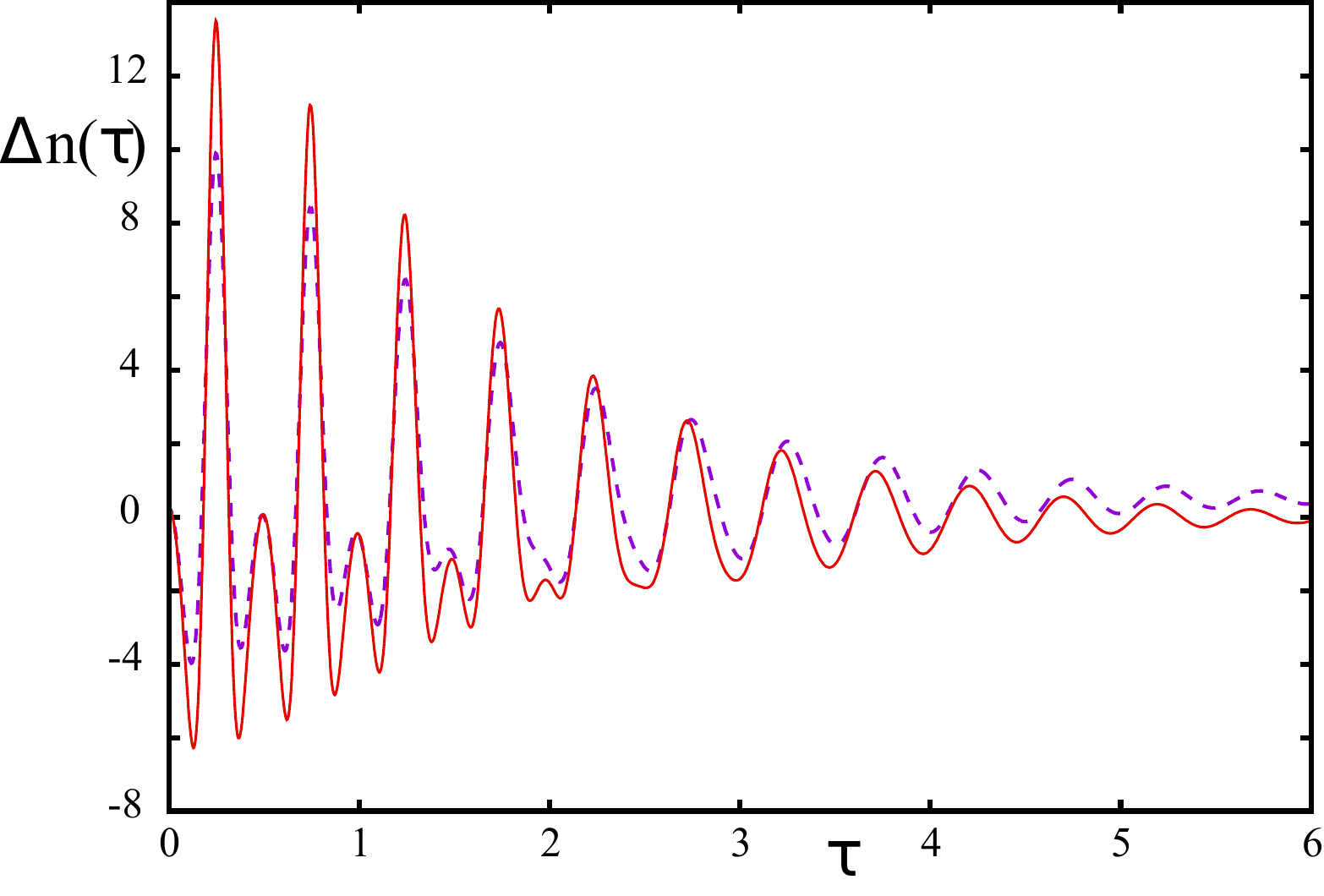}
\includegraphics[width=\linewidth]{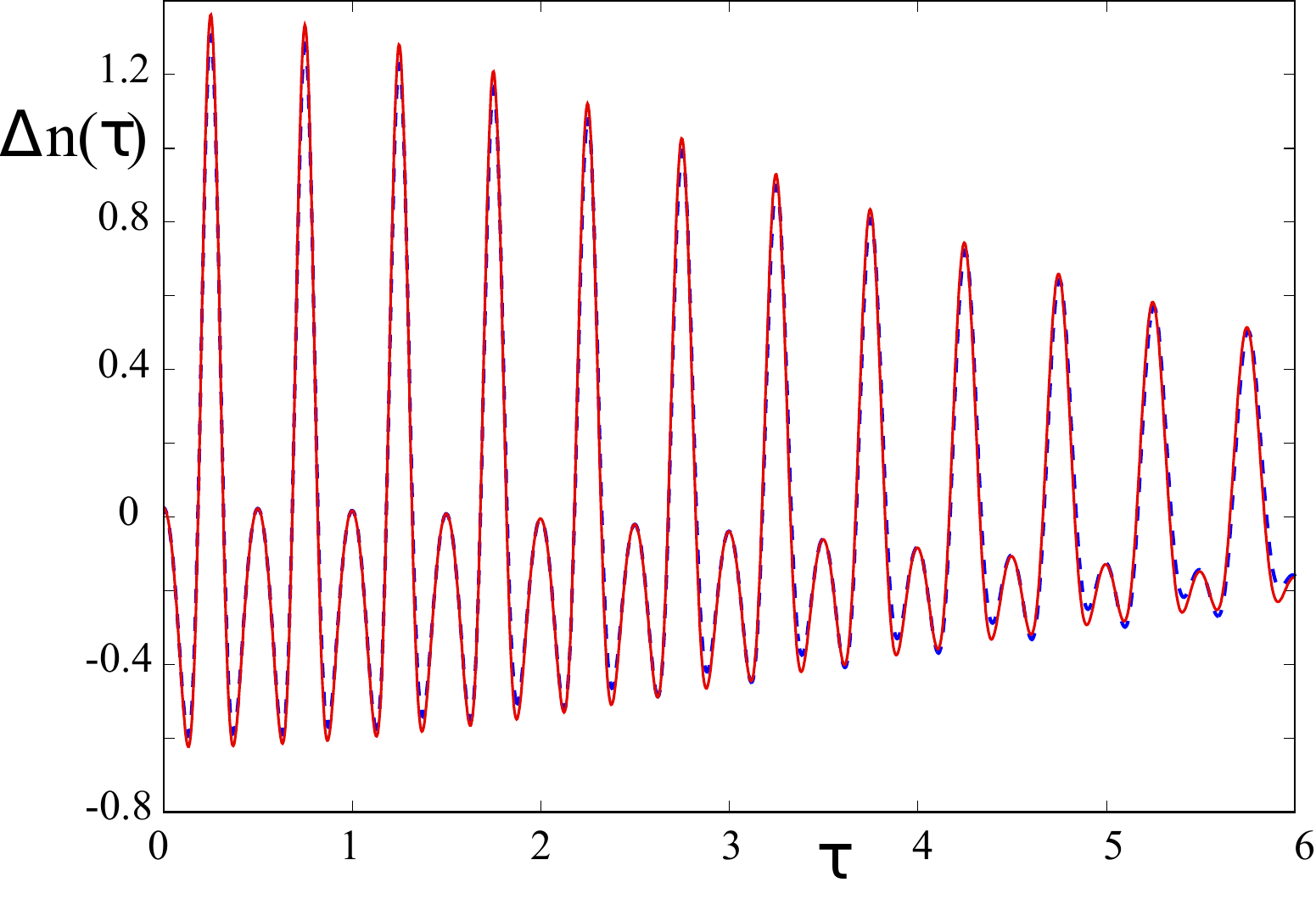}
\caption{$\Delta n(\tau)$ versus the $O(\hbar)$ correction term.  The upper panel contains the results for $\hbar=1$ and in the lower panel $\hbar=0.1$.  Not only does the difference decrease proportionally with $\hbar$, but its relative accuracy improves even more.  It is difficult to see any difference between $\Delta n(\tau)$ and the $O(\hbar)$ term in the lower panel.}
\label{fig3}
\end{figure}
Some small differences can be discerned between the two dimensional and one dimensional results in Fig.~\ref{fig2} with the one dimensional Monte Carlo slightly undershooting the minima.  By subtracting the $f_j$-term of the one dimensional Monte Carlo from the two dimensional Monte Carlo calculation, defining a $\Delta n(\tau)$, and comparing to the $O(\hbar)$-term of the one dimensional calculation, it is possible to magnify the differences so as to see them better.  In this way, the role of $\hbar$ in improving the results also becomes clear.  Figure~\ref{fig3} shows the differences for the first six periods of the motion for both $\hbar=1$ and $0.1$.  In the upper panel for $\hbar=1$, some differences can be seen with the $O(\hbar)$ correction slightly overshooting the behavior of $\Delta n(\tau)$.  In the lower panel, not only is $\Delta n(\tau)$ ten times smaller, but it is very difficult to see any differences between $\Delta n(\tau)$ and the $O(\hbar)$ correction. In fact, the absolute value of the maximum difference between the two and one dimensional results with the $O(\hbar)$ correction is $\approx 3.6$ for $\hbar=1$, but only $\approx 0.075$ for $\hbar=0.1$ (roughly $50$ times smaller).  Even so, for most times it is much smaller than that.  This is an excellent accuracy on the absolute scale of an approximate maximum value of $50$ for $n(\tau)$.  The pre-integrated Monte Carlo performs extremely well with the $O(\hbar)$ correction, and it is acceptable even without the correction.

\subsection{Application to two coupled pure quartic oscillators}
\label{2dqo}

\begin{figure}[ht]
    \includegraphics[trim = {1cm 0cm 1cm 2cm },width= \linewidth]{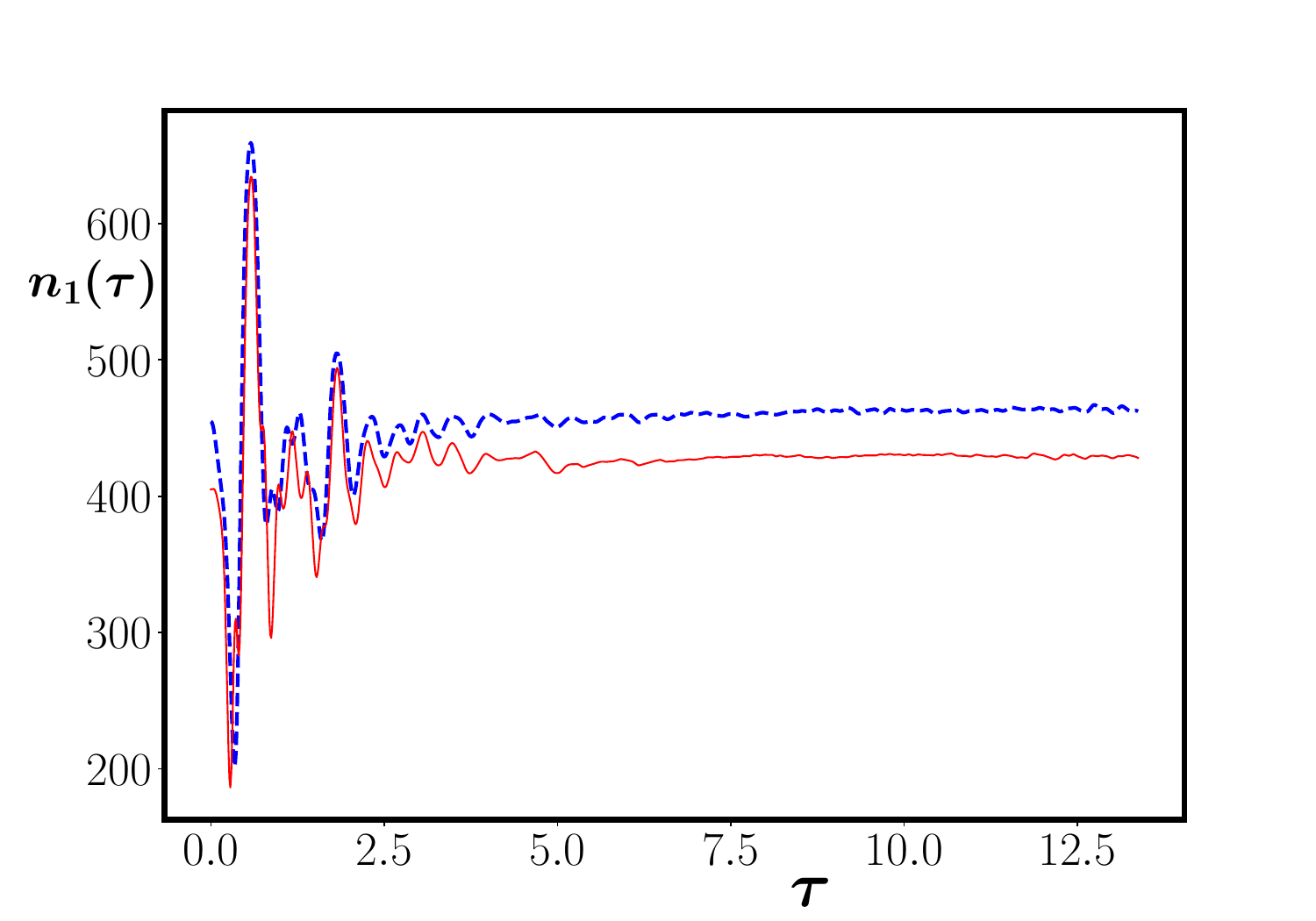}
    \includegraphics[trim = {1cm 1cm 1cm .45cm },width= \linewidth]{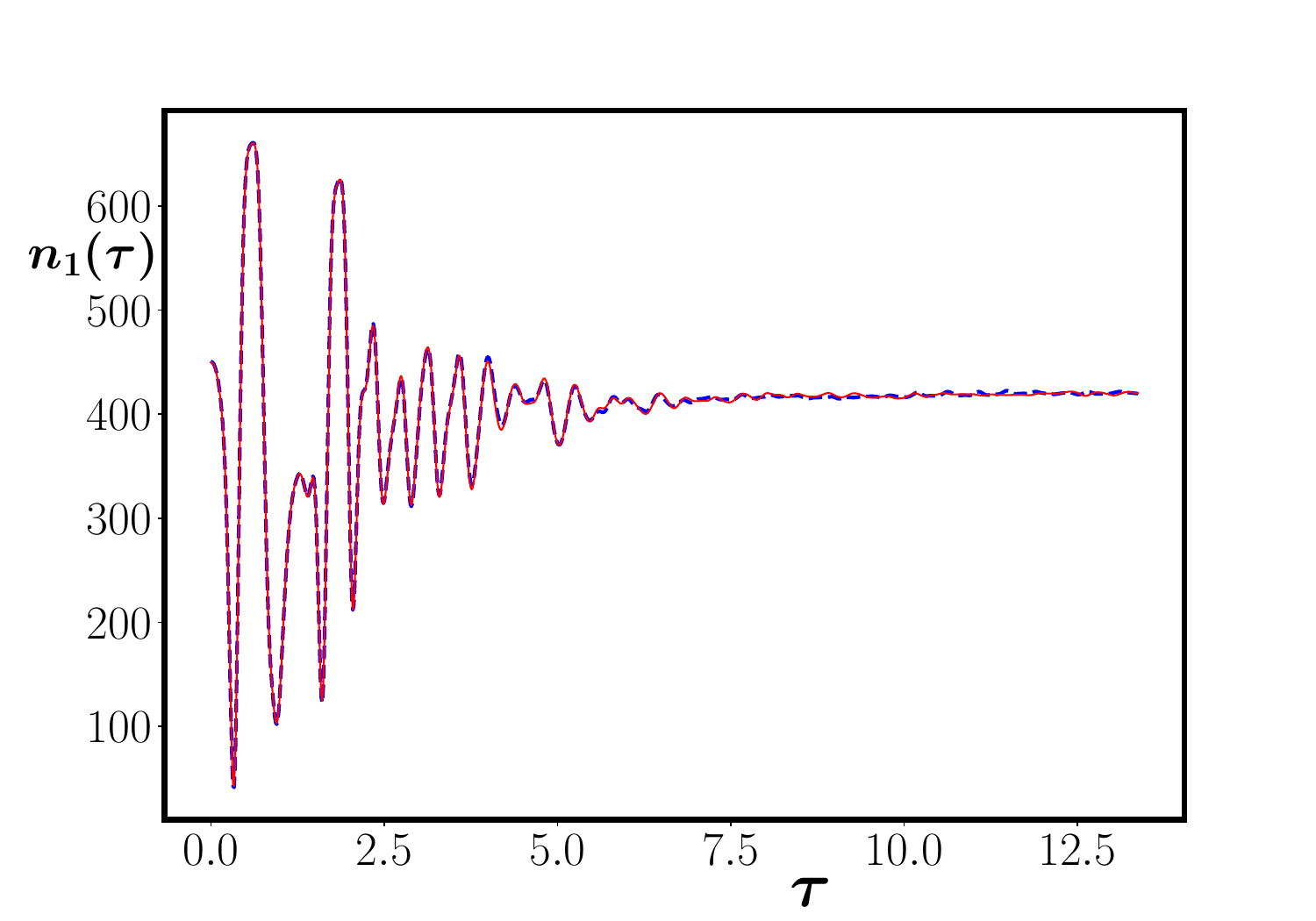}
    \caption{Expectation value of the function $f_W(\vec{p}, \vec{q}) = \frac{q_1^2 + p_1^2}{2}$, chosen to mimic a site occupancy calculation, using the chaotic case of a two-dimensional quartic oscillator. In the upper panel $\hbar = 10$ and in the lower panel $\hbar = 1$. The approximate two dimensional integral given by Eq.~\eqref{int3} (solid red) is compared to the exact four dimensional integral given by Eq.~\ref{int} (dashed blue). The timescale is $\tau = t/ \tau_{1,1}$ where $\tau_{1,1}$ is given by Eq.~\eqref{tau12} and acts as a "center of mass" period of motion for the two dimensional quartic oscillator. The initial conditions used are $(\vec{p}_\alpha, \vec{q}_\alpha) = (30, 40, 0, 0)$. } 
    \label{fig4}
\end{figure}

\begin{figure}[ht]
    \includegraphics[trim = {1cm 0cm 1cm 2cm },width= \linewidth]{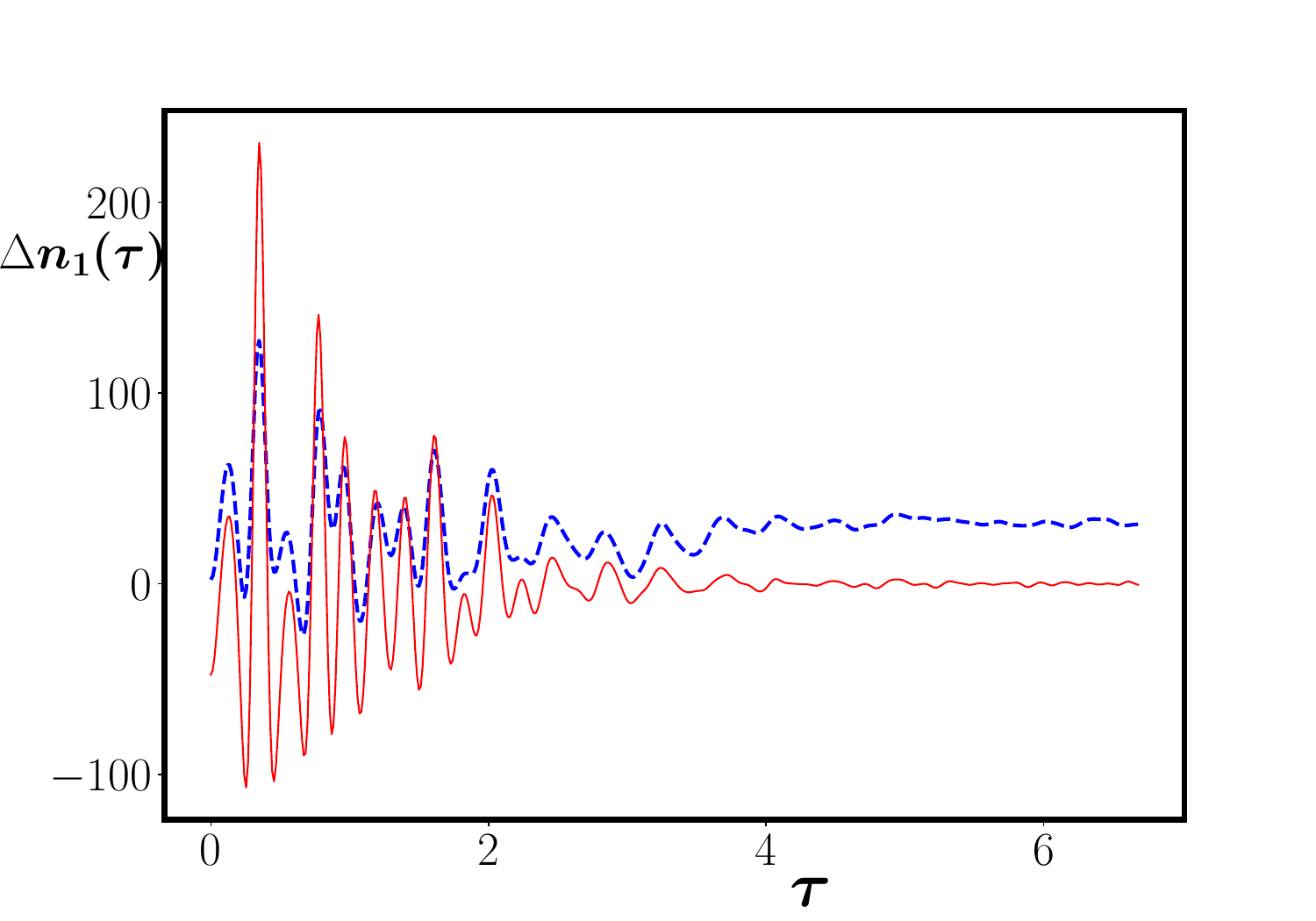}
    \includegraphics[trim = {1cm 1cm 1cm .45cm },width= \linewidth]{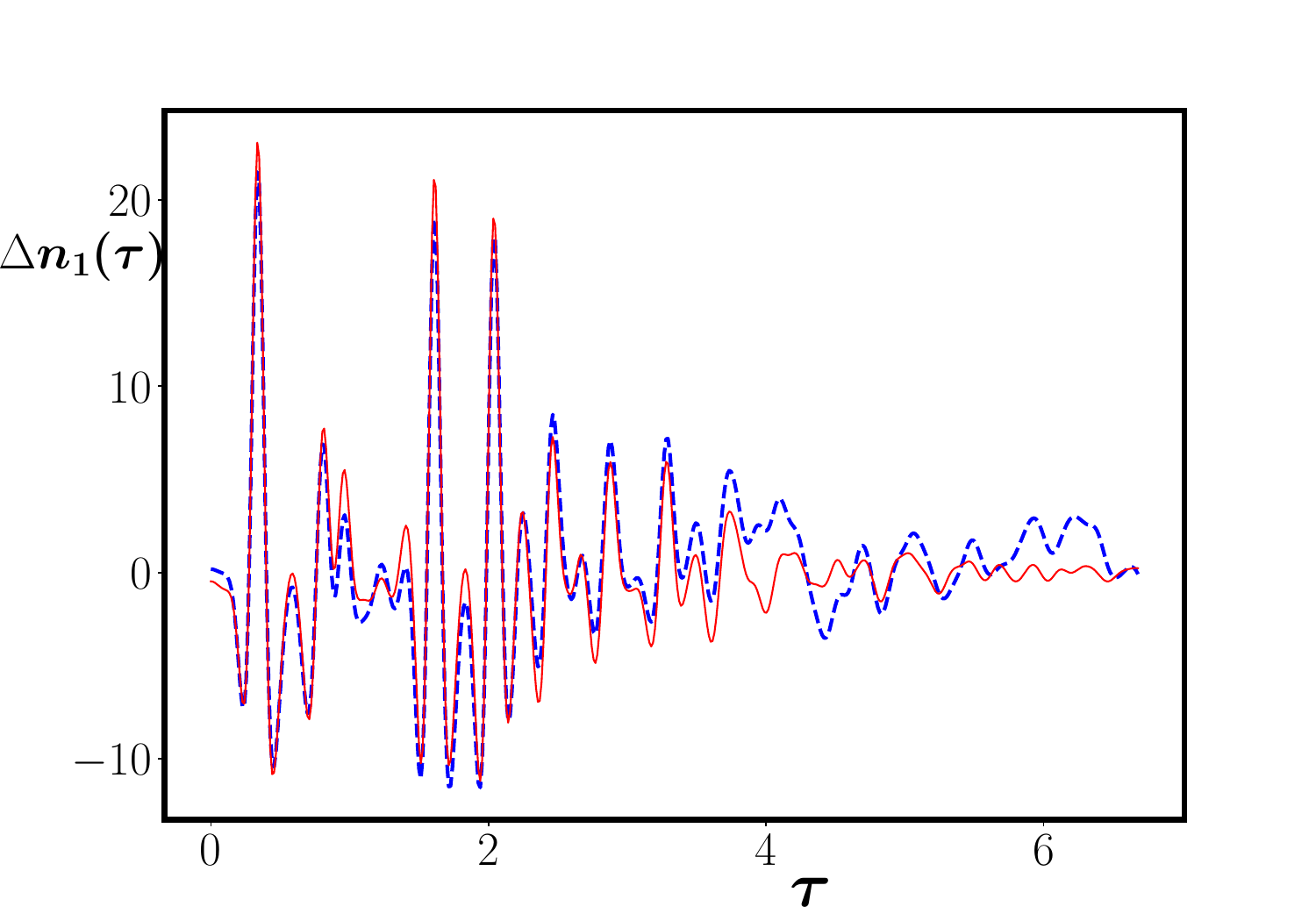}
    \caption{$\Delta n_1(\tau)$ is plotted versus the correction terms. The upper panel contains the results for $\hbar=10$ and in the lower panel $\hbar=1$. The correction term contributions seen in the plots decrease proportionally to $\hbar$.  The error between the Monte Carlo and the correction terms, as before, decreases even more. Note the difference in scale compared to Fig.~\ref{fig4}, which demonstrates that the approximation for $\hbar=1$ is highly accurate.}
    \label{fig5}
\end{figure}

The two coupled pure quartic oscillators discussed in Sec.~\ref{tcqo} provide an ideal example to test the procedure for a non-trivial chaotic system. The expectation value of the number operator of the first "site", 
\begin{equation}
\label{nop1}
f_W(p,q)= \frac{p_1^2+q_1^2}{2} \ ,
\end{equation}
provides a convenient example. The original four dimensional integral can be solved via Monte Carlo methods by sampling the Wigner distribution, $\rho_{\alpha}(\vec{p}_0, \vec{q}_0)$. The reduced dimensional Monte Carlo formula, Eq. \eqref{int3}, is a two dimensional integral along the constant of the motion and the unstable manifold. This formula can be evaluated by sampling initial conditions from the reduced dimensional Wigner function, Eq.~\eqref{reduced_wigner}. The initial coordinate transformation of the central orbit, $\mathbf{S}_{\alpha,0}$, is all that is needed for the sampling procedure. The transformation is given by Eq.~\eqref{smatrix} and is properly reordered by exchanging rows one and two and rows two and three. The reduced dimensional stability matrix for this case is:
\begin{align}
\mathbf{M}_t = \begin{pmatrix}
    1 & 0 \\ 0 & e^{-\lambda t}
\end{pmatrix}
\end{align}
The term associated with the stable manifold decays exponentially fast and is set to zero in this implementation. As noted previously, this leads to an error at very short times. However the error is of order $\hbar^{1/2}e^{-\lambda t}$ and does not greatly affect the results. The correction terms associated with the stable manifold then all vanish, leaving only functions of the first and second time derivatives of Eq.~\eqref{nop1} to be evaluated for the correction terms of the time coordinate. This evaluation is done simply with the use of Hamilton's equation applied to the Hamiltonian in Eq.~\eqref{ham}. The result of the reduced dimensionality formula, Eq.~\eqref{int3}, compared to the exact integral, Eq.~\eqref{int}, is given in Fig.~\ref{fig4}. The upper panel with $\hbar =10$ displays some inaccuracies, but the differences diminish greatly as $\hbar$ shrinks as seen in the lower panel for $\hbar=1$.  Note also that the time scale to equilibrate is not as strongly affected by shrinking $\hbar$ as is true for the one dimensional quartic oscillator (integrable case) results shown in the previous subsection.  Whereas for an integrable system, the time to equilibrate would be expected to scale as $\hbar^{-1/2}$, for the chaotic case, something closer to a $\ln(\hbar^{-1})$ scaling would be expected.
As done for the case of the one dimensional quartic oscillator, the differences can be magnified by plotting $\Delta n(\tau)$. In Fig.~\ref{fig5}, this is done for the case of $\hbar = 10$ and $\hbar = 1$. The discrepancy at $t=0$ also now becomes clear and it turns out not to be more significant than the error at longer times.

\section{Transport coefficients and return probabilities}
\label{transport}

Roughly the same approach as described for expectations values can be extended to transport coefficients. The same half of the variables can be integrated analytically before setting up the Monte Carlo calculation as for the expectation values. However, in addition, since contributions to the overlap only occur in localized regions of phase space at the beginning and endpoints, it is also possible to integrate out variables corresponding to all the constants of the motion.  The only variables which cannot be pre-integrated are associated with the unstable manifold.  Thus, for transport coefficients, the remaining number of variables for the Monte Carlo, $j$, can be as few as the number of positive Lyapunov exponents.  Finally, any variable related to a positive Lyapunov exponent so small that it cannot be responsible for generating dynamical recurrences over the time scale of interest, could be integrated as well. 

Ahead it is seen that the final expressions depend on the coordinate transformations at both $(\vec{q}_\alpha, \vec{p}_{\alpha})$ and $(\vec{q}_\beta, \vec{p}_{\beta})$. In addition, for the case where the constants of the motion are also pre-integrated, the expressions depend on the shearing rates of each shearing degree of freedom. The integrals over each constant of the motion will be replaced by a summation over the return times associated with its respective degree of freedom. In this section, the derivation for the integration of the same variables as for the expectation values is presented. Then it is shown how to extend this derivation to also integrate out the constants of the motion. A final remark: for return probabilities, in the first case when there is no summation, the initial decay has to be handled separately, but can be integrated over all $2D$ variables. A derivation is presented in Appendix~\ref{initdec}.

\subsection{Derivation of reduced dimensionality formulas for transport coefficients}
\label{1approx}

The derivation of the transport coefficients in the classical Wigner method proceeds in similar fashion to that of the expectation values. The integration variables can again be shifted and the coordinates transformed so that the integral in Eq.~\eqref{wmtc} becomes:
\begin{align}
\label{wmtc3}
{\cal C}_{\alpha\beta}(t) = (2\pi\hbar)^D
\int d\delta\vec{X}_0  \ 
\rho_\alpha(\delta\vec{X}_0)\rho_\beta(\vec{p}_t(\delta\vec{X}_0), \vec{q}_t(\delta\vec{X}_0)) 
\end{align}
with the Wigner function
\begin{equation}
\begin{aligned}
\rho_\alpha(\delta \vec{X}_0) = & \frac{1}{(\pi\hbar)^{D}}\text{exp}\Bigr{[}-(\delta \vec{X}_0)\cdot \frac{\mathbf{A'_\alpha}}{\hbar} \cdot (\delta \vec{X}_0) \Bigr{]} \\
& \text{with} \hspace{.3cm} \mathbf{A}'_\alpha = (\mathbf{S'}^{-1}_{\alpha,0})^T\mathbf{A}_{\alpha} \mathbf{S'}^{-1}_{\alpha,0}
\end{aligned}
\end{equation}
The main difference between transport coefficients and expectations values is the contributions to the integral at a time $t$ are localized in phase space around the point $(\vec{p}_{\beta}, \vec{q}_{\beta})$; i.e., all the variations are in the argument of the exponential, beginning and end. Thus, there is no expansion of $f_W$. Instead, the full argument of the exponential can be treated in a parallel way as done in Eqs.~\eqref{int2}-\eqref{int1}.  However, there are a few new considerations.  First note that the localization around the point $(\vec{p}_{\beta}, \vec{q}_{\beta})$ allows the reordered transformation at a time $t$ to be well approximated as,
\begin{align}
    \mathbf{S'}_{j,t} \approx \mathbf{S'}_{\beta}
\end{align}
and therefore the coordinate transformations are well approximated as completely independent of the trajectories to be used in the Monte Carlo and independent of time.  Second, it is necessary to decompose $\delta \vec{X}_t$ into two component parts because $(\vec{p}_{\beta}, \vec{q}_{\beta})$ is not related to, nor the endpoint of, any particular trajectory of interest.  Let
\begin{align}
\vec{p}_{t,j} & =\vec{p}_t(\delta \vec{X}^{2D-j}_0=\mathbf{0},\delta \vec{X}^j_0) \nonumber \\
\vec{q}_{t,j} & =\vec{q}_t(\delta \vec{X}^{2D-j}_0=\mathbf{0},\delta \vec{X}^j_0) \, 
\end{align}
which leads to an expression for the variations
\begin{align}
\delta \vec{X}_{\beta,t} & = \mathbf{S'}_{\beta} \cdot (\vec{p}_{t,j} - \vec{p}_{\beta}, \vec{q}_{t,j} - \vec{q}_{\beta}) \nonumber \\
\delta \vec{X}_t & = \mathbf{S'}_{\beta} \cdot (\vec{p}_t - \vec{p}_{t,j}, \vec{q}_t - \vec{q}_{t,j}) \ ,
\end{align}
where the trajectory endpoint $(\vec{p}_t, \vec{q}_t)$ also includes variations due to nonvanishing $\delta \vec{X}^{2D-j}_0$.  The special nature of this coordinate system means that
\begin{equation}
\delta \vec{X}_t = (\delta \vec{X}^{2D-j}_t, \delta \vec{X}^j_t=\mathbf{0})
\end{equation}
due to the block diagonalization of the stability matrix.  
With these definitions, the nonvanishing part of $\delta \vec{X}_t$, i.e., $\delta \vec{X}^{2D-j}_t$, can be replaced by the identical diagonal stability matrix multiplication of the initial variations, $\mathbf{M}_t \delta \vec{X}^{2D-j}_0$ just as before.  The full argument of the exponential can thus  be expressed as
\begin{widetext}
\begin{align}
& -(\delta \vec{X}^{2D-j}_0, \delta \vec{X}^j_0)^T \cdot \frac{\mathbf{A'_\alpha}}{\hbar} \cdot 
 (\delta \vec{X}^{2D-j}_0, \delta \vec{X}^j_0)
 - \left[(\mathbf{M}_t \cdot \delta \vec{X}^{2D-j}_0 , \mathbf{0})^T + \delta \vec{X}_{\beta,t}^T\right] \cdot \frac{\mathbf{A'_\beta}}{\hbar} \cdot \left[(\mathbf{M}_t \cdot \delta \vec{X}^{2D-j}_0 , \mathbf{0}) + \delta \vec{X}_{\beta,t}\right]
\end{align}
\end{widetext}
After collecting terms and performing some algebra, instead of the $\mathbf{A'_{11}}$ block of Eq.~\eqref{cov}, here,
\begin{equation}
\label{eq:Aab}
\mathbf{A'_{\alpha\beta}}= \mathbf{A'}_{\alpha,11} + \mathbf{M}_t^T \cdot  \mathbf{A'}_{\beta,11}\cdot \mathbf{M}_t
\end{equation}
and instead of the shift in Eq.~\eqref{eq:shift}, the necessary shift is
\begin{align}
& \delta \vec{X}^{2D-j}_0 \rightarrow \delta \vec{X}^{2D-j}_0  + \mathbf{A'^{-1}_{\alpha\beta}}\cdot \nonumber \\
& \left[\mathbf{A'_{\alpha,12}} \cdot \delta \vec{X}^j_0 + \mathbf{M}_t^T \cdot \left(\mathbf{A'_{\beta,11}}\cdot\delta\vec{X}^{2D-j}_{\beta,t} + \mathbf{A'_{\beta,12}}\cdot\delta\vec{X}^j_{\beta,t} \right)\right] \ ,
\end{align}
which now depends on initial and final point variations, but only those due to $\delta \vec{X}^j_0$.  The result after performing the $2D-j$ dimensional integral can be put into a slightly more compact form by extending the $\mathbf{A'^{-1}_{\alpha\beta}}$ and $\mathbf{M}_t$ matrices into the full space as follows,
\begin{align}
{\cal A^{-1}_{\alpha\beta}} & = \begin{pmatrix}
\mathbf{A'^{-1}_{\alpha\beta}} & \mathbf{0} \\
\mathbf{0} & \mathbf{0}
\end{pmatrix}
\nonumber \\
{\cal M}_t & = \begin{pmatrix}
\mathbf{M}_t & \mathbf{0} \\
\mathbf{0} & \mathbf{0} 
\end{pmatrix} \ .
\end{align}
The remaining transport coefficient integrals can in this way be expressed as
\begin{widetext}
\begin{align}
{\cal C}_{\alpha\beta}(t) & = \frac{2^D}{(\pi\hbar)^{j/2}\sqrt{\text{Det}\left(\mathbf{A'_{\alpha\beta}}\right)}}  \int d\delta \vec{X}^{j}_0 \exp\left[-(\mathbf{0},\delta \vec{X}^j_0)^T \cdot \left( \frac{\mathbf{A'_{\alpha}}-\mathbf{A'_{\alpha}}\cdot {\cal A}^{-1}_{\alpha\beta}\cdot \mathbf{A'_{\alpha}}}{\hbar} \right) \cdot (\mathbf{0},\delta \vec{X}^j_0)\right.\nonumber \\
& \left. - (\delta \vec{X}_{\beta,t})^T \cdot \left( \frac{\mathbf{A'_{\beta}}-\mathbf{A'_{\beta}}\cdot \cal{M}_t \cdot{ \cal A^{-1}_{\alpha\beta}}\cdot {\cal M}^T_t\mathbf{A'_{\beta}}}{\hbar} \right) \cdot \delta \vec{X}_{\beta,t} -2\times (\mathbf{0},\delta \vec{X}^j_0)^T \cdot  \frac{\mathbf{A'_{\alpha}} \cdot {\cal A^{-1}_{\alpha\beta}}\cdot {\cal M}_t^T \cdot \mathbf{A'_{\beta}}}{\hbar}\cdot \delta \vec{X}_{\beta,t} \right] \ ,
\end{align}
\end{widetext}
which may or may not quite be set up for a Monte Carlo calculation.  In the many cases where all the $\exp(-\lambda_j t)$ factors in $\mathbf{M}_t$ can be replaced by zero, there is no time-dependence in the $\mathbf{A'^{-1}_{\alpha\beta}}$ matrix; see Eq.~\eqref{eq:Aab}.  Then it is possible to use directly the factor $\mathbf{A'_{\alpha}}-\mathbf{A'_{\alpha}} \cdot \mathbf{A'^{-1}_{\alpha\beta}}\cdot \mathbf{A'_{\alpha}}$ as the weighting for choosing initial conditions.  However, more generally the time-dependence can be separated out using the Woodbury matrix identity~\cite{Woodbury50}, which gives
\begin{align}
\label{eq:wb}
\mathbf{A'^{-1}_{\alpha\beta}} & = \mathbf{A'^{-1}_{\alpha,11}} - \mathbf{A'^{-1}_{\alpha,11}}\cdot \mathbf{M}_t^T \cdot \nonumber \\
&\left(\mathbf{A'^{-1}_{\beta,11}} + \mathbf{M}_t \cdot \mathbf{A'^{-1}_{\alpha,11}}\cdot \mathbf{M}_t^T\right)^{-1}\cdot \mathbf{M}_t \cdot \mathbf{A'^{-1}_{\alpha,11}}
\end{align}
Using the first term on the right side, $\mathbf{A'^{-1}_{\alpha,11}}$ allows for isolating $\rho'_\alpha(\delta \vec{X}^j_0)$ of Eq.~\eqref{reduced_wigner} as the weight function for choosing initial conditions, exactly as for the expectation values.  The second term on the right-hand-side contributes to the time-dependent weighting of the trajectories after they are calculated.  Making use of Eqs.~\eqref{eq:det22}, \eqref{reduced_wigner}, and \eqref{eq:wb} gives
\begin{widetext}
\begin{align}
\label{final_tc}
{\cal C}_{\alpha\beta}(t) & = \frac{2^D}{\sqrt{\text{Det}(\mathbf{1} +\mathbf{A^{'-1}_{\alpha, 11}}\mathbf{M}_t^T\mathbf{A'_{\beta, 11}}\mathbf{M}_t)}}  \int d\delta \vec{X}^{j}_0\ \rho'_\alpha(\delta\vec{X}^{j}_0) \ \times \nonumber \\
& \exp\left[-(\delta \vec{X}^j_0)^T \cdot \left( \frac{\mathbf{A'_{\alpha,21}}\cdot \mathbf{A'^{-1}_{\alpha,11}}\cdot \mathbf{M}_t^T \cdot (\mathbf{A'^{-1}_{\beta,11}} + \mathbf{M}_t \cdot \mathbf{A'^{-1}_{\alpha,11}}\cdot \mathbf{M}_t^T)^{-1} \cdot \mathbf{M}_t \cdot \mathbf{A'^{-1}_{\alpha,11}}\cdot \mathbf{A'_{\alpha,12}}}{\hbar} \right) \cdot \delta \vec{X}^j_0\right. \nonumber \\
& \left. - (\delta \vec{X}_{\beta,t})^T \cdot \left( \frac{\mathbf{A'_{\beta}}-\mathbf{A'_{\beta}}\cdot {\cal M}_t \cdot {\cal A^{-1}_{\alpha\beta}}\cdot {\cal M}^T_t\mathbf{A'_{\beta}}}{\hbar} \right) \cdot \delta \vec{X}_{\beta,t} -2\times (\mathbf{0},\delta \vec{X}^j_0)^T \cdot  \frac{\mathbf{A'_{\alpha}} \cdot {\cal A^{-1}_{\alpha\beta}}\cdot {\cal M}_t^T \cdot \mathbf{A'_{\beta}}}{\hbar}\cdot \delta \vec{X}_{\beta,t} \right] \ ,
\end{align}
\end{widetext}
Similar to the case of expectation values, the selection of initial conditions is identical with this form of the equation, and the stability matrix elements corresponding to coordinates along the stable manifolds vanish to a good approximation unless the stability exponent is too small.  However, for the case of $\alpha = \beta$, at $t=0$ the integral would not be properly normalized since $\mathbf{M}_{t=0} = \mathbb{1}$. This issue can be remedied by using a separate approximation for the initial decay, see Appendix~\ref{initdec}. For transport coefficients this is not necessary as the integral will be initially zero if $(\vec{p}_\alpha, \vec{q}_\alpha)$ and $(\vec{p}_\beta, \vec{q}_\beta)$ are chosen to not be in the same localized region of phase space. 

\subsection{Integrating the constants of motion}

For transport coefficients, it is possible to pre-integrate the constants of the motion in addition to those integrated out in the previous section. The derivation follows the same steps leading to Eq.~\eqref{final_tc}, but the coordinates $\delta\vec{X}_0^{j}$ after this further integration are only along the unstable manifold ($j$ here is the dimensionality of the unstable manifold). The definition of the stability matrix $\mathbf{M}_t$ must also be changed to include the shearing rates. Each stable degree of freedom has a $2\times 2$ block in the stability matrix in the form of Eq.~\eqref{shear}. The shearing rates only have to be evaluated once for the central trajectory since the variations along the unstable manifolds do not affect these values. The quantity $\delta \vec{X}_{\beta,t}$ depends on the trajectory $(\vec{p}_{j,t}, \vec{q}_{j,t})$, and the contributions to the integral only occur during the times of closest approach of the trajectory $(\vec{p}_{j,t}, \vec{q}_{j,t})$ to the point in phase space $(\vec{p}_{\beta}, \vec{q}_{\beta})$.
For a one degree of freedom system this would simply be related to the period of the trajectory. In a more general system, these times are a function of the return times associated with each stable degree of freedom. It is labeled by an abuse of notation $\tau_{i}$, but it should be understood that this means the following: 
\begin{equation} 
\tau_{i} = \tau(\tau_{i_1},...,\tau_{i_n})
\end{equation}
Where $\tau_{i_1},...,\tau_{i_n}$ labels the return times associated with the $n$ constants of the motion. The solution to the integrable case of the two dimensional quartic oscillator gives an example of how this $\tau_i$ function is implemented in practice. A Taylor expansion of $\delta \vec{X}_{\beta, t}$ can be done about these points in time:
\begin{align}
    \label{time_expand}
    &\delta \vec{X}_{\beta,t} = \delta \vec{X}_{\beta,\tau_i} + \vec{1}(t-\tau_i)
\end{align}
where $\vec{1}$ is a $2D$ dimensional vector with $1$ for the time coordinate and $0$ for all other coordinates. 

Each integral along the constant of the motion is to be replaced by a summation. For example, in the simple case of a one dimensional quartic oscillator the sum is over the recurrences that occur after each period of the motion. For a more complicated system, such as the two dimensional chaotic quartic oscillator, the distance of closest approach to the final wave packet centroid would have to be found numerically and each recurrence summed. The final form of the integral is found by inserting Eq.~\eqref{time_expand} into Eq.~\eqref{final_tc2} , evaluating the stability matrix at the times $\tau_i$, and summing over the return times associated with each stable degree of freedom,
\begin{widetext}
\begin{align}
\label{final_tc2}
{\cal C}_{\alpha\beta}(t) & =\sum_{i_1}...\sum_{i_n} \frac{2^D}{\sqrt{\text{Det}(\mathbf{1} +\mathbf{A^{'-1}_{\alpha, 11}}\mathbf{M}_{\tau_i}^T\mathbf{A'_{\beta, 11}}\mathbf{M}_{\tau_i})}}  \int d\delta \vec{X}^{j}_0\ \rho'_\alpha(\delta\vec{X}^{j}_0) \ \times \nonumber \\[7pt]
& \exp\left[ -(\delta \vec{X}^j_0)^T \cdot \left( \frac{\mathbf{A'_{\alpha,21}}\cdot \mathbf{A'^{-1}_{\alpha,11}}\cdot \mathbf{M}_{\tau_i}^T \cdot (\mathbf{A'^{-1}_{\beta,11}} + \mathbf{M}_{\tau_i} \cdot \mathbf{A'^{-1}_{\alpha,11}}\cdot \mathbf{M}_{\tau_i}^T)^{-1} \cdot \mathbf{M}_{\tau_i} \cdot \mathbf{A'^{-1}_{\alpha,11}}\cdot \mathbf{A'_{\alpha,12}}}{\hbar} \right) \cdot \delta \vec{X}^j_0\right. \nonumber \\
& \left. - (\delta \vec{X}_{\beta,\tau_i} + \vec{1}(t-\tau_i))^T \cdot \left( \frac{\mathbf{A'_{\beta}}-\mathbf{A'_{\beta}}\cdot {\cal M}_{\tau_i} \cdot {\cal A}^{-1}_{\alpha\beta}\cdot {\cal M}^T_{\tau_i}\mathbf{A'_{\beta}}}{\hbar} \right) \cdot (\delta\vec{X}_{\beta,\tau_i} + \vec{1}(t-\tau_i)) \nonumber \right.\\
& \left. -2\times (\mathbf{0},\delta \vec{X}^j_0)^T \cdot  \frac{\mathbf{A'_{\alpha}} \cdot {\cal A^{-1}_{\alpha\beta}}\cdot {\cal M}_{\tau_i}^T \cdot \mathbf{A'_{\beta}}}{\hbar}\cdot (\delta\vec{X}_{\beta,\tau_i}+\vec{1}(t-\tau_i)) \right] \
\end{align}
\end{widetext}
The integral has now been reduced from a $2D$ dimensional integral to a $j$ dimensional integral over the unstable manifold. It can be solved via Monte Carlo methods by sampling the distribution $\rho'_\alpha(\delta \vec{X}^j_0)$. The sum can be started at $t=0$ with the initial stability matrix being $\mathbf{M}_{t=0} = \mathbb{1}$, and this properly captures the initial decay for return probabilities. By the time of the first recurrence it can be assumed that the stability component along the stable manifold is zero. The proliferation of sums that come with stable degrees of freedom is admittedly a drawback to the solution of Eq. (\ref{final_tc2}). That being said, this approximation was able to be successfully applied to both integrable and chaotic cases of the two dimensional quartic oscillator.

\subsection{Application to two coupled pure quartic oscillators}

In this section, the reduced dimensionality formulas are tested on two coupled quartic oscillators for different dynamical regimes. The case of return probabilities, i.e., $\alpha = \beta$, is sufficiently general to test the main ideas and is considered here. 

\subsubsection{Chaotic case: $\lambda =-.60$}

\begin{figure}[ht]
\center
\includegraphics[trim = {2cm 1cm 3cm 2cm}, width= \linewidth]{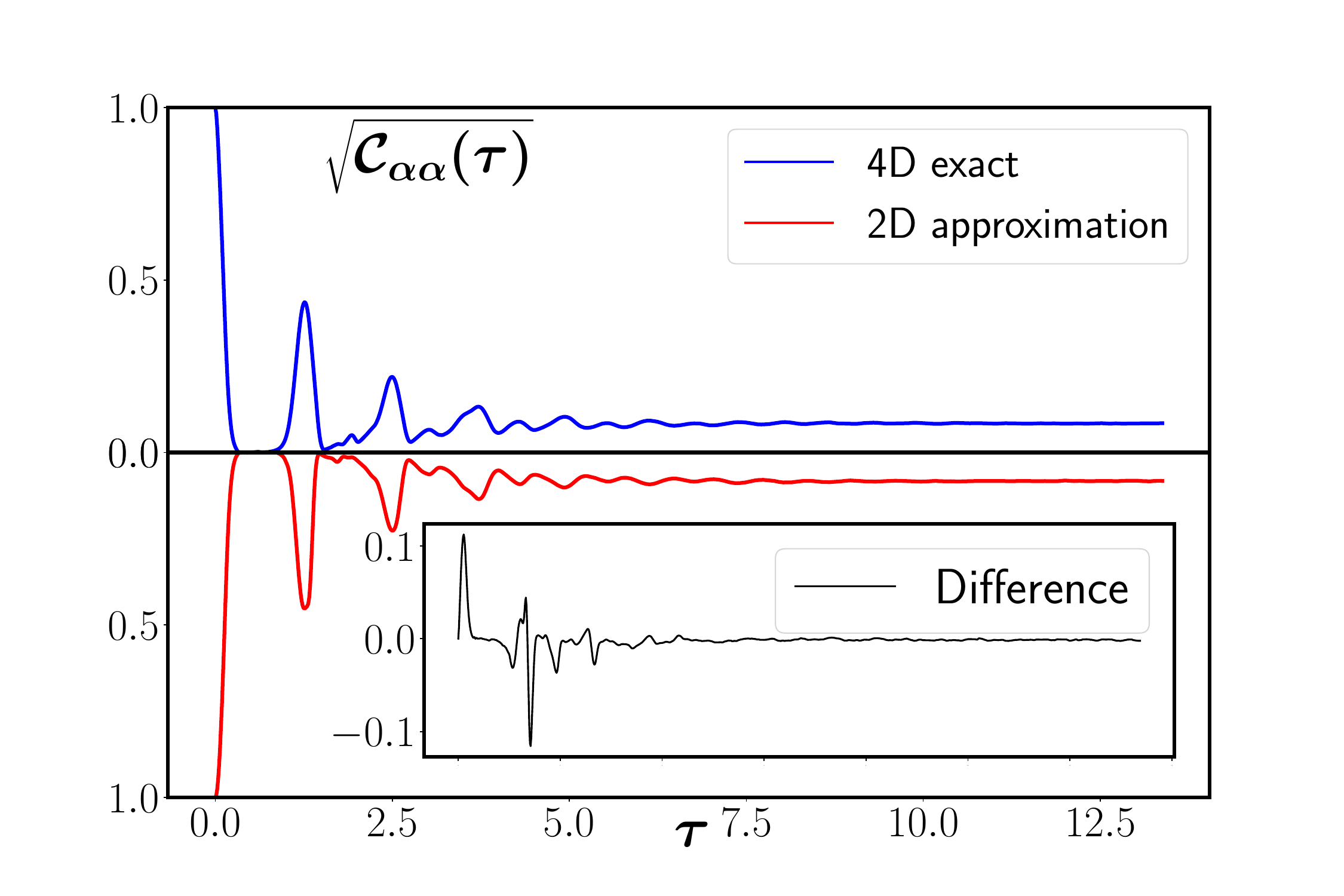}
\includegraphics[trim = {2cm 1cm 3cm 0cm}, width= \linewidth]{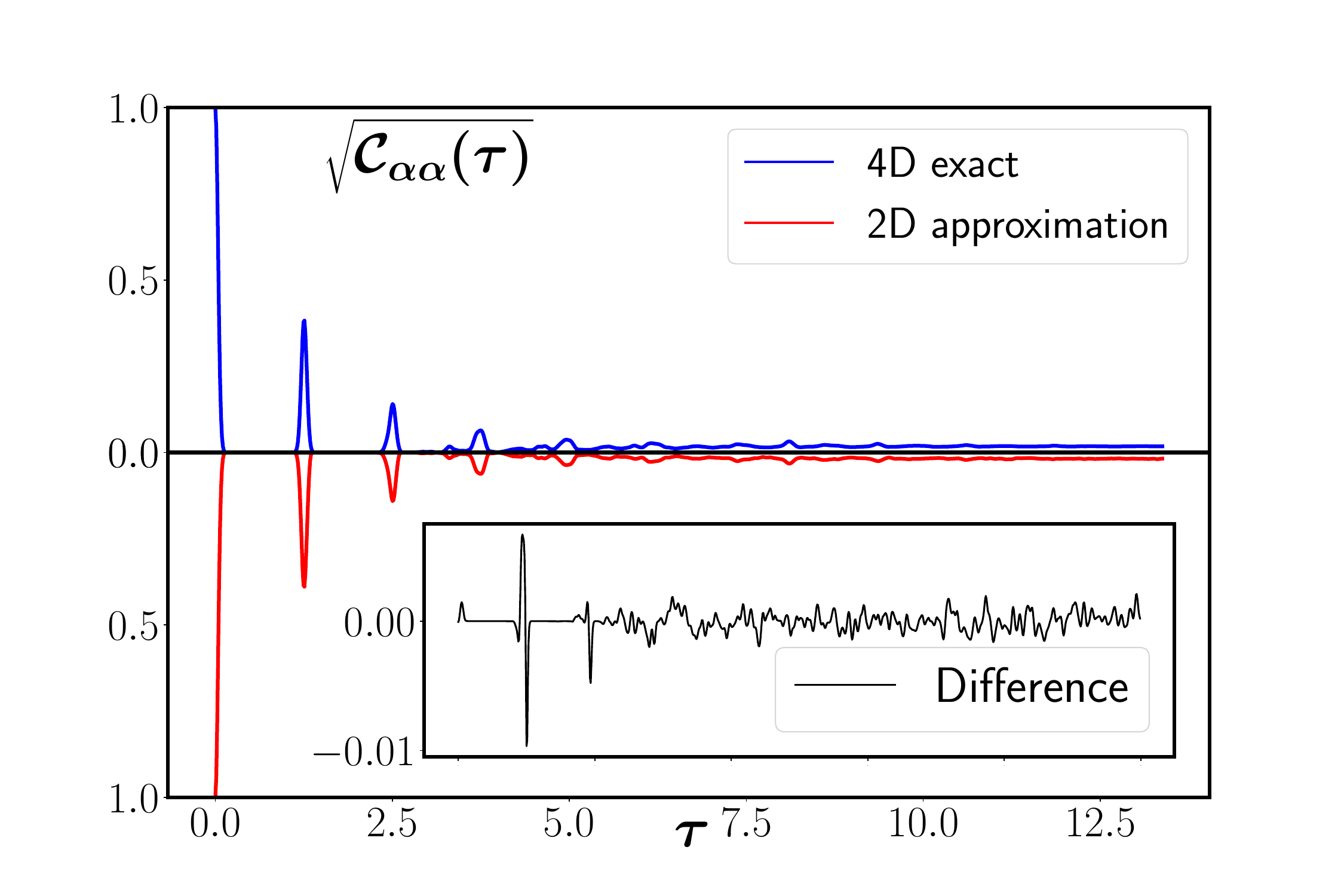}
\caption{Comparison of $\sqrt{\cal{C}_{\alpha\alpha}(\tau)}$ given by the square root of Eq.~\eqref{wmtc3} and the approximation given by  the square root of Eq.~\eqref{final_tc} for the case of $\lambda = -.60$. The approximation is a two-dimensional Monte Carlo using only the unstable manifold and the direction perpendicular to the energy surface. In the upper panel $\hbar=10$ and the lower panel $\hbar=1$. The error between the panels is seen to scale by a factor of $\hbar$. initial conditions used were $(\vec{p}_\alpha, \vec{q}_\alpha) = (30, 40, 0, 0)$ and the timescale is given by $t/\tau_{1,1}$. }
\label{fig7}
\end{figure}

For the chaotic case it is possible to reduce the original four dimensional integral to a one dimensional integral along the unstable manifold. However, the simpler formula Eq.~\eqref{final_tc}, can also be used to reduce the calculation to a two dimensional integral. This formula does not involve a summation and is therefore easier to implement numerically and may be more practical, depending on the circumstances. Here, results from both formulas are shown. The details of implementing Eq.~\eqref{final_tc} for the case of return probabilities are the same as described in the expectation values section. The results for $\hbar=10$ and $\hbar=1$ are shown in Fig.~\ref{fig7} where Eq.~\eqref{final_tc} is seen to be an excellent approximation. 

\begin{figure}[ht]
 \center
    \includegraphics[trim = {2cm 1cm 3cm 2cm}, width= \linewidth]{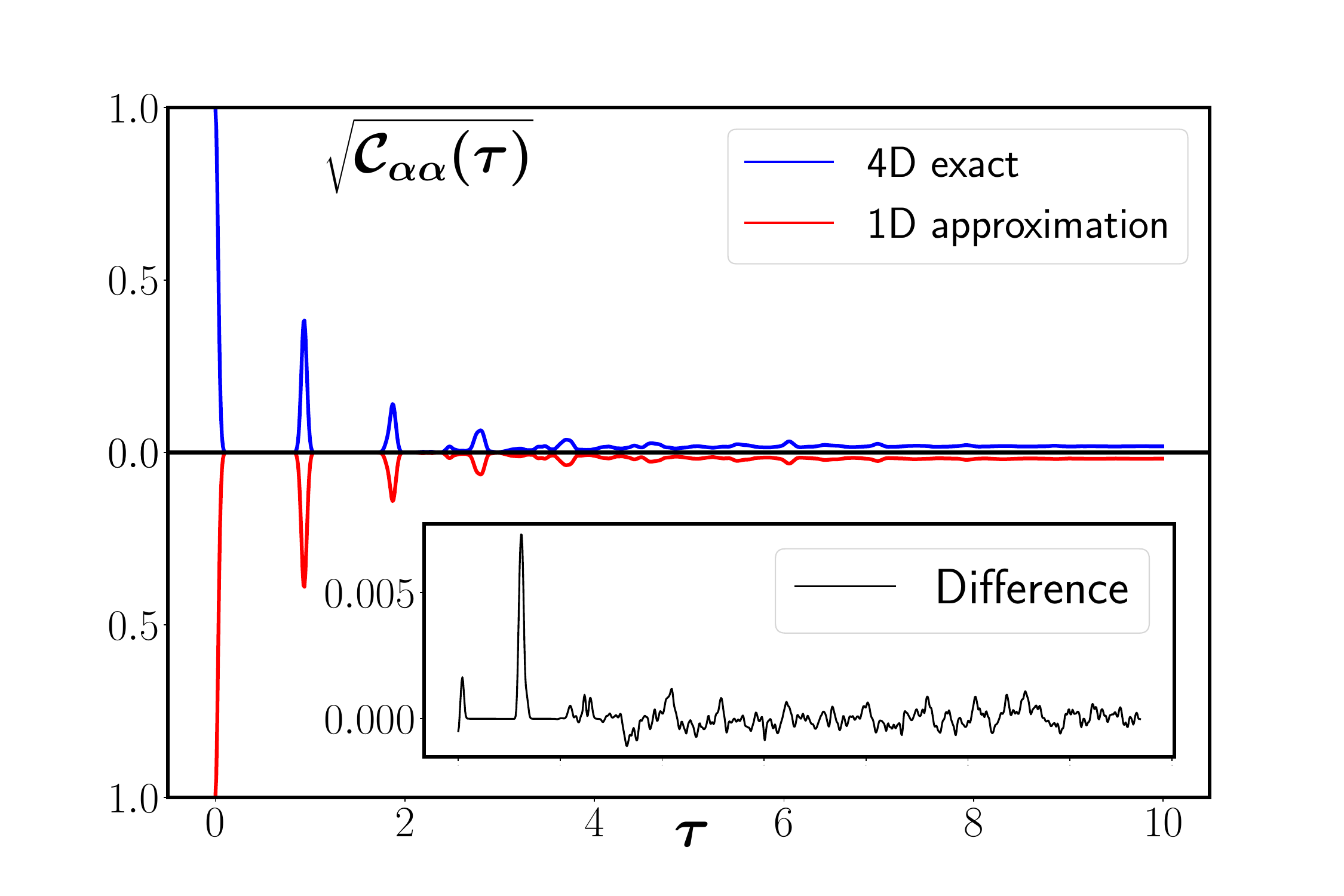}
    \caption{Comparison of $\sqrt{\cal{C}_{\alpha\alpha}(\tau)}$ given by the square root of Eq.~\eqref{wmtc3} and the approximation given by the square root of Eq.~\eqref{final_tc2} for the case of $\lambda = -.60$. The approximation is a single dimensional Monte Carlo using only the unstable manifold. The initial conditions used are $(\vec{p}_\alpha, \vec{q}_\alpha) = (30, 40, 0, 0)$ with $\hbar = 1$.  \label{fig8}}
\end{figure}

To integrate over the energy coordinate there needs to be a sum over each recurrence of the orbit. These recurrences occur when the trajectory makes its closest approach to the final wave packet centroid and has to be identified numerically. This is implemented by recognizing the greatest contribution to the integral comes when the function,
\begin{equation}
\label{min}
f(\vec{q}_{j,t}, \vec{p}_{j.t}) = (\vec{p}_{j,t} - \vec{p}_\beta, \vec{q}_{j,t} - \vec{q}_{\beta}) \cdot \mathbf{A_\beta} \cdot (\vec{p}_{j,t} - \vec{p}_\beta, \vec{q}_{j,t} - \vec{q}_\beta),
\end{equation} 
is minimized. The times at which the local minima occur are the $\tau_i$ described in the previous section and each one needs to be summed over. Since here there is only one constant of the motion, $\tau_i = \tau_{i_1}$ . In practice, a criteria can be created to throw away minima that are not small enough to contribute to the integral, i.e., $5\sigma$ from the final wave packet centroid. The minima change as a function of the component along the unstable manifold and thus have to be calculated for each sampled trajectory. It is necessary to know the stability matrix, $\mathbf{M}_{t}$, in the subspace being pre-integrated. Here it is a $3\times 3$ matrix with a $2\times2$ block corresponding to $(\delta E, \delta t)$ and $1\times1$ block corresponding to the direction along the stable manifold. 
\begin{equation}
\mathbf{M}_t = \begin{pmatrix} 1 & 0 & 0 \\ 
\omega'_\alpha t & 1 & 0 \\ 0 & 0 & e^{-\lambda t} \end{pmatrix} 
\end{equation}
To identify the shearing rate, the stability matrix in the $(\vec{p}_\alpha,\vec{q}_\alpha)$ coordinate system has to be found numerically and transformed to the new coordinate system using Eq.~\ref{lh} with the transformation Eq.~\ref{matrix3}. The shearing rate only needs to be found once before running the Monte Carlo. The stability component along the stable manifold decays exponentially fast and can be set to zero by the time of the first recurrence without harming the approximation. With this information, the evaluation of Eq.~\eqref{final_tc2} for the chaotic case of the two coupled quartic oscillators is a straightforward numerical calculation of a single integral using Monte Carlo methods and summing over the times of closest approach. The result compared to the four dimensional Monte Carlo is shown in Fig.~\ref{fig8}. Comparing the differences seen in Figs.~\ref{fig7} and \ref{fig8}, the single dimension approximation is at least as good as the excellent double integral approximation, perhaps even slightly better.

\subsubsection{Integrable case: $\lambda = 0$}

\begin{figure}[ht]
\centering
    \includegraphics[trim = {2cm 1cm 3cm 2cm}, width= \linewidth]{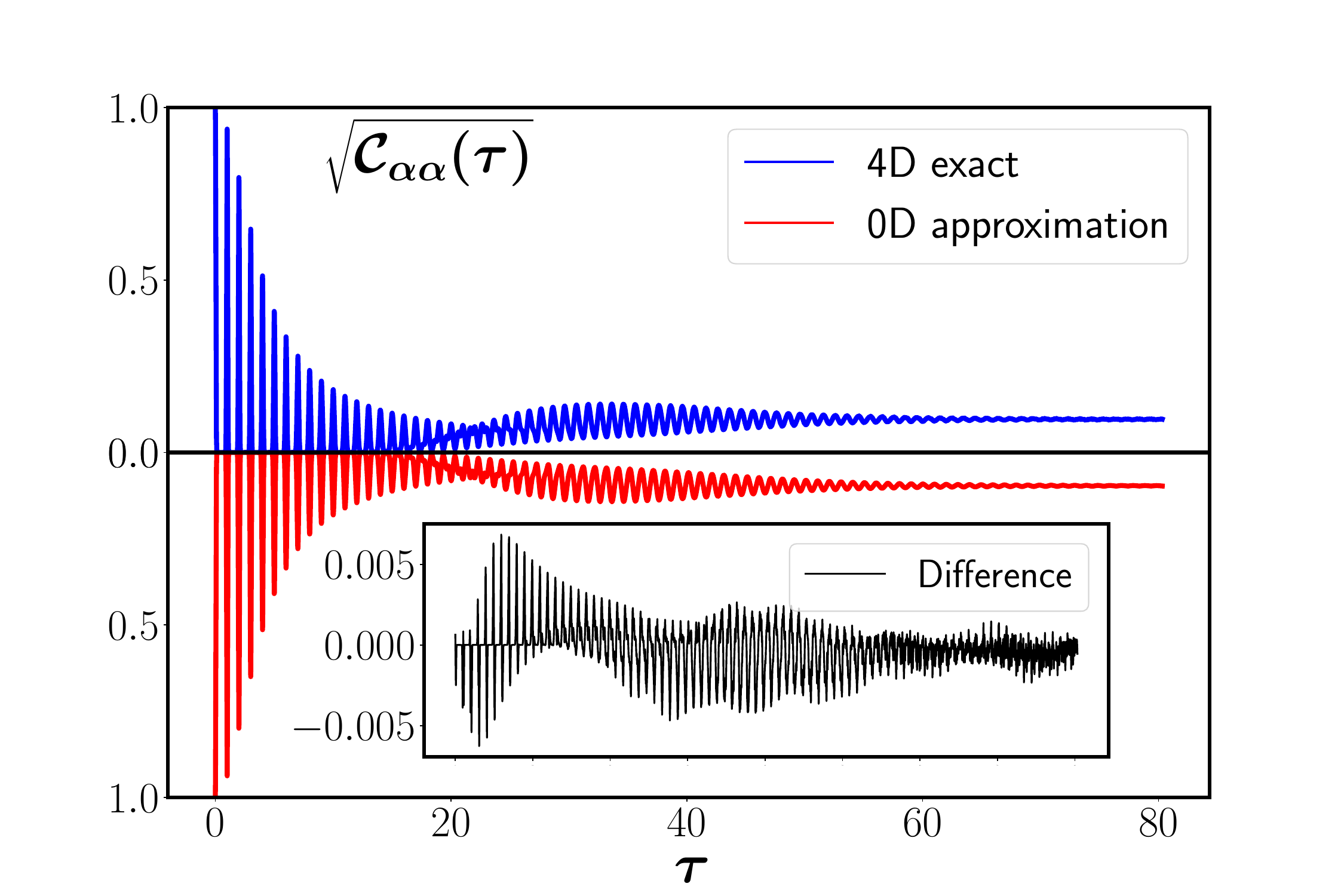}
    \caption{Comparison of $\sqrt{\cal{C}_{\alpha\alpha}(\tau)}$ given by the square root of Eq.~\eqref{wmtc3} and the approximation given by the square root of Eq.~\eqref{int_sp} for the case of $\lambda = 0$. The exact case is a four dimensional integral solved with Monte Carlo methods, whereas the approximation is just an evaluation of a double sum. The initial conditions used are $(\vec{p}_\alpha, \vec{q}_\alpha) = (30, 40, 0, 0)$ with $\hbar = 1$.  The approximation is plotted reflected as otherwise the two curves lie on top of each and cannot be distinguished.  The inset shows the small differences between the two approaches.} 
    \label{fig6}
\end{figure}

For the case of $\lambda =0$ it is possible to integrate analytically all of the coordinates. This case is not solved by treating the problem as two uncoupled quartic oscillators. For consistency with the chaotic case, the same analytic transformation, Eq.~\ref{matrix3} at $\lambda =0$ is used, so that the time and energy coordinates are that of the combined system and not of the individual oscillators. This approach helps elucidate how the sum over the return times associated with a stable degree of freedom works, since for this case $\tau_i = \tau(\tau_1, \tau_2)$ where $\tau_1$ and $\tau_2$ are the periods of the individual oscillators. The complete set of formulas are derived in Appendix~\ref{2approx}, and using these results it is possible to do all four integrals by evaluating a double sum. The result for return probabilities is shown in Fig.~\ref{fig6}. The double sum (zero-dimensional Monte Carlo) is quite accurate as the differences between the four dimensional Monte Carlo and fully pre-integrated version are less than a percent.

\subsubsection{Mixed case: $\lambda = -.35$}

\begin{figure}[ht]
   \center
    \includegraphics[trim = {2cm 1cm 3cm 2cm},clip, width= \linewidth]{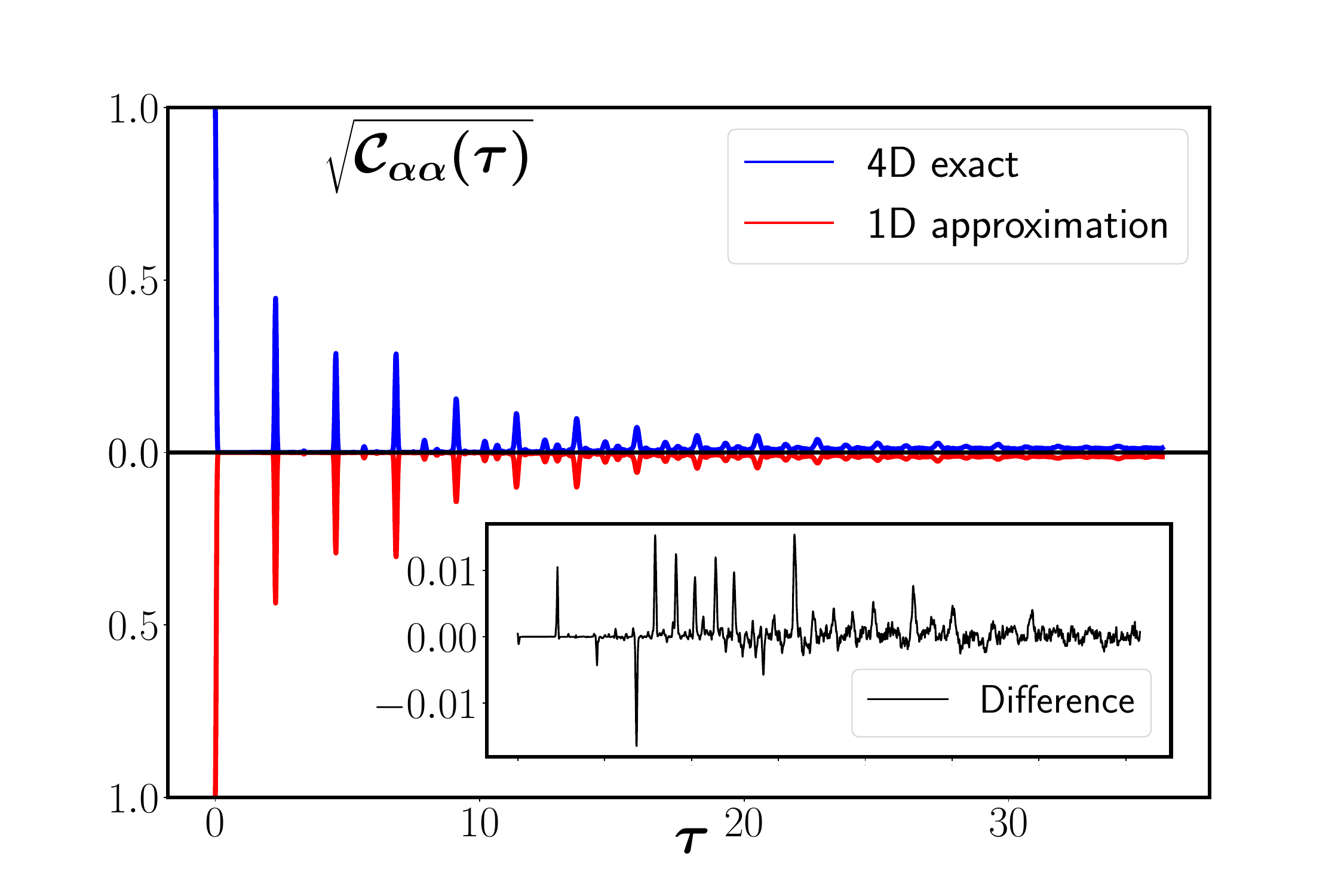}
    \caption{Comparison of $\sqrt{\cal{C}_{\alpha\alpha}(\tau)}$ given by the square root of Eq.~\eqref{wmtc3} and the approximation given by the square root of Eq.~\eqref{final_tc2} for the case of $\lambda = -.35$ in a chaotic region of phase space.  The approximation is a single dimensional Monte Carlo using only the unstable manifold. The initial conditions are $(\vec{p}_\alpha, \vec{q}_\alpha) = (30, 80, 0, 0)$ with $\hbar = 1$. \label{fig9} }
\end{figure}

The mixed case has both chaotic regions of phase space and regular regions where the motion is stable (shearing or rotational).  Here, a chaotic region is selected for illustration. To identify the appropriate region of phase space, the surface of sections from Ref.~\cite{Bohigas93} can be used. In principle, if the initial conditions of the central trajectory are chosen to be inside a regular island then a reduction to a zero dimensional integral is possible. However, in practice this is a considerable challenge due to the presence of unknown constants of the motion and would necessitate additional numerical methods. This is left to the subject of potential future research as is discussed in the conclusion. In a chaotic region the same methods as used for the $\lambda = -.60$ case must apply. Although, the situation is more complicated due to transport barriers in the dynamics (see \cite{Bohigas93}). To guarantee the region of phase space is chaotic, the stability exponents can be analyzed to ensure exponential behavior. The results are shown in Fig.~\ref{fig9} and are seen to only be slightly less accurate than for the $\lambda = -.60$ case.\\

\section{Approximation limitations}

\subsubsection{Wigner transform and manifold curvature}

The results of the previous sections relied on linearizations of the dynamics throughout the local regions of phase space where the contributions to the integrals are appreciable.  There are choices of central trajectory initial conditions or shape parameters (e.g., $\mathbf{A_\alpha}$ matrix) for which the linearization is not as good of an approximation as other choices.  Most obviously if the energy of the central trajectory is too small (or $\hbar$ too large), curvature corrections become more important. There are also regions of phase space where there is significant curvature of the energy surface, or, in fact, any of the invariant sets in the dynamics (related to other constants of the motion or the unstable/stable manifolds) through the contributing region of phase space.  In such locations, any linearization would be less accurate and appropriate.  Another example would be the interface between regular and chaotic motion in a system with a mixed phase space (the fractal boundary of a regular island).  

\begin{figure}[ht]
    \includegraphics[width= \linewidth, trim= {3cm 0cm 3cm 1cm}]{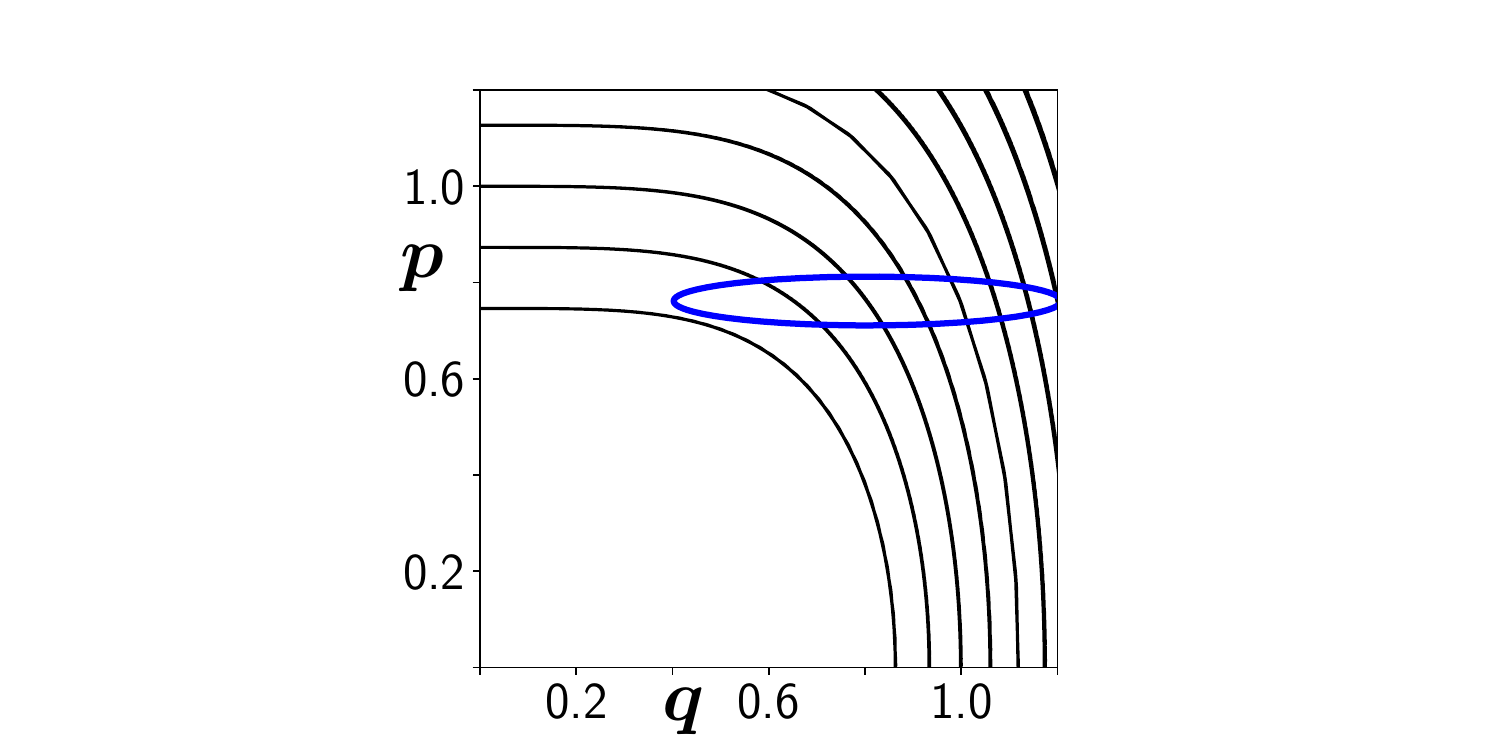}
    \includegraphics[width= \linewidth, trim ={3cm 1cm 3cm 0cm}]{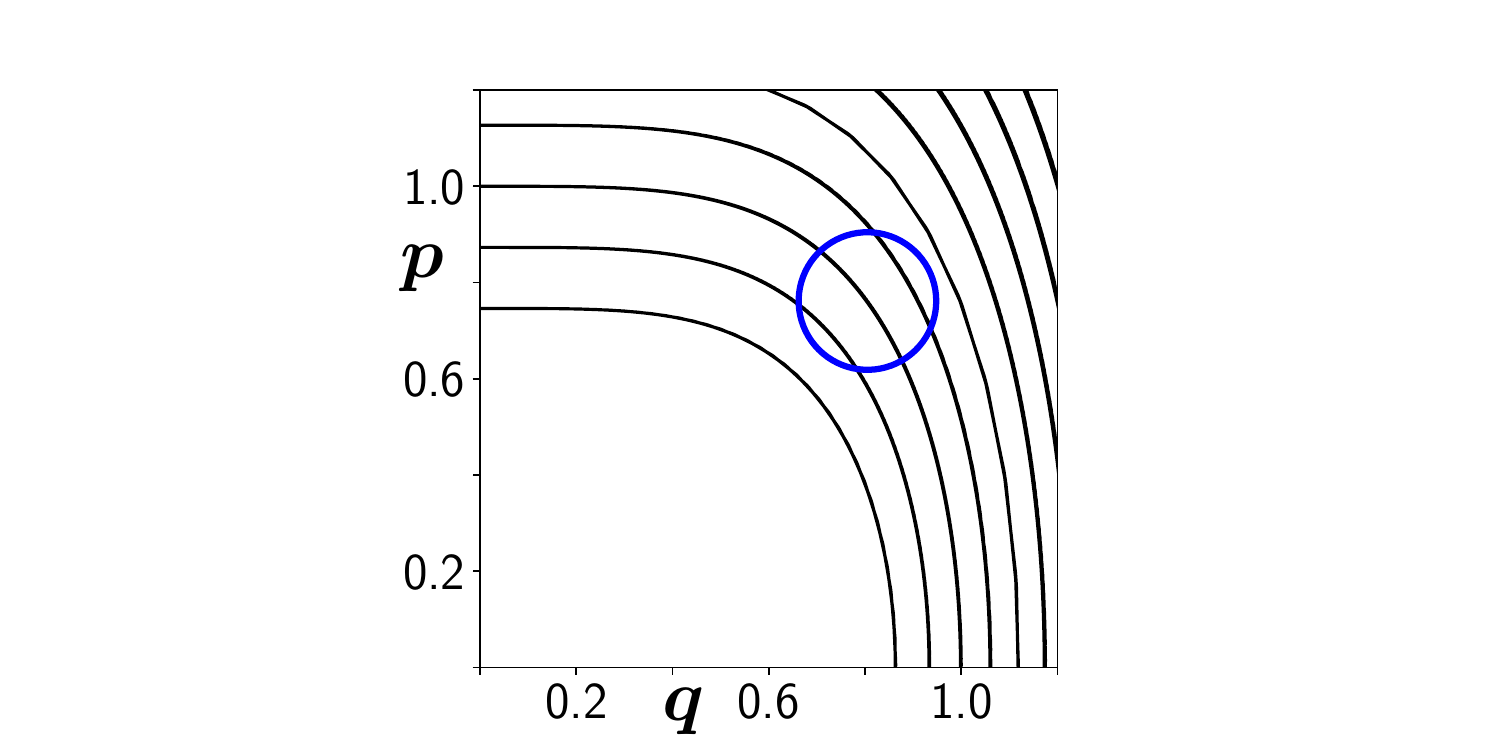}
    \caption{Hamiltonian contours of a one dimensional quartic oscillator are plotted with the $2\sigma$ ellipse of the Wigner transform. The $q$ and $p$ axes have been scaled by dividing by the maxima of $q_\alpha$ and $p_\alpha$, respectively. In the top panel, $\mathbf{A_{\alpha}}$ is taken to be the identity matrix. Due to the location of the Wigner transform, there is significant curvature of the energy surface through the ellipse and reliance on a linearization of the trajectories leads to an approximation with greater inaccuracy; see Figs.~\ref{fig11} and \ref{fig12} for its effects on return probabilities and expectation values, respectively. In the bottom panel, $\mathbf{A}_{\alpha}$ has been dilated by a factor of $p_{\alpha,\text{max}} / q_{\alpha, \text{max}}$ to compact the ellipse into a circle and the linearization procedure for both expectation values and return probabilities is now much improved. The initial conditions used are $(p_\alpha, q_\alpha) = (30,4)$.  \label{fig10} }
   
\end{figure}

\begin{figure}[ht]
    \includegraphics[width= \linewidth, trim = {1cm 0cm 1cm 2cm}]{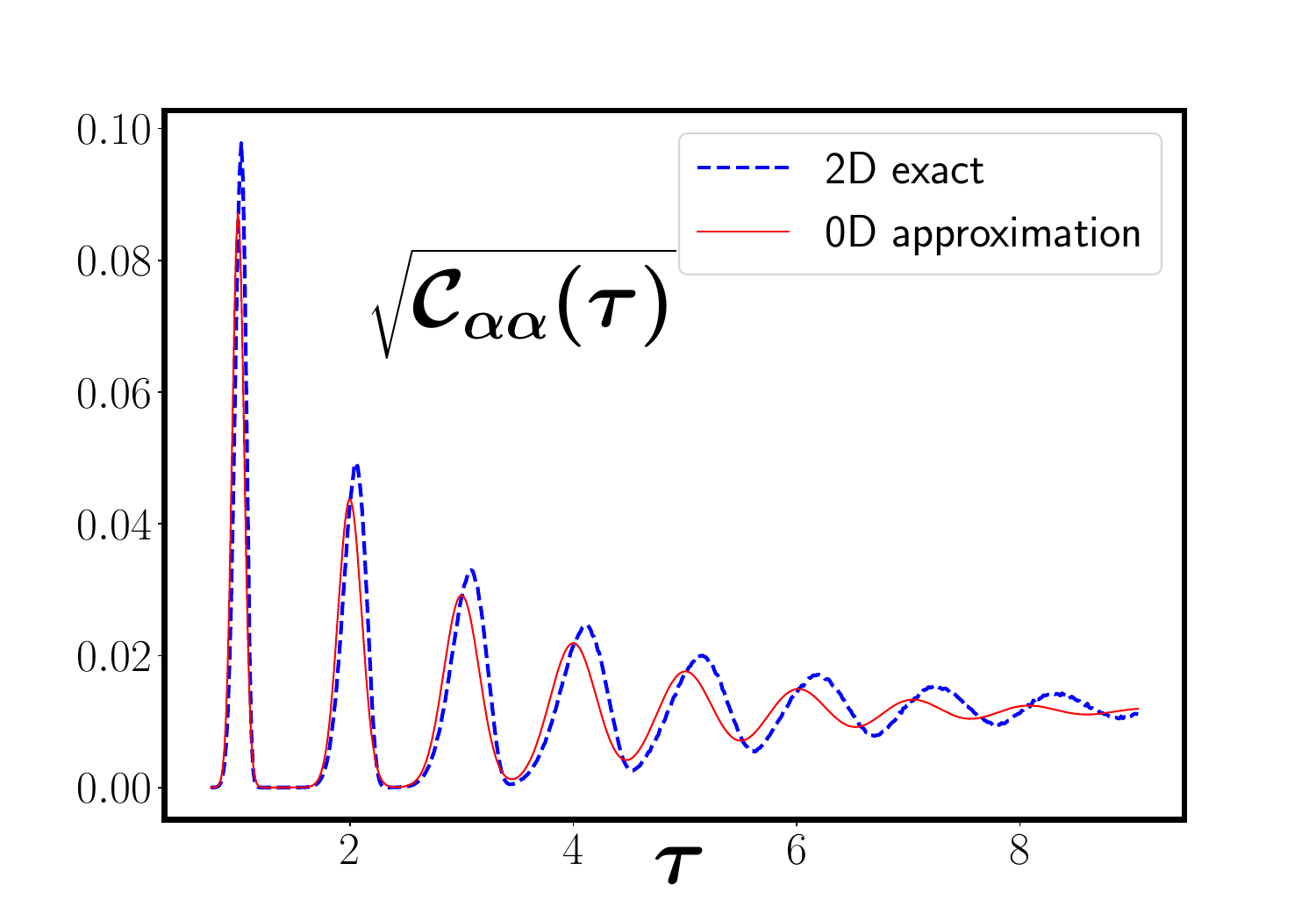}
     \includegraphics[width= \linewidth, trim = {1cm 0cm 1cm 0cm}]{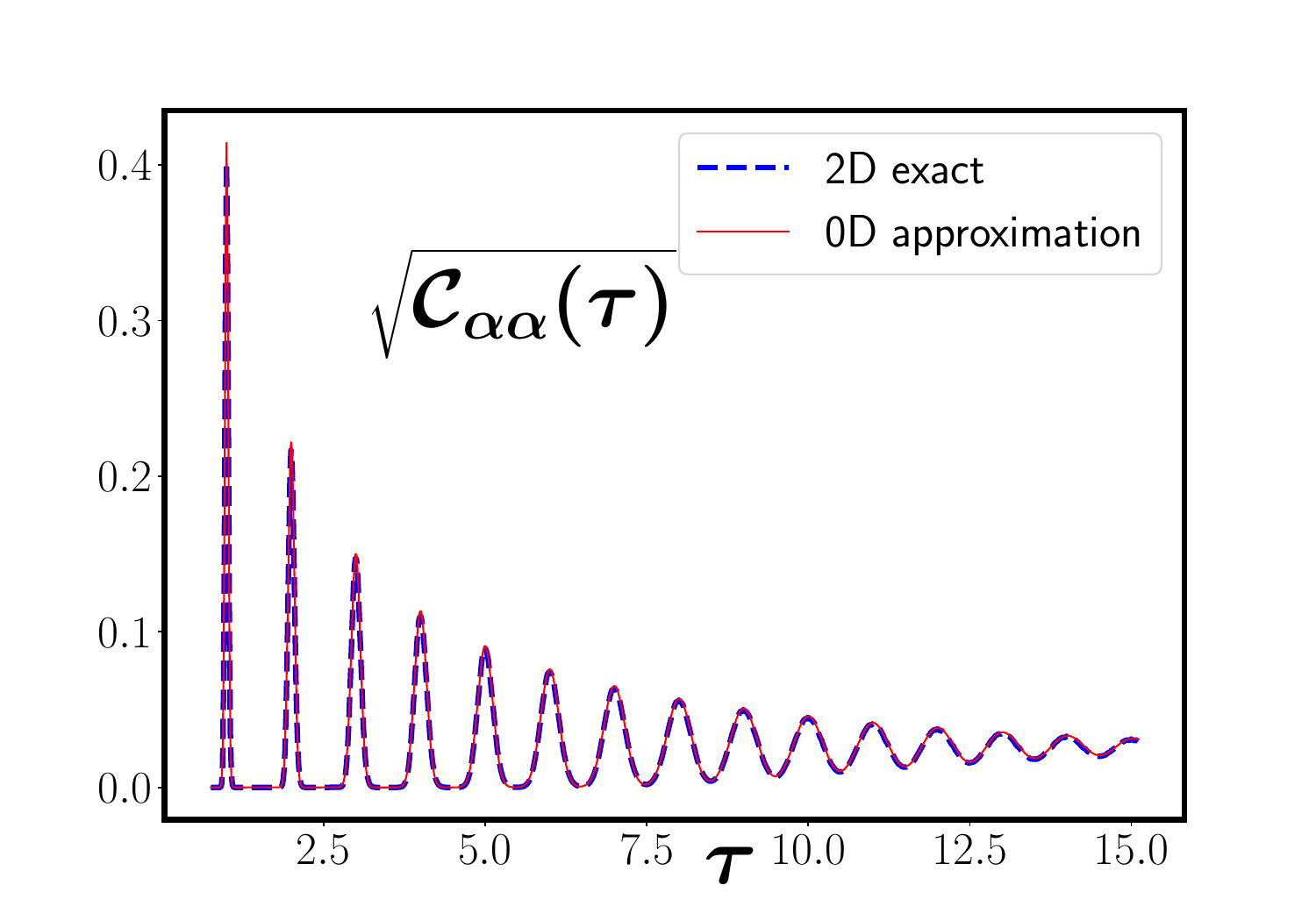}
      \caption{Return probabilities of a one dimensional quartic oscillator. The exact results from a Monte Carlo simulation are compared to an approximation analogous to Eq. (\ref{int_sp}). The return probabilities for the upper and lower panels are calculated with the Wigner transforms corresponding to the upper and lower panels of Fig.~\ref{fig10} respectively. The initial conditions used are $(p_\alpha, q_\alpha) = (30,4)$, with $\hbar = 1$.
    }
    \label{fig11}
\end{figure}

\begin{figure}
    \includegraphics[width= \linewidth, trim = {1cm 0cm 1cm 2cm}]{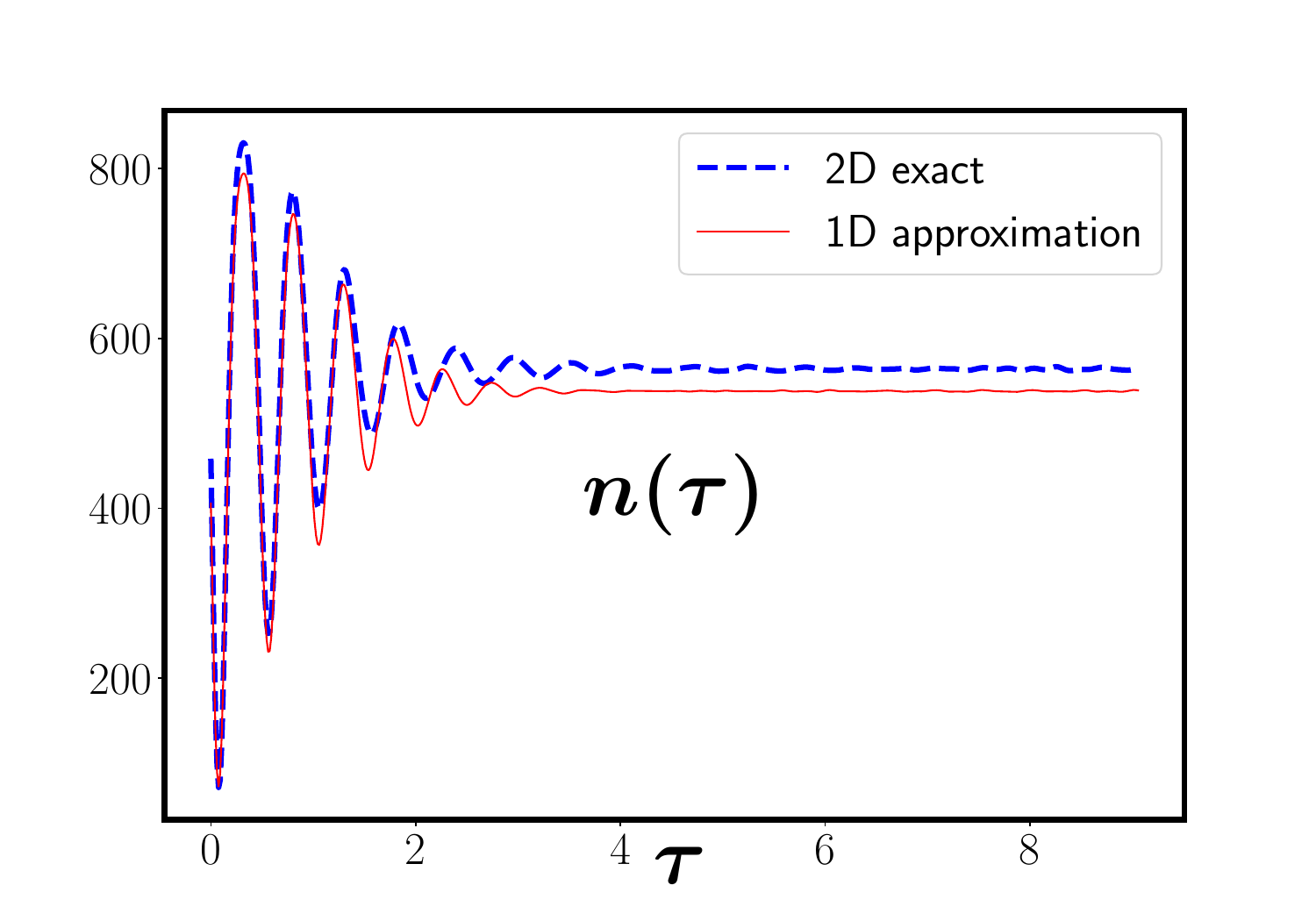}
    \includegraphics[width= \linewidth, trim = {1cm 0cm 1cm 0cm}]{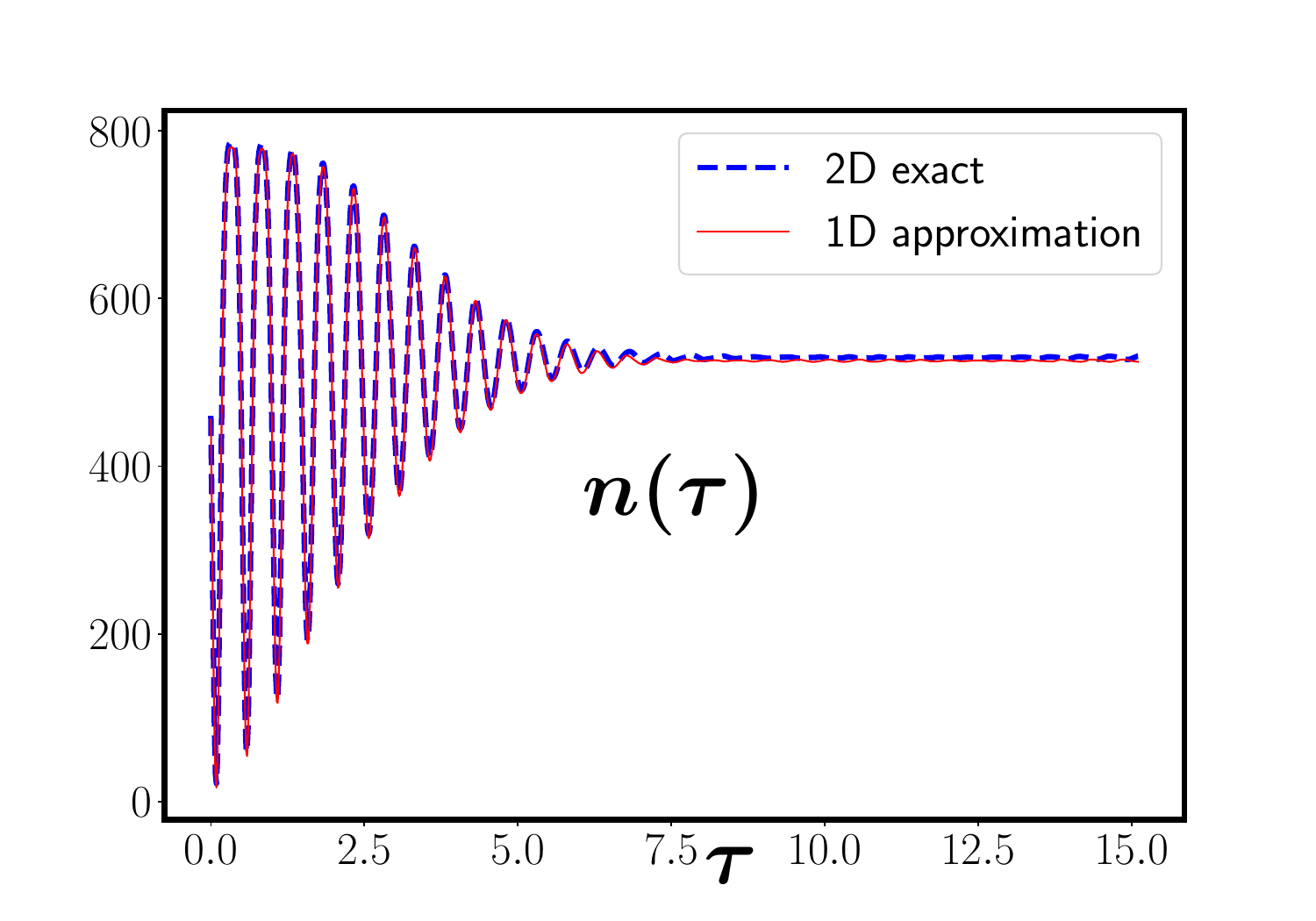}
    \caption{Expectation value of Eq.~\eqref{nop} for a one dimensional quartic oscillator. The exact results from a Monte Carlo simulation are compared to the approximation Eq. (\ref{int3}). The expectation value for the upper and lower panels are calculated with the Wigner transforms corresponding to the upper and lower panels of Fig.~\ref{fig10}, respectively. The initial conditions used are $(p_\alpha, q_\alpha) = (30,4)$, with $\hbar = 1$.
    }  
    \label{fig12}
\end{figure}

It is possible to use the Wigner transform density to assess how good or bad a particular choice may be. The region where the integral contributes can be visualized as an elliptical projection of a certain number of $\sigma$, say two or three, of the Wigner transform. By plotting the Hamiltonian contours (or other constants of the motion or unstable/stable manifolds) crossing through the ellipse, it becomes clear where the linearization fares worse. Under some circumstances, it may be possible to alter the choice of $\mathbf{A_\alpha}$, thus altering the eccentricity of the elliptical projection or shift the central coordinates $(\vec{p}_\alpha, \vec{q}_\alpha)$ to a different phase space location (and $(\vec{p}_\beta, \vec{q}_\beta)$) in order to reduce curvature corrections.  As a simple illustration, this procedure is carried out for a single pure quartic oscillator and the return probabilities and expectation value of Eq.~\eqref{nop} for two cases are calculated, one in which there is clearly more curvature of the energy surface within the Wigner transform elliptical projection than the other.  The elliptical projections are illustrated in Fig.~\ref{fig10}, the two $\sqrt{\cal{C}_{\alpha\alpha} (\tau)}$ in Fig.~\ref{fig11}, and the two $n(\tau)$ in Fig.~\ref{fig12}. The linearization approximation clearly improves as seen in the $\sqrt{\cal{C}_{\alpha\alpha} (\tau)}$ and $n(\tau)$ comparisons for the case in which the Wigner transform more closely resembles a circle and there is less energy surface curvature inside the contour.

\subsubsection{Influence of unstable/stable manifolds}

As explained in Sec.~\ref{ideal} (see Eqs.~\eqref{direct} and \eqref{direct2}), the coordinate transformations in the subspaces of the unstable/stable manifolds diagonalize the unstable blocks of the stability matrices only after a convergence time $t_c$. The initial decay of the return probabilities for the chaotic case, Fig.~\ref{fig7}, is not an issue since it can be evaluated in a separate calculation;  see Appendix~\ref{initdec}. In Fig.~\ref{fig8}, the initial decay is properly handled by taking the sum to start at $t=0$ and setting $\mathbf{M}_0 = \mathbb{1}$.  Generally speaking, the first recurrence might well occur before $t_c$ and this explains why the first recurrence is seen to have the largest error. However, the approximation still captures the essential behavior.  In general, the integrals calculated do not exhibit very much sensitivity to the choices of directions for the unstable/stable manifolds. First, the result for the integral over the stable manifold does not depend on its final specific direction and due to the powers of $\hbar^{1/2}\exp(-\lambda_j t)$ scaling of the correction terms, they are just dropped here.  Second, for the unstable manifolds any variations from the true directions collapse exponentially back on to the correct directions, except if the directions are, to an extremely high precision, in exactly the wrong directions (e.g., an unstable manifold chosen perfectly to be a stable manifold or chosen as the gradient of the Hamiltonian). In this sense, the chaos in the dynamics helps to correct any error in the approximations of the coordinate transformations. 

\section{Conclusion}

The idea of decomposing phase space into composite degrees of freedom is well known, and an explicit realization of a block diagonal stability matrix using a nontrivial dynamical coordinate transformation is provided here. It can potentially be a very practical and useful technique; here it facilitates reducing the number of degrees of freedom that must be sampled in a classical Monte Carlo calculation. The idea of decomposing phase space is presented in~\cite{Gaspard98}, but the coordinate transformation there does not result in a block diagonal stability matrix. In~\cite{Hummel22B}, time independent coordinate transformations motivated by symmetry arguments were used to block diagonalize the stability matrix, but this lacks the additional difficulties of dynamical transformations. Furthermore, the construction of a local coordinate system, manifesting the form of a shearing block in the corresponding subspace of the stability matrix is also given. Even for the simple one dimensional quartic oscillator presented in Eq.~\eqref{1dquartic}, it is not clear how to construct the linear coordinate transformation from $(\delta q, \delta p) \rightarrow (\delta I, \delta \theta)$ without the method described in Sec.~\ref{derive}. The rigorous construction of local time and energy coordinates shows the identification of the time direction is nontrivial, but could be technically quite useful, as the idea is widely used in semiclassical derivations~\cite{Haake10}. \

An important application of this phase space decomposition is the evaluation of integrals with Monte Carlo methods. A wide variety of observables in physical systems can be calculated by sampling initial conditions in phase space and propagating classical trajectories. The classical Wigner method, utilizing the Wigner transform of minimum uncertainty quantum states, provides an ideal framework, as this leads to Gaussian weighting functions which play an important role in many-body bosonic systems. Moreover, the calculations are essentially identical to those used in the truncated Wigner approximation, albeit they do not come from a mean field approximation to the dynamics. The results presented here demonstrate it is possible to pre-integrate at least half of the dimensions analytically for expectation values and transport coefficients, and in some cases even more dimensions.  Specifically, it has been shown directions along the stable manifolds and those conjugate to constants of the motion can always be integrated before setting up the Monte Carlo. Further pre-integrations are possible for constants of the motion with rotational dynamics locally, and the degrees of freedom corresponding to any of the constants of the motion can be integrated out for the case of return probabilities. The approximations can reproduce the exact Monte Carlo calculations with only minimal error. It has also been shown how to control the error in the approximation by appropriately shaping the elliptical projection of the Wigner transform. 

There are several directions of future research that are worth investigating. The analytic methods used to derive the local action angle coordinate system for the two dimensional coupled quartic oscillator need to be extended to more general Hamiltonians; specifically those with physical relevance such as the Bose-Hubbard model. The fact that it was shown to be possible for a non-trivial chaotic Hamiltonian is promising that it should be possible for more general Hamiltonians that lack homogeneous constants of the motion. It also might be possible to develop numerical methods that circumvent the need for the analytic techniques. Since the directions that define the local coordinate system in the subspace associated with constants of the motion are respected by the local linearized flow, it is conceivable that the necessary information is encoded in the stability matrix. This would be especially relevant for the case of unknown constants of the motion, e.g. inside an integrable island for a mixed phase space dynamics. The numerical techniques applied here in the subspace of the unstable degrees of freedom have been successfully applied in systems with more degrees of freedom and to the entire phase space in a search of saddles~\cite{Tomsovic18b}, without the rigorous decoupling of the time and energy coordinates. This suggests for large dimensional, strongly chaotic systems it might be possible to perform the integrations over the stable manifolds without removing the stable degrees of freedom. Another avenue of future research is to apply the integration techniques to semiclassical initial value representations, such as the Herman-Kluk propagator~\cite{Herman84, Miller01}. This would be similar to Kocia and Heller's work ~\cite{Kocia14b, Kocia15}, except with the full machinery of decoupling the phase space and proper identification of stable/unstable manifolds presented here. 

Further work may lead to streamlining techniques developed in this paper in order to make the application to Monte Carlo simulations less onerous. There are certain applications where reducing the dimensions of Monte Carlo simulations are more important. For instance, the aforementioned semiclassical initial value representations have oscillatory integrands, making Monte Carlo simulations less effective. In this case, going through the steps presented here in order to do the pre-integrations may be more advantageous. There are also situations involving systems in atomic physics where the relevant equations of motion lead to a discretized Gross-Pitaeveskii equation. It can be shown through a stability analysis that under certain conditions a vast majority of the degrees of freedom have vanishing Lyapunov exponents and behave locally as harmonic oscillators. In such a case, it would be possible to pre-integrate all but a small number of degrees of freedom, making the effort in applying the technique more consequential. The undertaking of these potential applications is reserved for future research.  
\appendix

\section{Stability matrix block diagonalization derivation for coupled quartic oscillators}
\label{transformation}

As described in Sec.~\ref{ideal}, the necessary transformation consists of first deriving the local canonical coordinate transformation that decouples stable from unstable degrees of freedom, and then diagonalizing the unstable block of the stability matrix. The directions that give the local energy and time coordinates can be found using the method described in Sec.~\ref{derive}. It is desirable to have a transformation that works for any value of $\lambda$. For the time and energy coordinates this is always the case, but the directions that give the unstable block are not uniquely defined so one has choose the directions that work for any value of $\lambda$. There is a simple trick that accomplishes this. Repeating the Hamiltonian here:
\begin{align}
H = \frac{p_1^2}{2m} + \frac{p_2^2}{2m}  + q^4_1 / b + q^4_2b + 2\lambda q_1^2q_2^2
\end{align}
In the case $\lambda = 0$ the Hamiltonian is separable and there are two constants, $E_1$ and $E_2$. Considering this case for a moment, it is clear that a coordinate transformation can be made from $(\delta p_1, \delta q_1) \rightarrow (\delta E_1, \delta t_1)$ and $(\delta p_2, \delta q_2) \rightarrow (\delta E_2, \delta t_2)$. The transformation matrix that does this is just be that of the single quartic oscillator given by Eq.~\eqref{1d}, with the correct constants in each separate subspace. This is given by:
\begin{align}
\label{matrix1}
\begin{pmatrix} \delta E_1 \\ \delta t_1 \\ \delta E_2 \\ \delta t_2 \end{pmatrix} = \begin{pmatrix} p_1 & 0 & 4q_1^3/b & 0  \\-\frac{q_1}{4E_1} & 0 & \frac{p_1}{2E_1} & 0 \\ 0 & p_2 & 0 & 4q_2^3b \\ 0 & -\frac{q_2}{4E_2} & 0 & \frac{p_2}{2E_2}  \end{pmatrix} \begin{pmatrix} \delta p_1 \\ \delta p_2 \\ \delta q_1 \\ \delta q_2 \end{pmatrix}
\end{align}
The coordinate defined by the total energy is simply: $\delta{E} = \delta{E_1} + \delta{E_2}$. The conjugate coordinate, $\delta t$, is then given by Eq.~\eqref{angle} applied to the new coordinate system: 
\begin{equation}
\label{time}
\delta t = \frac{E_1}{E}\delta t_1 + \frac{E_2}{E}\delta t_2
\end{equation} 
There needs to be two other local coordinates that, along with $\delta t$ and $\delta E$, satisfy the requirements of a canonical transformation, i.e. the Poisson brackets. Referring to this conjugate pair as $(\delta P, \delta Q)$, the straightforward choice is the following:
\begin{align}
\label{qp}
\delta P = & \frac{E_2}{E}\delta E_1 - \frac{E_1}{E}\delta E_2 \\[6pt]
\delta Q = & -\delta t_1 + \delta t_2 \nonumber
\end{align}
Or in matrix form:
\begin{align} 
\label{matrix2}
\begin{pmatrix} \delta E \\ \delta t \\ \delta P \\ \delta Q \end{pmatrix}
= \begin{pmatrix} 1 & 0 & 1 & 0 \\ 0 & \frac{E_1}{E} & 0 & \frac{E_2}{E} \\ \frac{E_2}{E} & 0 & -\frac{E_1}{E} & 0 \\ 0 & -1 & 0 & 1 \end{pmatrix} \begin{pmatrix} \delta E_1 \\ \delta t_1 \\ \delta E_2 \\ \delta t_2 \end{pmatrix}
\end{align}
The multiplication of the transformations in Eqs.~\eqref{matrix1} and \eqref{matrix2} give the complete transformation. The result for $\lambda \neq 0$ can be recovered with the following:
\begin{align}
E_1 \rightarrow & E_1  + \lambda q_1^2q_2^2 \nonumber \\
E_2 \rightarrow & E_2 + \lambda q_1^2q_2^2 \nonumber \\
\delta E_1 \rightarrow & \delta E_1 + 2\lambda q_1 q_2^2 \delta q_1 + 2\lambda q_1^2q_2 \delta q_2 \nonumber\\
\delta E_2 \rightarrow & \delta E_2 + 2\lambda q_1 q_2^2 \delta q_2+ 2\lambda q_1^2q_2 \delta q_2 
\end{align}
So that the full transformation is:
\begin{align}
\label{matrix3}
\mathbf{S_\alpha} & = \nonumber \\ 
& \begin{pmatrix} 1 & 0 & 1 & 0 \\ 0 & \frac{E_1}{E} & 0 & \frac{E_2}{E} \\ \frac{E_2}{E} & 0 & -\frac{E_1}{E} & 0 \\ 0 & -1 & 0 & 1 \end{pmatrix} \times \nonumber \\
&  \begin{pmatrix} p_1 & 0 & 4q_1^3/b + 2\lambda q_1 q_2^2 & 2\lambda q_1^2 q_2 \\-\frac{q_1}{4E_1} & 0 & \frac{p_1}{2E_1} & 0 \\ 0 & p_2 & 2\lambda q_1 q_2^2 & 4q_2^3b + 2\lambda q_1^2q_2 \\ 0 & -\frac{q_2}{4E_2} & 0 & \frac{p_2}{2E_2}  \end{pmatrix}
\end{align}
This transformation block diagonalizes the stability matrix: 
\begin{equation}
\mathbf{M}'_{\alpha,t} = \mathbf{S}_{\alpha,t} \mathbf{M}_{\alpha,t} \mathbf{S}^{-1}_{\alpha,0} = \begin{pmatrix}
\mathbf{M}^{sh}_{t} & \mathbf{0} \\ \mathbf{0} & \mathbf{M}^h_{t} 
\end{pmatrix}
\end{equation}
No more work is required for the shearing block, and $\mathbf{M}^{sh}_t$ is in the form of Eq. \ref{shear}:
\begin{equation}
\mathbf{M}^{sh}_t = \begin{pmatrix}
 1 & 0 \\ \omega'_{\alpha} t & 1 \end{pmatrix}
\end{equation}
In the case that $\lambda = 0$, $\mathbf{M}^h_t$ reduces to a second shearing degree of freedom. For a value of $\lambda$ that makes the dynamics chaotic, such as $\lambda = -.60$, $\mathbf{M}^h_t$ has no particular structure, but is completely hyperbolic and can be approximately diagonalized using the technique described in Sec.~\ref{ideal}. To be explicit, this looks like:
\begin{equation}
\begin{aligned}
&{} \mathbf{M}'_{\alpha,t} \rightarrow \mathbf{O}_t \mathbf{M}'_{\alpha,t} \mathbf{O}^T_0 =
\begin{pmatrix}
\mathbf{1} & \mathbf{0} \\ \mathbf{0} & \mathbf{O}_t
\end{pmatrix}
\begin{pmatrix}
\mathbf{M}^{sh}_{t} & \mathbf{0} \\ \mathbf{0} & \mathbf{M}_{h} 
\end{pmatrix}
\begin{pmatrix}
\mathbf{1} & \mathbf{0} \\ \mathbf{0} & \mathbf{O}^T_0
\end{pmatrix} \\
& = \begin{pmatrix}
\mathbf{M}^{sh}_{t} & \mathbf{0} \\ \mathbf{0} & \sqrt{\mathbf{\Lambda}_t} 
\end{pmatrix}
\hspace{.5cm} (t > t_c)
\end{aligned}
\end{equation}
where $\mathbf{\Lambda}_t$ $\mathbf{O}_t$, $\mathbf{O}^T_0$, and $t_c$ are given in Sec.~\ref{ideal}. The complete transformations used for the chaotic case throughout the paper is finally,
\begin{equation}
\label{smatrix}
\mathbf{S}_{\alpha,0} \rightarrow \mathbf{O}_0\mathbf{S}_{\alpha,0}, \hspace{.5cm} \mathbf{S}_{\alpha,t} \rightarrow \mathbf{O}_t\mathbf{S}_{\alpha,t}
\end{equation}
For the integrable case this is not necessary and Eq.
~\eqref{matrix3} is sufficient.

\section{Convenient coordinates and the stability matrix}
\label{cc}

Under some circumstances, it may be useful to transform to convenient coordinate systems described here.  Given that the symmetric matrices $\mathbf{A}_\alpha$ and $\mathbf{A}_\beta$ are positive definite and unit determinant, they can be decomposed as
\begin{equation}
\label{decomp1}
\mathbf{A}_\alpha = \mathcal{A}^T_\alpha \mathcal{A}_\alpha
\end{equation}
and likewise for $\mathbf{A}_\beta$.  Choosing 
\begin{equation}
\label{decomp2}
\mathcal{A}_\alpha = \begin{pmatrix}
 \mathbf{a} &     \mathbf{a}  \mathbf{d} \\
 \mathbf{0} &  \left(\mathbf{a}^{-1}\right)^T
\end{pmatrix}
\end{equation}
where the definitions $ \mathbf{a}^T \mathbf{a}= \mathbf{c}^{-1}$ along with the other symmetric matrix $\mathbf{d}$ defined in Eq.~\eqref{symmatrix} guarantees the symplectic property
\begin{equation}
\mathcal{A}^T_\alpha \mathbf{\Omega} \mathcal{A}_\alpha = \mathbf{\Omega} \qquad\qquad \mathbf{\Omega} = \begin{pmatrix}
 \mathbf{0} &    - \mathbb{1} \\
 \mathbb{1} &   \mathbf{0}
\end{pmatrix}
\end{equation}
The matrices $\mathcal{A}_\alpha$ and $\mathcal{A}_\beta$ thus represent linear canonical transformations.  Denoting
\begin{equation}
    \begin{pmatrix} \delta \vec{p} \\ \delta \vec{q} \end{pmatrix}_\alpha = \begin{pmatrix} \vec{p} - \vec{p}_\alpha \\ \vec{q} - \vec{q}_\alpha \end{pmatrix}
\end{equation}
and similarly for $\beta$, the transformed primed coordinates
\begin{equation}
    \begin{pmatrix} \delta \vec{p} \\ \delta \vec{q} \end{pmatrix}_\alpha^\prime = \mathcal{A}_\alpha \cdot \begin{pmatrix} \delta \vec{p} \\ \delta \vec{q} \end{pmatrix}_\alpha
\end{equation}
when divided $\sqrt{\hbar}$ are unitless and measure distances in units of standard deviations.  The primed coordinates render the density a simpler, hyperspherical shape;
\begin{equation}
\rho_\alpha(\vec{p}^\prime, \vec{q}^\prime) = \left(\pi\hbar\right)^{-D}\exp\left[-{\begin{pmatrix} \delta \vec{p} \\ \delta \vec{q} \end{pmatrix}_\alpha^\prime}^T \cdot \frac{\mathbb{1}}{\hbar} \cdot \begin{pmatrix} \delta \vec{p} \\ \delta \vec{q} \end{pmatrix}_\alpha^\prime\right]
\end{equation}
and the phase space volume element is still the component product of the column (row) vector.

As the stability matrix is of principal concern, recall that it is given by
\begin{equation}
\begin{pmatrix}
\delta \vec{p}_t \\ \delta \vec{q}_t
\end{pmatrix}_\gamma = \mathbf{M}_{\gamma,t} \begin{pmatrix}
\delta \vec{p}_0 \\ \delta \vec{q}_0
\end{pmatrix}_\gamma = \begin{pmatrix} \mathbf{M_{11}} & \mathbf{M_{12}} \\ \mathbf{M_{21}} & \mathbf{M_{22}} \end{pmatrix}_{\gamma,t} \begin{pmatrix}
\delta \vec{p}_0 \\ \delta \vec{q}_0
\end{pmatrix}_\gamma
\end{equation}
where
\begin{equation}
\begin{pmatrix}
\delta \vec{p}_0 \\ \delta \vec{q}_0
\end{pmatrix}_\gamma = \begin{pmatrix} \vec{p}_0 -  \vec{p}_\gamma\\ \vec{q}_0 -  \vec{q}_\gamma
\end{pmatrix}
\end{equation}
and 
\begin{equation}
\begin{pmatrix}\delta \vec{p}_t \\ \delta \vec{q}_t \end{pmatrix}_\gamma = \begin{pmatrix}
 \vec{p}_t(\vec p_0, \vec q_0) -  \vec{p}_t(\vec p_\gamma, \vec q_\gamma) \\ \vec{q}_t(\vec p_0, \vec q_0) - \vec{q}_t(\vec p_\gamma, \vec q_\gamma)
\end{pmatrix} \ .
\end{equation}
The initial condition, $(\vec p_0, \vec q_0)$, is necessarily in the immediate neighborhood of a reference trajectory's initial condition, $(\vec p_\gamma, \vec q_\gamma)$; in fact, ahead evaluating Eq.~\eqref{wmexp}, both must also be found in the immediate neighborhood of $(\vec p_\alpha, \vec q_\alpha)$ or the contribution effectively vanishes. 

It is necessary to understand how the stability matrix transforms under this change of coordinates.  For the case of expectation values, it suffices to calculate an $\mathbf{M}_{\gamma,t}^\prime$ using the transformed Hamiltonian, $H(\vec p~^\prime, \vec q~^\prime)$.  Note the simplification in the case that the wave packet's symmetric matrix $\mathbf{b}_\alpha$ is real. Using the decomposition defined in Eq.~\eqref{decomp2}, $\mathbf{b}_\alpha = \mathbf{a}^{-1} \left(\mathbf{a}^{-1}\right)^T$, the transformed Hamiltonian is given by $H(\vec p~^\prime, \vec q~^\prime) = H\left(\left[\mathbf{a}^{-1}\right]^T\cdot \vec p, \mathbf{a}\cdot\vec q\right)$.

This approach is not optimal for transport coefficients in which the local coordinates near the initial and final densities are, generally speaking, distinct.  Accounting for the fact, that no trajectory leads to a contribution if its initial condition is not sufficiently close to $(\vec p_\alpha, \vec q_\alpha)$ and its final point at time $t$ sufficiently close to $(\vec p_\beta, \vec q_\beta)$, it is reasonable to consider primed coordinates
\begin{equation}
    \begin{pmatrix} \delta \vec{p}_0 \\ \delta \vec{q}_0 \end{pmatrix}_\gamma^\prime = \mathcal{A}_\alpha \cdot \begin{pmatrix} \delta \vec{p}_0 \\ \delta \vec{q}_0 \end{pmatrix}_\gamma \qquad    
    \begin{pmatrix} \delta \vec{p}_t \\ \delta \vec{q}_t \end{pmatrix}_\gamma^\prime = \mathcal{A}_\beta \cdot \begin{pmatrix} \delta \vec{p}_t \\ \delta \vec{q}_t \end{pmatrix}_\gamma
\end{equation}
The primed stability matrix can then be calculated as
\begin{equation}
\label{mprime}
\mathbf{M}'_{\gamma,t} = {\cal A}_{\beta} \mathbf{M}_{\gamma,t} {\cal A}^{-1}_{\alpha}
\end{equation}
where $\mathbf{M}_{\gamma,t}$ is calculated in the original coordinates without transforming the Hamiltonian, and the second relation holds in the case of real $\mathbf{b}_\alpha$ and $\mathbf{b}_\beta$.

\section{Formulas for the integrable case $(\lambda = 0)$}
\label{2approx}

Here the formulas needed to evaluate the return probabilities of the integrable case are derived. Since all of the coordinates have been pre-integrated, Eq.~\eqref{final_tc2} immediately simplifies as all the coordinates associated with the $j$ directions can be set to zero. Furthermore, the index on the matrices are now unnecessary, since they encompass the dimensionality of the entire system. The approximation, Eq.\eqref{final_tc2}, for the case of return probabilities, then reduces to 
\begin{widetext}
\begin{align}
\label{int_sp}
{\cal C}_{\alpha\alpha}(t) & =\sum_{i_1}\sum_{i_2} \frac{4}{\sqrt{\text{Det}(\mathbf{1} +\mathbf{A^{'-1}_{\alpha}}\mathbf{M}_{\tau_{i_1,i_2}}^T\mathbf{A'_{\alpha}}\mathbf{M}_{\tau_{i_1,i_2}})}}  \ \times \nonumber \\[7pt]
& \exp\left[- (\delta \vec{X}_{\alpha,\tau_{i_1,i_2}} + \vec{1}(t-\tau_{i_1,i_2}))^T \cdot \left( \frac{\mathbf{A'_{\alpha}}-\mathbf{A'_{\alpha}}\cdot \mathbf{M}_{\tau_{i_1,i_2}} \cdot\mathbf{A^{-1}_{\alpha\alpha}}\cdot \mathbf{M}^T_{\tau_{i_1,i_2}}\mathbf{A'_{\alpha}}}{\hbar} \right) \cdot (\delta\vec{X}_{\alpha,\tau_i} + \vec{1}(t-\tau_{i_1,i_2})) \right] \
\end{align}
\end{widetext}

To apply this formula the vector $\delta \vec{X}_{\alpha,\tau_{i_1,i_2}}$, the stability matrix $\mathbf{M}_{\tau_{i_1,i_2}}$, and $\tau_{i_1,i_2}$ need to be known. For starters the quantity $\delta \vec{X}_{\alpha,\tau_{i_1,i_2}} + \vec{1}(t - \tau_{i_1,i_2})$ can be related to the desired local coordinate system, worked out in Appendix~\ref{transformation}, given by Eq.~\eqref{matrix2}
\begin{equation}
\label{vector}
\delta \vec{X}_{\alpha,\tau_{i_1,i_2}} + \vec{1}(t - \tau_{i_1,i_2}) =
\begin{pmatrix} \delta E \\ \delta t \\ \delta P \\ \delta Q \end{pmatrix}
\end{equation}
The component associated with the time coordinate is just $\delta t = (t - \tau_{i_1,i_2})$, which leads to setting the component of $\delta \vec{X}_{\alpha,\tau_{i_1,i_2}}$ along the time direction to zero. The component associated with energy coordinate is also zero, since $\delta E = E - E_\alpha = 0$ when $E$ is evaluated on the $\alpha$ centroid. The last two components can be found using Eq.~\eqref{qp},
\begin{align}
\delta P = & \frac{E_2}{E}\delta E_1 - \frac{E_1}{E}\delta E_2 \\[6pt]
\delta Q = & -\delta t_1 + \delta t_2 \nonumber
\end{align}
Here $\delta E_1 = E_1 - E_{1,\alpha}$ and $E_1$ is the energy of the uncoupled quartic oscillator, similarly for $\delta E_2$. $\delta P$ is then zero because the energy is being evaluated on the centroid $\alpha$. Continuing, $\delta t_1 = t - \tau_1i_1$ where $\tau_1$ is the period of first uncoupled oscillator evaluated on the central trajectory. This gives $\delta Q = -\tau_2i_2 + \tau_1i_1$. To find $\tau_{i_1,i_2}$ note that, from Eq. (\ref{time}),
\begin{align}
\delta t &{} = \frac{E_1}{E}\delta t_1 + \frac{E_2}{E}\delta t_2 = t - \frac{E_1}{E}\tau_1i_1 - \frac{E_2}{E}\tau_2i_2 \\[7pt]
\label{tau12}
& = t - \tau_{i_1,i_2} \rightarrow \tau_{i_1,i_2} = \frac{E_1}{E}\tau_1i_1 + \frac{E_2}{E}\tau_2i_2.
\end{align}
The components of the vector defined by (\ref{vector}) are then,
\begin{equation}
\delta \vec{X}_{\alpha,\tau_{i_1,i_2}} + \vec{1}(t - \tau_{i_1,i_2}) = \begin{pmatrix} 0 \\ t - \tau_{i_1,i_2} \\ 0 \\ -\tau_2i_2 + \tau_1i_1 \end{pmatrix}
\end{equation}
The stability matrix in this coordinate system is given by transforming the stability matrix in the original $(\vec{p}_\alpha,\vec{q}_\alpha)$ coordinate system, using Eq.~\eqref{lh}, here denoted $\mathbf{M}_t$,
\begin{equation}
\label{stability3}
\mathbf{M}_{t}= \begin{pmatrix} 1 & 0 & 0 & 0 \\ \omega_{\alpha,1}'t & 1 & 0& 0 \\ 0 & 0 & 1 & 0 \\ 0 & 0 & \omega'_{\alpha,2} t & 1\end{pmatrix} 
\end{equation}
Note these are not the shearing rates of the uncoupled oscillators. To find them, the stability matrix has to be numerically found in the $(\vec{p}_{\alpha}, \vec{q}_{\alpha})$ coordinate system and then transformed to the new coordinate system using Eq.~\eqref{matrix3}.  These are all the formulas needed in the main text needed to produce Fig.~\ref{fig6}. 

\section{Return probabilities: initial decay}
\label{initdec}

For diagonal transport coefficients, i.e., return (survival) probabilities for which $(\vec p_\alpha, \vec q_\alpha) = (\vec p_\beta, \vec q_\beta)$ and $\mathbf{A_\alpha} = \mathbf{A_\beta}$, the coefficient, ${\cal C}_{\alpha\alpha}(t)$, from Eq.~\eqref{wmtc3} starts at unity and decays.  This is such a short time process that all $2D$ integrals can be performed to an excellent approximation and it is unnecessary to enter into dynamical considerations of stability or instability.  Neither is it necessary to consider coordinate transformations such as $\mathbf{S_\alpha}$.  All that is needed is a single appropriate reference trajectory for any given fixed time.

An excellent approximation arises by mimicking to some extent that which happens in a saddle point approximation of $|\braket{\alpha | \alpha(t)}|^2$. At a given time $t$ during the initial decay, the real trajectory that most closely resembles the complex saddle point trajectory is one which arrives at the phase space point $(\vec p_\alpha, \vec q_\alpha)$ at $t/2$.  That gives 
\begin{align}
\label{diffdefs}
\delta \vec{X}_{\alpha,f} & = \begin{pmatrix}
\vec{p}_f -\vec{p}_\alpha \\
\vec{q}_f -\vec{q}_\alpha
\end{pmatrix} = \begin{pmatrix}
\vec{p}_\alpha(t/2) -\vec{p}_\alpha \\
\vec{q}_\alpha(t/2) -\vec{q}_\alpha
\end{pmatrix} \nonumber \\
\delta \vec{X}_{\alpha,i} & = \begin{pmatrix}
\vec{p}_i - \vec{p}_\alpha\\
\vec{q}_i - \vec{q}_\alpha
\end{pmatrix} = \begin{pmatrix}
\vec{p}_\alpha(-t/2) -\vec{p}_\alpha \\
\vec{q}_\alpha(-t/2) -\vec{q}_\alpha
\end{pmatrix}
\end{align}
In the $\hbar\rightarrow 0$ limit, for which the derivatives in Hamilton's equations are constant for the duration of the decay, the initial condition and final endpoint end up approximately equidistant from $(\vec p_\alpha, \vec q_\alpha)$, but in opposite directions. In order to evaluate the integrals for ${\cal C}_{\alpha\alpha}(t)$, it is necessary to consider initial conditions in the immediate neighborhood of the reference trajectory.  Defining the variations as follows gives
\begin{align}
\delta \vec{X}_0 & = \begin{pmatrix}
\vec{p}_0 -\vec{p}_i \\
\vec{q}_0 -\vec{q}_i
\end{pmatrix} \nonumber \\
\delta \vec{X}_t & = \begin{pmatrix}
\vec{p}_t -\vec{p}_f \\
\vec{q}_t -\vec{q}_f
\end{pmatrix} 
\end{align}
where the full $2D$-dimensional stability matrix relates the initial and final variations,
\begin{equation}
\delta \vec{X}_t  = \mathbf{M}_t \cdot \delta \vec{X}_0\ .
\end{equation}
Ahead, either the numerically calculated $\mathbf{M}_t$ can be used or its short time approximation
\begin{align}
\mathbf{M}_t & \approx  \mathbb{1} + \mathbf{K_\alpha} t \nonumber\\
 & = \mathbb{1} + \left.\begin{pmatrix}
 -\frac{\partial^2 H}{\partial\vec{q}\partial\vec{p} } & 
 -\frac{\partial^2 H}{\partial\vec{q}\partial\vec{q} } \nonumber \\
\frac{\partial^2 H}{\partial\vec{p}\partial\vec{p} } & 
  \frac{\partial^2 H}{\partial\vec{q}\partial\vec{p} } \end{pmatrix}\right|_{\vec{p}_\alpha,\vec{q}_\alpha} \times t 
\end{align}
With this reference trajectory and the variation definitions, locally, the integration differential can be shifted to $d\delta \vec{X}_i$.  Now, the arguments of the two Gaussian exponentials, not including the inverse of $-\hbar$, can be expressed as
\begin{equation}
(\delta \vec{X}_0+\delta \vec{X}_{\alpha,i})^T \cdot {\bf A}_\alpha \cdot (\delta \vec{X}_0+\delta \vec{X}_{\alpha,i}) \end{equation}
for the initial $\rho_\alpha$ and
\begin{equation}
(\delta \vec{X}_t+\delta \vec{X}_{\alpha,f})^T \cdot {\bf A}_\alpha \cdot (\delta \vec{X}_t+\delta \vec{X}_{\alpha,f}) \end{equation}
for the final $\rho_\alpha$. Using the stability matrix, the quadratic term becomes:
\begin{align}
    \delta \vec{X}_0^T \cdot (\mathbf{A_\alpha} + {\bf M}_t^T  \cdot \mathbf{A_\alpha} \cdot \mathbf{M_t})\cdot \delta \vec{X}_0
\end{align}
and the linear term,
\begin{align}
    & 2(\delta  \vec{X}_{\alpha,i} \cdot \mathbf{A_\alpha} +\mathbf{M}^T_t \cdot \mathbf{A_\alpha} \cdot \delta \vec{X}_{\alpha, f})^T \cdot \delta \vec{X}_{0} 
\end{align}
as well two terms quadratic in $\delta \vec{X}_{\alpha,i}$ and $\delta \vec{X}_{\alpha,f}$,
\begin{align}
    \delta \vec{X}^T_{\alpha,i }\cdot \mathbf{A}_\alpha \cdot \vec{X}_{\alpha, i} + \delta \vec{X}^T_{\alpha,f}\cdot \mathbf{A}_{\alpha} \cdot \vec{X}_{\alpha, f}
\end{align}
The linear terms in $\delta \vec{X}_0$ can be accounted for by the shift
\begin{widetext}
\begin{align}
\delta \vec{X}_0 \rightarrow \delta \vec{X}_0  + (\mathbf{A}_{\alpha} + \mathbf{M}^T_t \cdot \mathbf{A}_{\alpha} \cdot \mathbf{M}_t)^{-1} \cdot (\delta  \vec{X}_{\alpha,i} \cdot \mathbf{A_\alpha} +\mathbf{M}^T_t \cdot \mathbf{A_\alpha} \cdot \delta \vec{X}_{\alpha, f})
\end{align}
\end{widetext}
The leading quadratic terms in $\delta \vec{X}_0$ determine the prefactor, which is given by
\begin{equation}
{\cal A} = \frac{2^D}{\left[\text{Det}\left( {\bf A}_\alpha  + {\bf M}_t^T \cdot {\bf A}_\alpha \cdot {\bf M}_t \right)\right]^{1/2}}
\end{equation}
and which gives unity as it must in the limit of $t\rightarrow 0$ where ${\bf M}_t = \mathbb{1}$.
Performing the integration gives four terms that can be simplified by using the Woodbury matrix identity, such that the final integral can be expressed as 
\begin{equation} 
\label{eq:TWA2}
{\cal C}_{\alpha\alpha}(t) =  {\cal A} \exp\left(-\frac{1}{\hbar}{\cal S} \right)
\end{equation}
with 
\begin{align}
    {\cal S} &= \Big{[}\delta \vec{X}^T_{\alpha, f} - \delta{X}^T_{\alpha, i} \cdot \mathbf{M}_t^T \Big{]} \cdot \\ \nonumber
    &\Big{(}\mathbf{A}^{-1}_{\alpha} + \mathbf{M}_t\cdot \mathbf{A}^{-1}_\alpha \cdot \mathbf{M}^T_t\Big{)}^{-1} \cdot \Big{[}\delta \vec{X}_{\alpha, f} - \mathbf{M}_t \cdot \delta{X}_{\alpha, i} \Big{]} 
\end{align}

\bibliographystyle{abbrv}
\bibliography{general_ref,classicalchaos,extreme,furtherones,manybody,molecular,nano,oceanacoustics,quantumchaos,rmtmodify}

\bibliographystyle{unsrt}
\end{document}